\documentclass[12pt]{article}
\usepackage{amssymb,amscd,array}
\catcode `\@=11
\@addtoreset{equation}{section}

\newtheorem{thm}{Theorem}[section]

\newtheorem{prop}[thm]{Proposition}

\def\qed{\blacksquare}
\newcommand{\be}{\begin{equation}}
\newcommand{\ee}{\end{equation}}
\newcommand{\bea}{\begin{eqnarray}}
\newcommand{\eea}{\end{eqnarray}}
\newcommand{\R}{\mathbb{R}}

\newcommand{\C}{\mathbb{C}}

\begin{document}
\begin{titlepage}

\begin{center}
{\bf \Large{Wick Theorem and Hopf Algebra Structure in Causal Perturbative Quantum Field Theory \\}}
\end{center}
\vskip 1.0truecm
\centerline{D. R. Grigore, 
\footnote{e-mail: grigore@theory.nipne.ro}}
\vskip5mm
\centerline{Department of Theoretical Physics,}
\centerline{Institute for Physics and Nuclear Engineering ``Horia Hulubei"}
\centerline{Bucharest-M\u agurele, P. O. Box MG 6, ROM\^ANIA}

\vskip 2cm
\bigskip \nopagebreak
\vskip 1cm
\begin{abstract}
\noindent
We consider the general framework of perturbative quantum field theory for the pure Yang-Mills model. We give a more precise version of the 
Wick theorem using Hopf algebra notations for chronological products and not for Feynman graphs. Next we prove that Wick expansion 
property can be preserved for all cases in order 
$
n = 2.
$
However, gauge invariance is broken for chronological products of Wick submonomials.
\end{abstract}
%\newpage\setcounter{page}1

\end{titlepage}

\section{Introduction}

The most natural way to arrive at the Bogoliubov axioms of perturbative quantum field theory (pQFT) is by analogy with non-relativistic 
quantum mechanics \cite{Gl}, \cite{H}, \cite{D}, \cite{DF}: in this way one arrives naturally at Bogoliubov axioms 
\cite{BS}, \cite{EG}, \cite{Sc1}, \cite{Sc2}. We prefer the formulation from \cite{DF} and as presented in \cite{algebra}; 
for every set of monomials 
$ 
A_{1}(x_{1}),\dots,A_{n}(x_{n}) 
$
in some jet variables (associated to some classical field theory) one associates the operator-valued distributions
$ 
T^{A_{1},\dots,A_{n}}(x_{1},\dots,x_{n})
$  
called chronological products; it will be convenient to use another notation: 
$ 
T(A_{1}(x_{1}),\dots,A_{n}(x_{n})). 
$ 

The Bogoliubov axioms, presented in Section \ref{Bogoliubov} express essentially some properties of the scattering matrix understood as a 
formal perturbation
series with the ``coefficients" the chronological products: 
\begin{itemize}
\item 
(skew)symmetry properties in the entries 
$ 
A_{1}(x_{1}),\dots,A_{n}(x_{n}) 
$;
\item
Poincar\'e invariance; 
\item
causality; 
\item
unitarity; 
\item
the ``initial condition" which says that
$
T(A(x)) 
$
is a Wick monomial.
\end{itemize}

So we need some basic notions on free fields and Wick monomials. One can supplement these axioms by requiring 
\begin{itemize}
\item 
power counting;
\item
Wick expansion property. 
\end{itemize}

It is a highly non-trivial problem to find solutions for the Bogoliubov axioms, even in the simplest case of a real scalar field. 

There are, at least to our knowledge, three rigorous ways to do that; for completeness we remind them following \cite{ano-free}:
\begin{itemize}
\item 
{\it Hepp axioms} \cite{H};
\item
{\it Polchinski flow equations} \cite{P}, \cite{S};
\item
{\it the causal approach} due to Epstein and Glaser \cite{EG}, \cite{Gl} which we prefer. 
\end{itemize}

The procedure of Epstein and Glaser is a recursive construction for the basic objects
$ 
T(A_{1}(x_{1}),\dots,A_{n}(x_{n}))
$
and reduces the induction procedure to a distribution splitting of some distributions with causal support.  
In an equivalent way, one can reduce the induction procedure to the process of extension of distributions \cite{PS}. 

An equivalent point of view uses retarded products \cite{St1} instead of chronological products. For gauge models one has to deal with 
non-physical fields (the so-called ghost fields) and impose a supplementary axiom namely  gauge invariance, which guarantees that the 
physical states are left invariant by the chronological products.

In the next section we give the main prerequisites we need, following essentially \cite{algebra}. Then will present a more 
precise version of the Wick theorem and emphasize the Hopf structure in Section \ref{wick thm}. This Hopf structure differs 
from the Hopf structure introduced in \cite{K1} which is based on the forest formula of Zimmermann. Our Hopf structure is
based only on the basic chronological products and it first appeared in \cite{ward}. Similar Hopf structure appeared in \cite{BFFO} 
and \cite{B}.

We will apply these results to the pure YM model in Section \ref{submonomials}. Kreimer version of Hopf algebra for gauge models 
was developped in \cite{K2}, \cite{K3}, \cite{Su1}, \cite{Su2}.

Next we discuss in detail the simplest case of 
$
n = 2
$
for the pure YM model. We will study the basic chronological products of the type
$
T(T(x_{1}),T(x_{2})) 
$
where 
$
T(x)
$
is the interaction Lagrangian, but also chronological products of the type
$
T(A_{1}(x_{1}),A_{2}(x_{2}))
$
where 
$
A_{1}, A_{2}
$
are Wick submonomials.

All these chronological products can be split in the loop and tree contributions. 
First, we prove in Section \ref{loop} that the loop contributions do not produce anomalies. This point can be cleared by direct 
computations. Then, in Section \ref{tree} we investigate the tree contributions. As it is known, they produce anomalies; 
for the basic chronological products of the type
$
T(T(x_{1}),T(x_{2})) 
$ 
these anomalies can be removed by finite renormalizations, but we will prove that this is not true for the associated chronological 
products of the type
$
T(A_{1}(x_{1}),A_{2}(x_{2})). 
$
However, it is important to establish if there is a clever finite renormalization saving gauge invariance for the basic chronological products of the type
$
T(T(x_{1}),T(x_{2})) 
$ 
and Wick expansion property also. This point can be proved in a very compact way starting from a redefinition of the chronological
product
$
T(v_{a\mu,\nu}(x_{1}),v_{b\rho,\sigma}(x_{1}))
$
where 
$
v_{a\mu}
$
are some jet variable for the vector fields describing the gluons and 
$
v_{a\mu,\nu}
$
the associated derivative jet variables. This idea appeared in \cite{DKS} and \cite{ASD}.
\newpage
\section{Perturbative Quantum Field Theory\label{pQFT}}
There are two main ingrediants in the contruction of a perturbative quantum field theory (pQFT): the construction of the Wick monomials 
and the Bogoliubov axioms. For a pQFT of Yang-Mills theories one needs one more ingrediant, namely the introduction of ghost fields and
gauge charge.

\subsection{Wick Products\label{wick prod}}

We follow the formalism from \cite{algebra}. We consider a classical field theory on the Minkowski space
$
{\cal M} \simeq \R^{4}
$
(with variables
$
x^{\mu}, \mu = 0,\dots,3
$
and the metric $\eta$ with 
$
diag(\eta) = (1,-1,-1,-1)
$)
described by the Grassmann manifold 
$
\Xi_{0}
$
with variables
$
\xi_{a}, a \in {\cal A}
$
(here ${\cal A}$ is some index set) and the associated jet extension
$
J^{r}({\cal M}, \Xi_{0}),~r \geq 1
$
with variables 
$
x^{\mu},~\xi_{a;\mu_{1},\dots,\mu_{n}},~n = 0,\dots,r;
$
we denote generically by
$
\xi_{p}, p \in P
$
the variables corresponding to classical fields and their formal derivatives and by
$
\Xi_{r}
$
the linear space generated by them. The variables from
$
\Xi_{r}
$
generate the algebra
$
{\rm Alg}(\Xi_{r})
$
of polynomials.

To illustrate this, let us consider a real scalar field in Minkowski space ${\cal M}$. The first jet-bundle extension is
$$
J^{1}({\cal M}, \R) \simeq {\cal M} \times \R \times \R^{4}
$$
with coordinates 
$
(x^{\mu}, \phi, \phi_{\mu}),~\mu = 0,\dots,3.
$

If 
$
\varphi: \cal M \rightarrow \R
$
is a smooth function we can associate a new smooth function
$
j^{1}\varphi: {\cal M} \rightarrow J^{1}(\cal M, \R) 
$
according to 
$
j^{1}\varphi(x) = (x^{\mu}, \varphi(x), \partial_{\mu}\varphi(x)).
$

For higher order jet-bundle extensions we have to add new real variables
$
\phi_{\{\mu_{1},\dots,\mu_{r}\}}
$
considered completely symmetric in the indexes. For more complicated fields, one needs to add supplementary indexes to
the field i.e.
$
\phi \rightarrow \phi_{a}
$
and similarly for the derivatives. The index $a$ carries some finite dimensional representation of
$
SL(2,\C)
$
(Poincar\'e invariance) and, maybe a representation of other symmetry groups. 
In classical field theory the jet-bundle extensions
$
j^{r}\varphi(x)
$
do verify Euler-Lagrange equations. To write them we need the formal derivatives defined by
\be
d_{\nu}\phi_{\{\mu_{1},\dots,\mu_{r}\}} \equiv \phi_{\{\nu,\mu_{1},\dots,\mu_{r}\}}.
\ee

We suppose that in the algebra 
$
{\rm Alg}(\Xi_{r})
$
generated by the variables 
$
\xi_{p}
$
there is a natural conjugation
$
A \rightarrow A^{\dagger}.
$
If $A$ is some monomial in these variables, there is a canonical way to associate to $A$ a Wick 
monomial: we associate to every classical field
$
\xi_{a}, a \in {\cal A}
$
a quantum free field denoted by
$
\xi^{\rm quant}_{a}(x), a \in {\cal A}
$
and determined by the $2$-point function
\be
<\Omega, \xi^{\rm quant}_{a}(x), \xi^{\rm quant}_{b}(y) \Omega> = - i~D_{ab}^{(+)}(x - y)\times {\bf 1}.
\label{2-point}
\ee
Here 
\be
D_{ab}(x) = D_{ab}^{(+)}(x) + D_{ab}^{(-)}(x)
\ee
is the causal Pauli-Jordan distribution associated to the two fields; it is (up to some numerical factors) a polynomial
in the derivatives applied to the Pauli-Jordan distribution. We understand by 
$
D^{(\pm)}_{ab}(x)
$
the positive and negative parts of
$
D_{ab}(x)
$.
From (\ref{2-point}) we have
\be
[ \xi_{a}(x), \xi_{b}(y) ] = - i~ D_{ab}(x - y) \times {\bf 1} 
\ee
where by 
$
[\cdot, \cdot ]
$
we mean the graded commutator. 

The $n$-point functions for
$
n \geq 3
$
are obtained assuming that the truncated Wightman functions are null: see \cite{BLOT}, relations (8.74) and (8.75) and proposition 8.8
from there. The definition of these truncated Wightman functions involves the Fermi parities
$
|\xi_{p}|
$
of the fields
$
\xi_{p}, p \in P.
$

Afterwards we define
$$
\xi^{\rm quant}_{a;\mu_{1},\dots,\mu_{n}}(x) \equiv \partial_{\mu_{1}}\dots \partial_{\mu_{n}}\xi^{\rm quant}_{a}(x), a \in {\cal A}
$$
which amounts to
\be
[ \xi_{a;\mu_{1}\dots\mu_{m}}(x), \xi_{b;\nu_{1}\dots\nu_{n}}(y) ] =
(-1)^{n}~i~\partial_{\mu_{1}}\dots \partial_{\mu_{m}}\partial_{\nu_{1}}\dots \partial_{\nu_{n}}D_{ab}(x - y )\times {\bf 1}.
\label{2-point-der}
\ee
More sophisticated ways to define the free fields involve the GNS construction. 

The free quantum fields are generating a Fock space 
$
{\cal F}
$
in the sense of the Borchers algebra: formally it is generated by states of the form
$
\xi^{\rm quant}_{a_{1}}(x_{1})\dots \xi^{\rm quant}_{a_{n}}(x_{n})\Omega
$
where 
$
\Omega
$
the vacuum state.
The scalar product in this Fock space is constructed using the $n$-point distributions and we denote by
$
{\cal F}_{0} \subset {\cal F}
$
the algebraic Fock space.

One can prove that the quantum fields are free, i.e.
they verify some free field equation; in particular every field must verify Klein Gordon equation for some mass $m$
\be
(\square + m^{2})~\xi^{\rm quant}_{a}(x) = 0
\label{KG}
\ee
and it follows that in momentum space they must have the support on the hyperboloid of mass $m$. This means that 
they can be split in two parts
$
\xi^{\rm quant (\pm)}_{a}
$
with support on the upper (resp. lower) hyperboloid of mass $m$. We convene that 
$
\xi^{\rm quant (+)}_{a} 
$
resp.
$
\xi^{\rm quant (-)}_{a} 
$
correspond to the creation (resp. annihilation) part of the quantum field. The expressions
$
\xi^{\rm quant (+)}_{p} 
$
resp.
$
\xi^{\rm quant (-)}_{p} 
$
for a generic
$
\xi_{p},~ p \in P
$
are obtained in a natural way, applying partial derivatives. For a general discussion of this method of constructing free fields, see 
ref. \cite{BLOT} - especially prop. 8.8.
The Wick monomials are leaving invariant the algebraic Fock space.
%\newpage
The definition for the Wick monomials is contained in the following Proposition.

\begin{prop}
The operator-valued distributions
$
N(\xi_{q_{1}}(x_{1}),\dots,\xi_{q_{n}}(x_{n}))
$
are uniquely defined by:

\be
N(\xi_{q_{1}}(x_{1}),\dots,\xi_{q_{n}}(x_{n}))\Omega = \xi_{q_{1}}^{(+)}(x_{1})\dots  \xi_{q_{n}}^{(+)}(x_{n})\Omega
\ee

\bea
[ \xi_{p}(y), N(\xi_{q_{1}}(x_{1}),\dots,\xi_{q_{n}}(x_{n})) ]
\nonumber\\
= \sum_{m=1}^{n} \prod_{l <m} (-1)^{|\xi_{p}||\xi_{q_{l}}|}~[ \xi_{p}(y), \xi_{q_{m}}(x_{m})] 
~N(\xi_{q_{1}}(x_{1}),\dots,\hat{m},\dots,\xi_{q_{n}}(x_{n}))
\nonumber\\
= - i~\sum_{m=1}^{n} \prod_{l <m} (-1)^{|\xi_{p}||\xi_{q_{l}}|}~D_{pq_{m}}(y - x_{m})
~N(\xi_{q_{1}}(x_{1}),\dots,\hat{m},\dots,\xi_{q_{n}}(x_{n}))
\eea

\be
N(\emptyset) = I.
\ee

The expression
$
N(\xi_{q_{1}}(x_{1}),\dots,\xi_{q_{n}}(x_{n}))
$
is (graded) symmetrical in the arguments.
\end{prop}

The expressions
$
N(\xi_{q_{1}}(x_{1}),\dots,\xi_{q_{n}}(x_{n}))
$
are called {\it Wick monomials}. There is an alternative definition based on the splitting of the fields into the creation and annihilation 
part for which we refer to \cite{algebra}.

It is a non-trivial result of Wightman and G\aa rding \cite{WG} that in
$
N(\xi_{q_{1}}(x_{1}),\dots,\xi_{q_{n}}(x_{n}))
$
one can collapse all variables into a single one and still gets an well-defined expression: 
\begin{prop}
The expressions
\be
W_{q_{1},\dots,q_{n}}(x) \equiv N(\xi_{q_{1}}(x),\dots,\xi_{q_{n}}(x))
\ee
are well-defined. They verify:

\be
W_{q_{1},\dots,q_{n}}(x) \Omega = \xi_{q_{1}}^{(+)}(x)\dots  \xi_{q_{n}}^{(+)}(x)\Omega
\ee

\bea
[ \xi^{(\epsilon)}_{p}(y), W_{q_{1},\dots,q_{n}}(x)  ] = 
%\nonumber\\
- i~\sum_{m=1}^{n} \prod_{l <m} (-1)^{|\xi_{p}||\xi_{q_{l}}|}~D_{pq_{m}}^{(-\epsilon)}(y - x_{m})~W_{q_{1},\dots,\hat{m},\dots,q_{n}}(x)
\eea

\be
W(\emptyset) = I.
\ee
\end{prop}
We call expressions of the type
$
W_{q_{1},\dots,q_{n}}(x) 
$
{\it Wick monomials}.  By
\be
|W| \equiv \sum_{l=1}^{n} |\xi_{q_{l}}|
\ee
we mean the Fermi number of $W$. We define the derivative
\be
{\partial \over \partial \xi_{p}}W_{q_{1},\dots,q_{n}}(x) \equiv 
\sum_{s=1}^{n}~\prod_{l < s}~(-1)^{|\xi_{p}||\xi_{q_{l}}|}~\delta_{pq_{s}}~W_{q_{1},\dots,\hat{q_{s}},\dots,q_{n}}(x)
\ee
and we have a generalization of the preceding Proposition.
\begin{prop}
Let
$
W_{j} = W_{q^{(j)}_{1},\dots,q^{(j)}_{r_{j}}},~j = 1,\dots,n
$
be Wick monomials. Then the expression
$
N(W_{1}(x_{1}),\dots,W_{n}(x_{n}))
$
is well-defined through

\be
N(W_{1}(x_{1}),\dots,W_{n}(x_{n}))\Omega = \prod_{j=1}^{n} \prod_{l=1}^{r_{j}} \xi_{q^{(j)}_{l}}^{(+)}(x_{j})\Omega
\ee

\bea
[ \xi_{p}(y), N(W_{1}(x_{1}),\dots,W_{n}(x_{n}))  ] = 
\nonumber\\
- i~\sum_{m=1}^{n} \prod_{l <m} (-1)^{|\xi_{p}||W_{l}|}~\sum_{q}~D_{pq}(y - x_{m})~
N(W_{1}(x_{1}),\dots,{\partial \over \partial \xi_{q}}W_{m}(x_{m}),\dots,W_{n}(x_{n}))
\label{normal}
\eea

\be
N(W_{1}(x_{1}),\dots,W_{n}(x_{n}),{\bf 1}) = N(W_{1}(x_{1}),\dots,W_{n}(x_{n})) 
\ee

\be
N(W(x)) = W(x).
\ee

The expression
$
N(W_{1}(x_{1}),\dots,W_{n}(x_{n}))
$
is symmetric (in the Grassmann sense) in the entries
$
W_{1}(x_{1}),\dots,W_{n}(x_{n}).
$
\label{N}
\end{prop}

One can prove that 
\be
[ N(A(x)), N(B(y)) ] = 0,\quad (x - y)^{2} < 0
\ee
where by
$
[ \cdot,\cdot]
$
we mean the graded commutator. This is the most simple case of causal support property.
Now we are ready for the most general setting. We define for any monomial
$
A \in {\rm Alg}(\Xi_{r})
$
the derivation
\be
\xi \cdot A \equiv (-1)^{|\xi| |A|}~{\partial \over \partial \xi}A
\label{derivative}
\ee
for all
$
\xi \in \Xi_{r}.
$
Here 
$|A|$ 
is the Fermi parity of $A$ and we consider the left derivative in the Grassmann sense. So, the product $\cdot$ is defined as an map
$
\Xi_{r} \times {\rm Alg}(\Xi_{r}) \rightarrow {\rm Alg}(\Xi_{r}).
$
An expression
$
E(A_{1}(x_{1}),\dots,A_{n}(x_{n}))
$
is called {\it of Wick type} iff verifies:

\bea
[ \xi_{p}(y), E(A_{1}(x_{1}),\dots,A_{n}(x_{n}))  ]
\nonumber\\
= \sum_{m=1}^{n} \prod_{l \leq m} (-1)^{|\xi_{p}||A_{l}|}~\sum_{q}~[ \xi_{p}(y), \xi_{q}(x_{m})]~
E(A_{1}(x_{1}),\dots,\xi_{q}\cdot A_{m}(x_{m}),\dots,A_{n}(x_{n}))
\nonumber\\
= - i~\sum_{m=1}^{n} \prod_{l \leq m} (-1)^{|\xi_{p}||A_{l}|}~\sum_{q}~D_{pq}(y - x_{m})~
E(A_{1}(x_{1}),\dots,\xi_{q}\cdot A_{m}(x_{m}),\dots,A_{n}(x_{n}))
\label{comm-wick}
\eea

\be
E(A_{1}(x_{1}),\dots,A_{n}(x_{n}),{\bf 1}) = E(A_{1}(x_{1}),\dots,A_{n}(x_{n})) 
\ee

\be
E(1) = {\bf 1}.
\ee
%\newpage

The expression
$
N(W_{1}(x_{1}),\dots,W_{n}(x_{n}))
$
from  Proposition \ref{N} is of Wick type. Then we easily have:
\begin{prop}
If
$
E(A_{1}(x_{1}),\dots,A_{k}(x_{k}))
$
and
$
F(A_{k+1}(x_{1}),\dots,A_{n}(x_{n}))
$
are expressions of Wick type, then
$
E(A_{1}(x_{1}),\dots,A_{k}(x_{k}))~F(A_{k+1}(x_{1}),\dots,A_{n}(x_{n}))
$
is also an expression of Wick type.
\end{prop}

We can also prove:
\begin{thm}
The following formula is true:
\bea
N(\xi_{p}(y),A_{1}(x_{1}),\dots,A_{n}(x_{n})) 
\nonumber\\
= \xi^{(+)}_{p}(y)~N(A_{1}(x_{1}),\dots,A_{n}(x_{n}))
+ \prod_{l \leq n} (-1)^{|\xi_{p}||A_{l}|}~N(A_{1}(x_{1}),\dots,A_{n}(x_{n}))~\xi^{(-)}_{p}(y)
\nonumber\\
+ i~\sum_{m=1}^{n} \prod_{l <m} (-1)^{|\xi_{p}||A_{l}|}~\sum_{q}~D^{(+)}_{pq}(y - x_{m})~
N(A_{1}(x_{1}),\dots,\xi_{q}\cdot A_{m}(x_{m}),\dots,A_{n}(x_{n}))
\eea
\end{thm}

\newpage
\subsection{Bogoliubov Axioms \label{Bogoliubov}}
Suppose the monomials
$
A_{1},\dots,A_{n} \in {\rm Alg}(\Xi_{r})
$
are self-adjoint:
$
A_{j}^{\dagger} = A_{j},~\forall j = 1,\dots,n
$
and of Fermi number
$
f_{i}.
$

The chronological products
$$ 
T(A_{1}(x_{1}),\dots,A_{n}(x_{n})) \equiv T^{A_{1},\dots,A_{n}}(x_{1},\dots,x_{n}) \quad n = 1,2,\dots
$$
are some distribution-valued operators leaving invariant the algebraic Fock space and verifying the following set of axioms:
\begin{itemize}
\item
{\bf Skew-symmetry} in all arguments:
\be
T(\dots,A_{i}(x_{i}),A_{i+1}(x_{i+1}),\dots,) =
(-1)^{f_{i} f_{i+1}} T(\dots,A_{i+1}(x_{i+1}),A_{i}(x_{i}),\dots)
\ee

\item
{\bf Poincar\'e invariance}: we have a natural action of the Poincar\'e group in the
space of Wick monomials and we impose that for all 
$g \in inSL(2,\C)$
we have:
\be
U_{g} T(A_{1}(x_{1}),\dots,A_{n}(x_{n})) U^{-1}_{g} =
T(g\cdot A_{1}(x_{1}),\dots,g\cdot A_{n}(x_{n}))
\label{invariance}
\ee
where in the right hand side we have the natural action of the Poincar\'e group on
$
\Xi
$.

Sometimes it is possible to supplement this axiom by other invariance
properties: space and/or time inversion, charge conjugation invariance, global
symmetry invariance with respect to some internal symmetry group, supersymmetry,
etc.
\item
{\bf Causality}: if 
$
y \cap (x + \bar{V}^{+}) = \emptyset
$
then we denote this relation by
$
x \succeq y
$.
Suppose that we have 
$x_{i} \succeq x_{j}, \quad \forall i \leq k, \quad j \geq k+1$;
then we have the factorization property:
\be
T(A_{1}(x_{1}),\dots,A_{n}(x_{n})) =
T(A_{1}(x_{1}),\dots,A_{k}(x_{k}))~~T(A_{k+1}(x_{k+1}),\dots,A_{n}(x_{n}));
\label{causality}
\ee

\item
{\bf Unitarity}: We define the {\it anti-chronological products} using a convenient notation introduced
by Epstein-Glaser, adapted to the Grassmann context. If 
$
X = \{j_{1},\dots,j_{s}\} \subset N \equiv \{1,\dots,n\}
$
is an ordered subset, we define
\be
T(X) \equiv T(A_{j_{1}}(x_{j_{1}}),\dots,A_{j_{s}}(x_{j_{s}})).
\ee
Let us consider some Grassmann variables
$
\theta_{j},
$
of parity
$
f_{j},  j = 1,\dots, n
$
and let us define
\be
\theta_{X} \equiv \theta_{j_{1}} \cdots \theta_{j_{s}}.
\ee
Now let
$
(X_{1},\dots,X_{r})
$
be a partition of
$
N = \{1,\dots,n\}
$
where
$
X_{1},\dots,X_{r}
$
are ordered sets. Then we define the (Koszul) sign
$
\epsilon(X_{1},\dots,X_{r})
$
through the relation
\be
\theta_{1} \cdots \theta_{n} = \epsilon(X_{1}, \dots,X_{r})~\theta_{X_{1}} \dots \theta_{X_{r}}
\ee
and the antichronological products are defined according to
\be
(-1)^{n} \bar{T}(N) \equiv \sum_{r=1}^{n} 
(-1)^{r} \sum_{I_{1},\dots,I_{r} \in Part(N)}
\epsilon(X_{1},\dots,X_{r})~T(X_{1})\dots T(X_{r})
\label{antichrono}
\ee
Then the unitarity axiom is:
\be
\bar{T}(N) = T(N)^{\dagger}.
\label{unitarity}
\ee
\item
{\bf The ``initial condition"}:
\be
T(A(x)) =  N(A(x)).
\ee

\item
{\bf Power counting}: We can also include in the induction hypothesis a limitation on the order of
singularity of the vacuum averages of the chronological products associated to
arbitrary Wick monomials
$A_{1},\dots,A_{n}$;
explicitly:
\be
\omega(<\Omega, T^{A_{1},\dots,A_{n}}(X)\Omega>) \leq
\sum_{l=1}^{n} \omega(A_{l}) - 4(n-1)
\label{power}
\ee
where by
$\omega(d)$
we mean the order of singularity of the (numerical) distribution $d$ and by
$\omega(A)$
we mean the canonical dimension of the Wick monomial $W$.

\item
{\bf Wick expansion property}: In analogy to (\ref{comm-wick}) we require
\bea
[ \xi_{p}(y), T(A_{1}(x_{1}),\dots,A_{n}(x_{n})) ]
\nonumber\\
= \sum_{m=1}^{n}~\prod_{l \leq m} (-1)^{|\xi_{p}||A_{l}|}~\sum_{q}~[\xi_{p}(y), \xi_{q}(x_{m})]~
T(A_{1}(x_{1}),\dots,\xi_{q}\cdot A_{m}(x_{m}),\dots, A_{n}(x_{n}))
\nonumber\\
= - i~\sum_{m=1}^{n}~\prod_{l \leq m} (-1)^{|\xi_{p}||A_{l}|}~\sum_{q}~D_{pq}(y - x_{m} )~
T(A_{1}(x_{1}),\dots,\xi_{q}\cdot A_{m}(x_{m}),\dots, A_{n}(x_{n}))
\nonumber\\
\label{wick}
\eea

In fact we can impose a sharper form:
\bea
[ \xi_{p}^{(\epsilon)}(y), T(A_{1}(x_{1}),\dots,A_{n}(x_{n})) ]
\nonumber\\
= - i~\sum_{m=1}^{n}~\prod_{l \leq m} (-1)^{|\xi_{p}||A_{l}|}~\sum_{q}~D_{pq}^{(-\epsilon)}(y - x_{m} )~
T(A_{1}(x_{1}),\dots,\xi_{q}\cdot A_{m}(x_{m}),\dots, A_{n}(x_{n}))
\nonumber\\
\label{wick-e}
\eea
\end{itemize}

Up to now, we have defined the chronological products only for self-adjoint Wick monomials 
$
W_{1},\dots,W_{n}
$
but we can extend the definition for Wick polynomials by linearity. 

The chronological products
$
T(A_{1}(x_{1}),\dots,A_{n}(x_{n}))
$
are not uniquely defined by the axioms presented above. They can be modified with quasi-local expressions i.e. 
expressions localized on the big diagonal 
$
x_{1} = \cdots = x_{n};
$
such expressions are of the type
\be
N(A_{1}(x_{1}),\dots,A_{n}(x_{n})) = P_{j}(\partial)\delta(X)~W_{j}(X)
\ee
where
$
\delta(X) \equiv \delta(x_{1} - x_{n}) \dots, \delta(x_{n-1} - x_{n}), 
$
the expressions
$
P_{j}(\partial)
$
are polynomials in the partial derivatives and 
$
W_{j}(X)
$
are Wick polynomials. There are some restrictions on these quasi-local expressions such that the Bogoliubov axioms remain true.
One such consistency relations comes from Wick expansion property. Such redefinitions of the chronological products are extremely
important for Yang-Mills models, because gauge invariance can be preserved only if some redefinitions of this type are done. 

For simplicity let us consider that the the variables 
$
\xi_{a}
$
are commutative and the finite renormalizations are of the type:
\be
N(A_{1}(x_{1}),\dots,A_{n}(x_{n})) = \delta(X)~N(A_{1},\dots,A_{n})(x_{n})
\ee
where 
$
N(A_{1},\dots,A_{n})
$
are polynomials in the jet variables. Then the relation (\ref{wick}) from above is preserved if we require
\be
\xi_{p}\cdot N(A_{1},\dots,A_{n}) = \sum_{m=1}^{n}~N(A_{1},\dots,\xi_{p}\cdot A_{m},\dots, A_{n}).
\nonumber\\
\label{consistency}
\ee

In the general case some combinatorial complications, as for instance Fermi signs, do appear.
This consistency checks seems to be absent from the literature. We will investigate in the context of Yang-Mills models. 

The construction of Epstein-Glaser is based on a recursive procedure \cite{EG}. 
We can derive from these axioms the following result \cite{Sto1}.
\begin{thm}
One can fix the causal products such that the following formula is true
\bea
T(\xi_{p}(y), A_{1}(x_{1}),\dots,A_{n}(x_{n}))
\nonumber\\
= - i~\sum_{m=1}^{n}~\prod_{l \leq m} (-1)^{|\xi_{p}||A_{l}|}~\sum_{q}~D^{F}_{pq}(y - x_{m} )~
T(A_{1}(x_{1}),\dots,\xi_{q}\cdot A_{m}(x_{m}),\dots, A_{n}(x_{n}))
\nonumber\\
+ \xi^{(+)}_{p}(y)~T(A_{1}(x_{1}),\dots, A_{n}(x_{n}))
+ \prod_{l \leq n}~(-1)^{|\xi_{p}| f_{l}}~T(A_{1}(x_{1}),\dots, A_{n}(x_{n}))~\xi^{(-)}_{p}(y)
\label{linear}
\eea
where 
$
D^{F}_{pq}
$
is a Feynman propagator associated to the causal distribution
$
D_{pq}
$.
\end{thm}
Some times (\ref{linear}) - or  variants of it - is called the {\bf equation of motion} axiom \cite{DF}. 
\newpage
\subsection{Yang-Mills Fields\label{ym}}

First, we can generalize the preceding formalism to the case when some of the scalar fields
are odd Grassmann variables. One simply insert everywhere the Fermi sign. The next generalization is to arbitrary vector and spinorial
fields. If we consider for instance the Yang-Mills interaction Lagrangian corresponding to pure QCD \cite{algebra} then the jet variables 
$
\xi_{a}, a \in \Xi
$
are
$
(v^{\mu}_{a}, u_{a}, \tilde{u}_{a}),~a = 1,\dots,r
$
where 
$
v^{\mu}_{a}
$
are Grassmann even and 
$
u_{a}, \tilde{u}_{a}
$
are Grassmann odd variables. 

The interaction Lagrangian is determined by gauge invariance. Namely we define the {\it gauge charge} operator by
\be
d_{Q} v^{\mu}_{a} = i~d^{\mu}u_{a},\qquad
d_{Q} u_{a} = 0,\qquad
d_{Q} \tilde{u}_{a} = - i~d_{\mu}v^{\mu}_{a},~a = 1,\dots,r
\ee
where 
$
d^{\mu}
$
is the formal derivative. The gauge charge operator squares to zero:
\be
d_{Q}^{2} \simeq  0
\ee
where by
$
\simeq
$
we mean, modulo the equation of motion. Now we can define the interaction Lagrangian by the relative cohomology relation:
\be
d_{Q}T(x) \simeq {\rm total~divergence}.
\ee
If we eliminate the corresponding coboundaries, then a tri-linear Lorentz covariant 
expression is uniquely given by
\bea
T = f_{abc} \left( {1\over 2}~v_{a\mu}~v_{b\nu}~F_{c}^{\nu\mu}
+ u_{a}~v_{b}^{\mu}~d_{\mu}\tilde{u}_{c}\right)
\label{Tint}
\eea
where
\be
F^{\mu\nu}_{a} \equiv d^{\mu}v^{\nu}_{a} - d^{\nu}v^{\mu}_{a}, 
\quad \forall a = 1,\dots,r
\ee 
and 
$
f_{abc}
$
are real and completely anti-symmetric. (This is the tri-linear part of the usual QCD interaction Lagrangian from classical field theory.)

Then we define the associated Fock space by the non-zero $2$-point distributions are
\bea
<\Omega, v^{\mu}_{a}(x_{1}) v^{\nu}_{b}(x_{2})\Omega> = 
i~\eta^{\mu\nu}~\delta_{ab}~D_{0}^{(+)}(x_{1} - x_{2}),
\nonumber \\
<\Omega, u_{a}(x_{1}) \tilde{u}_{b}(x_{2})\Omega> = - i~\delta_{ab}~D_{0}^{(+)}(x_{1} - x_{2}),
\nonumber\\
<\Omega, \tilde{u}_{a}(x_{1}) u_{b}(x_{2})\Omega> = i~\delta_{ab}~D_{0}^{(+)}(x_{1} - x_{2}).
\label{2-massless-vector}
\eea
and construct the associated Wick monomials. Then the expression (\ref{Tint}) gives a Wick polynomial 
$
T^{\rm quant}
$
formally the same, but: 
(a) the jet variables must be replaced by the associated quantum fields; (b) the formal derivative 
$
d^{\mu}
$
goes in the true derivative in the coordinate space; (c) Wick ordering should be done to obtain well-defined operators. We also 
have an associated {\it gauge charge} operator in the Fock space given by
\bea
~[Q, v^{\mu}_{a}] = i~\partial^{\mu}u_{a},\qquad
\{ Q, u_{a} \} = 0,\qquad
\{Q, \tilde{u}_{a}\} = - i~\partial_{\mu}v^{\mu}_{a}
\nonumber \\
Q \Omega = 0.
\label{Q-vector-null}
\eea

Then it can be proved that
$
Q^{2} = 0
$
and
\be
~[Q, T^{\rm quant}(x) ] = {\rm total~divergence}
\label{gauge1}
\ee
where the equations of motion are automatically used because the quantum fields are on-shell.
From now on we abandon the super-script {\it quant} because it will be obvious from the context if we refer 
to the classical expression (\ref{Tint}) or to its quantum counterpart.

In (\ref{2-massless-vector}) we are using the Pauli-Jordan distribution
\be
D_{m}(x) = D_{m}^{(+)}(x) + D_{m}^{(-)}(x)
\ee
where
\be
D_{m}^{(\pm)}(x) =
\pm {i \over (2\pi)^{3}}~\int dp e^{- i p\cdot x} \theta(\pm p_{0}) \delta(p^{2} -
m^{2})
\ee
and
\be
D^{(-)}(x) = - D^{(+)}(- x).
\ee

We conclude our presentation with a generalization of (\ref{gauge1}). In fact, it can be proved that (\ref{gauge1}) implies
the existence of Wick polynomials
$
T^{\mu}
$
and
$
T^{\mu\nu}
$
such that we have:
\be
~[Q, T^{I} ] = i \partial_{\mu}T^{I\mu}
\label{gauge2}
\ee
for any multi-index $I$ with the convention
$
T^{\emptyset} \equiv T.
$
Explicitly:
\bea
T^{\mu} = f_{abc} \left( u_{a}~v_{b\nu}~F_{c}^{\nu\mu}
- {1\over 2}u_{a}~u_{b}~d^{\mu}\tilde{u}_{c}\right)
\label{Tmu-int}
\eea
and 
\bea
T^{\mu\nu} = {1\over 2}~f_{abc}~u_{a}~u_{b}~F_{c}^{\mu\nu}.
\label{Tmunu-int}
\eea

Finally we give the relation expressing gauge invariance in order $n$ of the perturbation theory. We define the operator 
$
\delta
$
on chronological products by:
\bea
\delta T(T^{I_{1}}(x_{1}),\dots,T^{I_{n}}(x_{n})) \equiv 
%\nonumber\\
\sum_{m=1}^{n}~( -1)^{s_{m}}\partial_{\mu}^{m}T(T^{I_{1}}(x_{1}), \dots,T^{I_{m}\mu}(x_{m}),\dots,T^{I_{n}}(x_{n}))
\label{derT}
\eea
with
\be
s_{m} \equiv \sum_{p=1}^{m-1} |I_{p}|,
\ee
then we define the operator
\be
s \equiv d_{Q} - i \delta.
\label{s-n}
\ee

Gauge invariance in an arbitrary order is then expressed by
\be
sT(T^{I_{1}}(x_{1}),\dots,T^{I_{n}}(x_{n})) = 0.
\label{brst-n}
\ee
This concludes the necessary prerequisites.
\newpage

\section{A More Precise Version of Wick Theorem\label{wick thm}}

For simplicity, we assume in this Section that the variables 
$
\xi_{a}
$
(see Subsection \ref{wick prod}) are commutative. 
We also use the summation convention over the dummy indices. 
We begin with the following elementary result.
\begin{thm}
Let us first consider the chronological products of the form 
$$
T(\xi_{p}(x_{1}), A_{2}(x_{1}),\dots, A_{n}(x_{n})).
$$
We define
\bea
T_{1}(\xi_{p}(x_{1}), A_{2}(x_{1}),\dots, A_{n}(x_{n}))
\equiv 
\nonumber\\
\xi_{p}^{(+)}(x_{1})~T(A_{2}(x_{1}),\dots, A_{n}(x_{n}))
+ T(A_{2}(x_{1}),\dots, A_{n}(x_{n}))~\xi_{p}^{(-)}(x_{1})
\label{11}
\eea
and
\be
T_{0} \equiv T - T_{1}.
\label{t10}
\ee
Then 
$
T_{0}
$
is of Wick type only in the entries 
$
A_{2},\cdots,A_{n}
$
i.e. it verifies:
\bea
[ \xi_{q}(y), T_{0}(\xi_{p}(x_{1}), A_{2}(x_{1}),\dots, A_{n}(x_{n})) ]
\nonumber\\
= \sum_{m=2}^{n}~[\xi_{q}(y),\xi_{r}(x_{m})]~T_{0}(\xi_{p}(x_{1}), A_{2}(x_{1}),\dots,\xi_{r}\cdot A_{m}(x_{m}),\dots, A_{n}(x_{n})).
\label{w1}
\eea
\label{tw1}
\end{thm}
The proof of (\ref{w1}) is elementary, by direct computation. In the general case of arbitray Grassmann variables we have to
consider a Grassmann sign i.e.
\bea
T_{1}(\xi_{p}(x_{1}), A_{2}(x_{1}),\dots, A_{n}(x_{n})) \equiv :\xi_{p}(x_{1})~T(A_{2}(x_{1}),\dots, A_{n}(x_{n})):
\nonumber\\
\xi_{p}^{(+)}(x_{1})~T(A_{2}(x_{1}),\dots, A_{n}(x_{n}))
+ (-1)^{f} T(A_{2}(x_{1}),\dots, A_{n}(x_{n}))~\xi_{p}^{(-)}(x_{1})
\label{t11a}
\eea
where
\be
f \equiv |\xi_{p}| \sum_{m=2}^{n} |A_{m}|.
\ee

We will use another notation emphasizing that we ``get out" of the chronological products one factor.
\be
T(\xi_{p}^{(1)}(x_{1}), A_{2}(x_{1}),\dots, A_{n}(x_{n})) \equiv T_{1}(\xi_{p}(x_{1}), A_{2}(x_{1}),\dots, A_{n}(x_{n})).
\ee
This new notations will prove useful when we will consider more general cases. We remark that from (\ref{t10}) we trivially have
a partial Wick theorem, only in the first variable 
$
A_{1}:
$
\be
T \equiv T_{0} + T_{1}.
\label{w11}
\ee
\newpage
Now we consider a more complicated case.
\begin{thm}
Let us first consider that
\be
A_{1} = {1\over 2}~f_{pq} \xi_{p}\xi_{q},\qquad f_{pq} = f_{qp}
\ee
and
$
A_{2},\dots,A_{n}
$
are arbitrary. We define
\bea
T_{1}(A_{1}(x_{1}),\dots, A_{n}(x_{n}))
\equiv 
\nonumber\\
f_{pq}~[ \xi_{p}^{(+)}(x_{1})~T_{0}(\xi_{q}(x_{1}),A_{2}(x_{2}), \dots, A_{n}(x_{n}))
+ T_{0}(\xi_{q}(x_{1}), A_{2}(x_{2}),\dots, A_{n}(x_{n}))~\xi_{p}^{(-)}(x_{1})]
\label{t21}
\eea
where
$
T_{0}
$
was defined above (\ref{t10}); we also define:
\bea
T_{2}(A_{1}(x_{1}),\dots, A_{n}(x_{n}))
\equiv 
%\nonumber\\
f_{pq}~\Bigl[ {1 \over 2}\xi_{p}^{(+)}(x_{1})\xi_{q}^{(+)}(x_{1})~T(A_{2}(x_{2}), \dots, A_{n}(x_{n}))
\nonumber\\
+ \xi_{p}^{(+)}(x_{1})T(A_{2}(x_{2}),\dots, A_{n}(x_{n}))~\xi_{q}^{(-)}(x_{1})
+ {1 \over 2}~T(A_{2}(x_{2}),\dots, A_{n}(x_{n}))~\xi_{p}^{(-)}(x_{1})\xi_{q}^{(-)}(x_{1})\Bigl]
\label{t22}
\eea
and
\be
T_{0} \equiv T - T_{1} - T_{2}.
\label{t20}
\ee
Then 
$
T_{0}
$
is of Wick type only in the entries 
$
A_{2},\cdots,A_{n}
$
i.e. it verifies
\bea
[ \xi_{q}(y), T_{0}(A_{1}(x_{1}), A_{2}(x_{1}),\dots, A_{n}(x_{n})) ]
\nonumber\\
= \sum_{m=2}^{n}~[\xi_{q}(y),\xi_{r}(x_{m})]~T_{0}(A_{1}(x_{1}), A_{2}(x_{1}),\dots,\xi_{r}\cdot A_{m}(x_{m}),\dots, A_{n}(x_{n})).
\label{w2}
\eea
\label{tw2}
\end{thm}

The proof is also elementary but more tedious. We also notice that (\ref{t20}) is a partial Wick theorem in the first variable
$
A_{1}:
$
\be
T \equiv T_{0} + T_{1} + T_{2}.
\label{w21}
\ee

We can use the more compact notations:
\bea
T(A_{1}^{(1)}(x_{1}),\dots, A_{n}(x_{n})) \equiv T_{1}(A_{1}(x_{1}),\dots, A_{n}(x_{n})) =
\nonumber\\
f_{pq}~:\xi_{p}(x_{1})~T_{0}(\xi_{q}(x_{1}),A_{2}(x_{2}), \dots, A_{n}(x_{n}):
\label{t21a}
\eea
and
\bea
T(A_{1}^{(2)}(x_{1}),\dots, A_{n}(x_{n})) \equiv T_{2}(A_{1}(x_{1}),\dots, A_{n}(x_{n})) =
\nonumber\\
{1 \over 2}~f_{pq}~: \xi_{p}(x_{1})\xi_{q}(x_{1})~T(A_{2}(x_{2}), \dots, A_{n}(x_{n}):
\label{t22a}
\eea
\newpage

Finally, we consider the more interesting case when the monomial
$
A_{1}
$
is tri-linear.
\begin{thm}
Let us first consider that
\be
A_{1} = {1\over 3!}~f_{pqr} \xi_{p}\xi_{q}\xi_{r},\qquad f_{pqr} = {\rm completely~ symmetric}
\ee
and
$
A_{2},\dots,A_{n}
$
are arbitrary. We define
\bea
T_{1}(A_{1}(x_{1}),\dots, A_{n}(x_{n}))
\equiv 
\nonumber\\
{1 \over 2}~f_{pqr}~[ \xi_{p}^{(+)}(x_{1})~T_{0}(\xi_{q}\xi_{r}(x_{1}),A_{2}(x_{2}), \dots, A_{n}(x_{n}))
\nonumber\\
+ T_{0}(\xi_{q}\xi_{r}(x_{1}), A_{2}(x_{2}),\dots, A_{n}(x_{n}))~\xi_{p}^{(-)}(x_{1})]
\label{t31}
\eea
where
$
T_{0}
$
was defined above (\ref{t20}); we also define:
\bea
T_{2}(A_{1}(x_{1}),\dots, A_{n}(x_{n}))
\equiv 
%\nonumber\\
f_{pqr} \Bigl[ {1 \over 2}~\xi_{p}^{(+)}(x_{1})\xi_{q}^{(+)}(x_{1})~T_{0}(\xi_{r}(x_{1}),A_{2}(x_{2}), \dots, A_{n}(x_{n}))
\nonumber\\
+ \xi_{p}^{(+)}(x_{1})T_{0}(\xi_{r}(x_{1}),A_{2}(x_{2}),\dots, A_{n}(x_{n}))~\xi_{q}^{(-)}(x_{1})
\nonumber\\
+ {1 \over 2}~T_{0}(\xi_{r}(x_{1}), A_{2}(x_{2}),\dots, A_{n}(x_{n}))~\xi_{p}^{(-)}(x_{1})\xi_{q}^{(-)}(x_{1})\Bigl]
\label{t32}
\eea
where
$
T_{0}
$
was defined above (\ref{t10}); finally
\bea
T_{3}(A_{1}(x_{1}),\dots, A_{n}(x_{n}))
\equiv 
%\nonumber\\
f_{pqr} \Bigl[ {1 \over 3!}~\xi_{p}^{(+)}(x_{1})\xi_{q}^{(+)}(x_{1})\xi_{r}^{(+)}(x_{1})~T(A_{2}(x_{2}), \dots, A_{n}(x_{n}))
\nonumber\\
+ {1 \over 2}~\xi_{p}^{(+)}(x_{1})\xi_{q}^{(+)}(x_{1})~T(A_{2}(x_{2}),\dots, A_{n}(x_{n}))~\xi_{r}^{(-)}(x_{1})
\nonumber\\
+ {1 \over 2}~\xi_{p}^{(+)}(x_{1})~T(A_{2}(x_{2}),\dots, A_{n}(x_{n}))~\xi_{q}^{(-)}(x_{1})\xi_{r}^{(-)}(x_{1})
\nonumber\\
\nonumber\\
+ {1 \over 3!}~T(A_{2}(x_{2}), \dots, A_{n}(x_{n}))~\xi_{p}^{(-)}(x_{1})\xi_{q}^{(-)}(x_{1})\xi_{r}^{(-)}(x_{1})\Bigl]
\label{t33}
\eea
and
\be
T_{0} \equiv T - T_{1} - T_{2} - T_{3}.
\label{t30}
\ee
Then 
$
T_{0}
$
is of Wick type only in the entries 
$
A_{2},\cdots,A_{n}
$
i.e. it verifies (\ref{w2}).
\label{tw3}
\end{thm}

The computations are long but straitforward. We have more compact notations:
\bea
T(A_{1}^{(1)}(x_{1}),\dots, A_{n}(x_{n})) \equiv T_{1}(A_{1}(x_{1}),\dots, A_{n}(x_{n})) =
\nonumber\\
{1 \over 2}~f_{pqr}~:\xi_{p}(x_{1})~T_{0}(\xi_{q}(x_{1})\xi_{r}(x_{1}),A_{2}(x_{2}), \dots, A_{n}(x_{n})):
\label{t31a}
\eea
\bea
T(A_{1}^{(2)}(x_{1}),\dots, A_{n}(x_{n})) \equiv T_{2}(A_{1}(x_{1}),\dots, A_{n}(x_{n})) =
\nonumber\\
{1 \over 2}~f_{pqr}~: \xi_{p}(x_{1})\xi_{q}(x_{1})~T(\xi_{r}(x_{1}),A_{2}(x_{2}), \dots, A_{n}(x_{n})):
\label{t32a}
\eea
and
\bea
T(A_{1}^{(3)}(x_{1}),\dots, A_{n}(x_{n})) \equiv T_{3}(A_{1}(x_{1}),\dots, A_{n}(x_{n})) =
\nonumber\\
{1 \over 3!}~f_{pqr}~: \xi_{p}(x_{1})\xi_{q}(x_{1})\xi_{r}(x_{1})~T(A_{2}(x_{2}), \dots, A_{n}(x_{n})):
\label{t33a}
\eea

From (\ref{t30}) we have a Wick theorem in the first entry:
\be
T \equiv T_{0} + T_{1} + T_{2} + T_{3}.
\label{w31}
\ee

We can iterate the arguments above in the entries 
$
A_{2},\dots,A_{n}
$
and obtain the following version of Wick theorem:
\be
T(A_{1}(x_{1}),\dots, A_{n}(x_{n})) = \sum T(A_{1}^{(k_{1})}(x_{1}),\dots, A_{n}^{(k_{n})}(x_{n}))
\label{w1-n}
\ee
where the sum runs over 
$
k_{1},\dots,k_{n} = 0,\dots,3
$
for 
$
A_{1},\dots,A_{n}
$
tri-linear. This formula can be written in a more transparent way if we use Hopf algebra notions. We define
\be
C_{p} \equiv \xi_{p}\cdot A
\ee
for an arbitrary Wick polynomial; then we define the co-product
\be
\Delta A \equiv 1 \otimes A + A \otimes 1 + \xi_{p} \otimes C_{p} + C_{p} \otimes \xi_{p}.
\ee
Now, if we use Sweedler notation
\be
\Delta A_{j} = \sum A_{j}^{(1)} \otimes A_{j}^{(2)}
\label{sweedler}
\ee
we can rewrite (\ref{w31}) as
\be
T(A_{1}(x_{1}),\dots, A_{n}(x_{n})) = \sum :A_{1}^{(1)}(x_{1})~T(A_{1}^{(2)}(x_{1})^{(0)},A_{2}(x_{2})\dots,A_{n}(x_{n})):
\label{w3a}
\ee
and if we use induction we arrive at: 
\be
T(A_{1}(x_{1}),\dots, A_{n}(x_{n})) = \sum T_{0}(A_{1}^{(2)}(x_{1}),\dots,A_{n}^{(2)}(x_{n}))~
: A_{1}^{(1)}(x_{1})\dots,A_{n}^{(1)}(x_{n}):
\label{w-hopf}
\ee
where
\be
T_{0}(A_{1}(x_{1}),\dots,A_{n}(x_{n})) \equiv T(A_{1}(x_{1})^{(0)},\dots,A_{n}(x_{n})^{(0)})
\ee
has no Wick property in any of the variables, i.e. it commutes with all variables so it must be the vacuum average:
\be
T_{0}(B_{1}(x_{1}),\dots, B_{n}(x_{n})) \equiv <\Omega, T(B_{1}(x_{1}),\dots, B_{n}(x_{n}))\Omega>.
\ee
In the general case of arbitrary Grassmann variables, we have to include the apropriate Fermi signs. 

We note in the end that the Hopf structure appearing in the preceding form of the Wick theorem does not involve Feynman graphs.
The Hopf structure is valid for the chronological product which are sums of (many) Feynman contributions but with better 
smoothness properties: the vacuum averages are well-behaved tempered distributions. 
\newpage

\section{Wick submonomials\label{submonomials}}
\subsection{The case of Pure Yang-Mills Theories\label{BC}}

We notice that in (\ref{Q-vector-null}) and in (\ref{gauge2}) we have a pattern of the type:
\be
d_{Q}A = {\rm total~divergence}.
\label{total-div}
\ee
This pattern remains true for Wick submonomials if we use the definition (\ref{derivative}). We consider the expressions
(\ref{Tint}), (\ref{Tmu-int}) and (\ref{Tmunu-int}) from the pure Yang-Mills case and define:
\bea
B_{a\mu} \equiv \tilde{u}_{a,\mu} \cdot T = - f_{abc} u_{b}~v_{c\mu}
\nonumber\\
C_{a\mu} \equiv v_{a\mu} \cdot T = f_{abc} (v_{b}^{\nu}~F_{c\nu\mu} - u_{b}~\tilde{u}_{c,\mu})
\nonumber\\
D_{a} \equiv u_{a} \cdot T = f_{abc} v^{\mu}_{b}~\tilde{u}_{c,\mu}
\nonumber\\
E_{a\mu\nu} \equiv v_{a\mu,\nu} \cdot T = f_{abc} v_{b\mu}~v_{c\nu}
\nonumber\\
C_{a\nu\mu} \equiv v_{a\nu} \cdot T_{\mu} = - f_{abc} u_{b}~F_{c\nu\mu}.
\label{sub1}
\eea
We also have
\bea
u_{a} \cdot T = - C_{a\mu}
\nonumber\\
v_{a\rho,\sigma} \cdot T_{\mu} = \eta_{\mu\sigma}~B_{a\rho} -  \eta_{\mu\rho}~B_{a\sigma}
\nonumber\\
u_{a} \cdot T_{\mu\nu} = - C_{a\mu\nu}
\label{sub2}
\eea
If we define
\be
B_{a} \equiv {1 \over 2}~f_{abc}~u_{b}~u_{c}
\label{B}
\ee
we also have
\bea
\tilde{u}_{a,\nu} \cdot T_{\mu} = \eta_{\mu\nu}~B_{a}
\nonumber\\
v_{a\rho,\sigma} \cdot T_{\mu\nu} = (\eta_{\mu\sigma}~\eta_{\nu\rho} - \eta_{\nu\sigma}~\eta_{\mu\rho})~B_{a}.
\label{sub3}
\eea

Then we try to extend the structure (\ref{total-div}) to the Wick submonomials defined above. We have:
\bea
d_{Q} B_{a}^{\mu} = i d^{\mu}B_{a}
\nonumber\\
d_{Q} C_{a}^{\mu} = i d_{\nu}C_{a}^{\mu\nu}
\nonumber\\
d_{Q} D_{a} = - i d_{\mu}C_{a}^{\mu}
\nonumber\\
d_{Q} E_{a}^{\mu\nu} = i (d^{\nu}B_{a}^{\mu} -  d^{\mu}B_{a}^{\nu} + C_{a}^{\mu\nu}) 
\nonumber\\
d_{Q} B_{a} = 0
\nonumber\\
d_{Q} C_{a}^{\mu\nu} = 0. 
\label{dQB}
\eea
So we see that the patern (\ref{total-div}) is broken only for 
$
E_{a}^{\mu\nu}.
$
We fix this in the following way. We have the formal derivative
\be
\delta A \equiv d_{\mu}A^{\mu}
\ee
used in the definition of gauge invariance (\ref{derT}) + (\ref{s-n}); we also define the derivative
$
\delta^{\prime}
$
by
\be
\delta^{\prime}E_{a}^{\mu\nu} = C_{a}^{\mu\nu}.
\label{dprime}
\ee
and $0$ for the other Wick submonomials (\ref{sub1}) and (\ref{B}). Finally
\be
s \equiv d_{Q} - i\delta, \qquad s^{\prime} \equiv s - i\delta^{\prime} = d_{Q} - i(\delta +\delta^{\prime}).
\ee
Then we have the structure
\be
s^{\prime} A = 0
\ee
for all expressions
$
A = T^{I}, 
$
$
B_{a\mu}, C_{a\mu},
$
etc. and also for the basic jet variables
$
v_{a\mu}, u_{a}, \tilde{u}_{a}.
$
\newpage
\subsection{Hopf Structure of the Yang-Mills pQFT\label{h-ym}}
We implement the Wick theorem in the form (\ref{w-hopf}) for the pure YM model. First we use theorem  \ref{tw3} for the expressions
$
T^{I}
$
defined in Subsection \ref{ym} and compute the expressions
$
T(T^{I}(x_{1})^{(k)}, A_{2},\dots,A_{n}). 
$
Proceeding as in subsection \ref{wick thm} we derive by direct computations the following expressions; the Wick submonomials play an important role.

- For the case 
$
k = 1
$
i.e. when we ``pull out" one factor from the first entry
$
T^{I}
$
we have:
\bea
T(T(x_{1})^{(1)}, A_{2}(x_{2}),\dots,A_{n}(x_{n})) =
\nonumber\\
: v_{a\mu}(x_{1})~T(C_{a}^{\mu}(x_{1})^{(0)}, A_{2}(x_{2}),\dots,A_{n}(x_{n})):
\nonumber\\
+ {1\over 2}~: F_{a\nu\mu}(x_{1})~T(E_{a}^{\mu\nu}(x_{1})^{(0)}, A_{2}(x_{2}),\dots,A_{n}(x_{n})):
\nonumber\\
+ : u_{a}(x_{1})~T(D_{a}(x_{1})^{(0)}, A_{2}(x_{2}),\dots,A_{n}(x_{n})):
\nonumber\\
+ : \partial_{\mu}\tilde{u}_{a}(x_{1})~T(B_{a}^{\mu}(x_{1})^{(0)}, A_{2}(x_{2}),\dots,A_{n}(x_{n})):
\label{t1}
\eea
\bea
T(T^{\mu}(x_{1})^{(1)}, A_{2}(x_{2}),\dots,A_{n}(x_{n})) =
\nonumber\\
+ : u_{a}(x_{1})~T(C_{a}^{\mu}(x_{1})^{(0)}, A_{2}(x_{2}),\dots,A_{n}(x_{n})):
\nonumber\\
+ : v_{a\nu}(x_{1})~T(C_{a}^{\nu\mu}(x_{1})^{(0)}, A_{2}(x_{2}),\dots,A_{n}(x_{n})):
\nonumber\\
+ : F_{a}^{\mu\nu}(x_{1})~T(B_{a\nu}(x_{1})^{(0)}, A_{2}(x_{2}),\dots,A_{n}(x_{n})):
\nonumber\\
- : \partial^{\mu}\tilde{u}_{a}(x_{1})~T(B_{a}(x_{1})^{(0)}, A_{2}(x_{2}),\dots,A_{n}(x_{n})):
\label{tmu1}
\eea
and
\bea
T(T^{\mu\nu}(x_{1})^{(1)}, A_{2}(x_{2}),\dots,A_{n}(x_{n})) =
\nonumber\\
- : u_{a}(x_{1})~T(C_{a}^{\mu\nu}(x_{1})^{(0)}, A_{2}(x_{2}),\dots,A_{n}(x_{n})):
\nonumber\\
+ : F_{a}^{\mu\nu}(x_{1})~T(B_{a}(x_{1})^{(0)}, A_{2}(x_{2}),\dots,A_{n}(x_{n})):
\label{tmunu1}
\eea

- For the case
$
k = 2
$
i.e. when we ``pull out" two factors from the first entry
$
T^{I}
$
we have:
\bea
T(T(x_{1})^{(2)}, A_{2}(x_{2}),\dots,A_{n}(x_{n})) =
\nonumber\\
: C_{a}^{\mu}(x_{1})~T(v_{a\mu}(x_{1})^{(0)}, A_{2}(x_{2}),\dots,A_{n}(x_{n})):
\nonumber\\
+ {1\over 2}~: E_{a}^{\mu\nu}(x_{1})~T(F_{a\nu\mu}(x_{1})^{(0)}, A_{2}(x_{2}),\dots,A_{n}(x_{n})):
\nonumber\\
- : D_{a}(x_{1})~T(u_{a}(x_{1})^{(0)}, A_{2}(x_{2}),\dots,A_{n}(x_{n})):
\nonumber\\
- : B_{a}^{\mu}(x_{1})~T(\tilde{u}_{a,\mu}(x_{1})^{(0)}, A_{2}(x_{2}),\dots,A_{n}(x_{n})):
\label{t2}
\eea
\bea
T(T^{\mu}(x_{1})^{(2)}, A_{2}(x_{2}),\dots,A_{n}(x_{n})) =
\nonumber\\
: C_{a}^{\mu}(x_{1})~T(u_{a}(x_{1})^{(0)}, A_{2}(x_{2}),\dots,A_{n}(x_{n})):
\nonumber\\
+ : C_{a}^{\nu\mu}(x_{1})~T(v_{a\nu}(x_{1})^{(0)}, A_{2}(x_{2}),\dots,A_{n}(x_{n})):
\nonumber\\
- : B_{a\nu}(x_{1})~T(F_{a}^{\nu\mu}(x_{1})^{(0)}, A_{2}(x_{2}),\dots,A_{n}(x_{n})):
\nonumber\\
- : B_{a}(x_{1})~T({\tilde{u}_{a},}^{\mu}(x_{1})^{(0)}, A_{2}(x_{2}),\dots,A_{n}(x_{n})):
\label{tmu2}
\eea
and
\bea
T(T^{\mu\nu}(x_{1})^{(1)}, A_{2}(x_{2}),\dots,A_{n}(x_{n})) =
\nonumber\\
: B_{a}(x_{1})~T(F_{a}^{\mu\nu}(x_{1})^{(0)}, A_{2}(x_{2}),\dots,A_{n}(x_{n})):
\nonumber\\
+ : C_{a}^{\mu\nu}(x_{1})~T(u(x_{1})^{(0)}, A_{2}(x_{2}),\dots,A_{n}(x_{n})):
\label{tmunu2}
\eea

- For the case 
$
k = 3
$
we trivially have
\be
T(T^{I}(x_{1})^{(3)}, A_{2}(x_{2}),\dots,A_{n}(x_{n})) =
: T^{I}(x_{1}) T(A_{2}(x_{2}),\dots,A_{n}(x_{n})):
\label{t3}
\ee
Here the expressions
$
T(A_{1}(x_{1})^{(0)}, A_{2}(x_{2}),\dots,A_{n}(x_{n}))
$
are of Wick type only in 
$
A_{2},\cdots,A_{n}
$
as explained in subsection \ref{wick thm}. We also note that there are various signs $-$ because we have a true Grassmann structure:
the variables
$
u_{a}, \tilde{u}_{a}
$
and the Wick submonomials
$
B_{a}^{\mu}, D_{a}, C_{a}^{\mu\nu}
$
are odd. We can write completely the preceding relations using the Hopf structure. We define the co-product:
\bea
\Delta T = 1 \otimes T + T \otimes 1 + v_{a\mu} \otimes C_{a}^{\mu} + C_{a}^{\mu} \otimes v_{a\mu}
+ {1\over 2}~F_{a\nu\mu} \otimes E_{a}^{\mu\nu} + {1\over 2}~E_{a}^{\mu\nu} \otimes F_{a\nu\mu}
\nonumber\\
+ u_{a} \otimes D_{a} - D_{a} \otimes u_{a} 
+ \tilde{u}_{a,\mu} \otimes B_{a}^{\mu} - B_{a}^{\mu} \otimes \tilde{u}_{a,\mu} 
\label{dT}
\eea
\bea
\Delta T^{\mu} = 1 \otimes T^{\mu} + T^{\mu} \otimes 1 + u_{a} \otimes C_{a}^{\mu} + C_{a}^{\mu} \otimes u_{a}
+ v_{a\nu} \otimes C_{a}^{\nu\mu} + C_{a}^{\nu\mu} \otimes v_{a\nu}
\nonumber\\
+ F^{\mu\nu}_{a} \otimes B_{a\nu}  + B_{a\nu} \otimes F_{a}^{\mu\nu} 
- \tilde{u}_{a}^{,\mu} \otimes B_{a} - B_{a} \otimes \tilde{u}_{a}^{,\mu} 
\label{dTmu}
\eea
\bea
\Delta T^{\mu\nu} = 1 \otimes T^{\mu\nu} + T^{\mu\nu} \otimes 1 - u_{a} \otimes C_{a}^{\mu\nu} + C_{a}^{\mu\nu} \otimes u_{a}
%\nonumber\\
+ F^{\mu\nu}_{a} \otimes B_{a}  + B_{a} \otimes F_{a}^{\mu\nu} 
\label{dTmunu}
\eea

Then the relations (\ref{t1}) - (\ref{t3}) can be written in the compact form
\be
T(A_{1}(x_{1}), A_{2}(x_{2}),\dots,A_{n}(x_{n})) = 
\sum :A_{1}^{(1)}(x_{1})~T_{0}(A_{1}^{(2)}(x_{1}), A_{2}(x_{2}),\dots,A_{n}(x_{n})): 
\label{w-hopf-1}
\ee
where
$
A_{1} = T, T^{\mu}, T^{\mu\nu},
$
the expressions
$
T_{0}(A_{1}^{(1)}(x_{1}), A_{2}(x_{2}),\dots,A_{n}(x_{n})) 
$
are of Wick type only in
$
A_{2},\dots,A_{n}
$
and we use Sweedler notation (\ref{sweedler}). Now we can use induction and iterate the procedure in the entries 
$
A_{2},\dots,A_{n}.
$
The final result is 
\be
T(A_{1}(x_{1}), A_{2}(x_{2}),\dots,A_{n}(x_{n})) =
\sum \epsilon_{n} T_{0}(A_{1}^{(2)}(x_{1}), \dots,A_{n}^{(2)}(x_{n}))~:A_{1}^{(1)}(x_{1}) \dots A_{n}^{(1)}(x_{n}): 
\label{w-hopf-n}
\ee
where 
$
A_{1},\cdots, A_{n} = T, T^{\mu}, T^{\mu\nu},
$
the expressions
\be
T_{0}(A_{1}^{(2)}(x_{1}),\dots,A_{n}^{(2)}(x_{n})) \equiv <\Omega, T(A_{1}^{(2)}(x_{1}),\dots,A_{n}^{(2)}(x_{n}))\Omega>
\ee
are vacuum averages and we use Sweedler notations for all entries. The sign is given by:
\be
\epsilon_{n} = ( - 1)^{s_{n}}, \qquad s_{n} = \sum_{p=1}^{n - 1} |A_{p+1}^{(1)}| \sum_{q = 1}^{p} |A_{q}^{(2)}|
\label{epsilon}
\ee
i.e. is the Fermi sign associated with the permutations of the odd factors from the preceding relation.

We can understand the previous sign better if we use the notion of co-multiplication on the product of algebras. 
Suppose that 
$
W_{1}, W_{2}
$
are two graded algebras and for
$
a_{j} \in W_{j},~j = 1,2
$
we have
\be
\Delta a_{j} = \sum a_{j}^{(1)} \otimes a_{j}^{(2)}
\ee
(in Sweedler notation) with 
$
a_{j}^{(1)}, a_{j}^{(2)}
$
of fixed grading. Then we can define the co-multiplication on 
$
W_{1} \otimes W_{2}
$
according to
\be
\Delta a_{1} \otimes a_{2} = \sum ( - 1)^{ |a_{2}^{(1)}| |a_{1}^{(2)}|} 
a_{1}^{(1)} \otimes a_{2}^{(1)} \otimes a_{1}^{(2)} \otimes a_{2}^{(2)}
\ee
We can iterate this definition to a product on
$
W_{1} \otimes \cdots \otimes W_{n}.
$
Suppose that all these algebras are identical
$
W_{1} = \cdots = W_{n} = {\cal W}
$
with the polynomial algebra in the jet variable of the pure YM model. If 
$
A_{1}, \dots, A_{n} \in {\cal W}
$
and we have
\be
\Delta A_{1} \otimes \dots \otimes A_{n} = \sum {\cal A}^{(1)} \otimes {\cal A}^{(2)}
\ee
with 
$
{\cal A}^{(1)}, {\cal A}^{(2)} \in {\cal W} \otimes \cdots \otimes {\cal W};
$
then the formula (\ref{w-hopf-n}) has the simple form
\be
T(A_{1}(x_{1}), A_{2}(x_{2}),\dots,A_{n}(x_{n})) = \sum T_{0}({\cal A}^{(2)})~: {\cal A}^{(1)}: 
\label{w-hopf-n-a}
\ee
where we have used the following compact notations: if
$
{\cal A} = a_{1} \otimes \cdots \otimes  a_{n}
$
then
\be
T_{0}({\cal A}) = T_{0}(a_{1}(x_{1}),\dots, a_{n}(x_{n}))
\ee
and
\be
:{\cal A}: = :a_{1}(x_{1})\dots a_{n}(x_{n}):.
\ee
\newpage
We try to define now expressions of the type
\be
s^{\prime}T(T^{I_{1}}(x_{1})^{(k)}, T^{I_{2}}(x_{2}),\dots,T^{I_{n}}(x_{n}))
\ee
in term of similar expressions involving Wick submonomials. 
%If we substitute (\ref{t1}), (\ref{tmu1}) and (\ref{tmunu1}) we obtain such relations. 
We generalize first the operator
$
\delta
$
by imposing, as suggested by (\ref{dQB})
\bea
\delta T(B_{a}^{\mu}(x_{1}), T^{I_{2}}(x_{2}),\dots,T^{I_{n}}(x_{n})) = 
\partial^{\mu}_{1}T(B_{a}(x_{1}), T^{I_{2}}(x_{2}),\dots,T^{I_{n}}(x_{n}))
\nonumber\\
- \sum_{m=2}^{n} (- 1)^{s_{1}(m)} \partial_{\mu}^{m}T(B_{a}^{\mu}(x_{1}),T^{I_{2}}(x_{2}),\dots,T^{I_{m}\mu}(x_{m}),\dots,T^{I_{n}}(x_{n}))
\eea
\bea
\delta T(C_{a}^{\mu}(x_{1}), T^{I_{2}}(x_{2}),\dots,T^{I_{n}}(x_{n})) = 
\partial_{\nu}^{1}T(C_{a}^{\mu\nu}(x_{1}), T^{I_{2}}(x_{2}),\dots,T^{I_{n}}(x_{n}))
\nonumber\\
+ \sum_{m=2}^{n} \partial_{\mu}^{m}T(C_{a}^{\mu}(x_{1}), T^{I_{2}}(x_{2}),\dots,T^{I_{m}\mu}(x_{m}),\dots,T^{I_{n}}(x_{n}))
\eea
\bea
\delta T(D_{a}(x), T^{I_{2}}(x_{2}),\dots,T^{I_{n}}(x_{n})) = 
- \partial_{\mu}^{1}T(C_{a}^{\mu}(x_{1}), T^{I_{2}}(x_{2}),\dots,T^{I_{n}}(x_{n}))
\nonumber\\
- \sum_{m=2}^{n} (- 1)^{s_{1}(m)} \partial_{\mu}^{m}T(D_{a}(x_{1}), T^{I_{2}}(x_{2}),\dots,T^{I_{m}\mu}(x_{m}),\dots,T^{I_{n}}(x_{n}))
\eea
\bea
\delta T(E_{a}^{\mu\nu}(x_{1}), T^{I_{2}}(x_{2}),\dots,T^{I_{n}}(x_{n})) = 
\partial^{\nu}_{1}T(B_{a}^{\mu}(x_{1}), T^{I_{2}}(x_{2}),\dots,T^{I_{n}}(x_{n})) - (\mu \longleftrightarrow \nu) 
\nonumber\\
+ \sum_{m=2}^{n} (- 1)^{s_{1}(m)}\partial_{\rho}^{m}T(E_{a}^{\mu\nu}(x_{1}),T^{I_{2}}(x_{2}),\dots,T^{I_{m}\rho}(x_{m}),\dots,T^{I_{n}}(x_{n}))
\eea
\bea
\delta T(B_{a}(x_{1}), T^{I_{2}}(x_{2}),\dots,T^{I_{n}}(x_{n})) = 
\nonumber\\
\sum_{m=2}^{n} (- 1)^{s_{1}(m)}\partial_{\rho}^{m}T(B_{a}(x_{1}),T^{I_{2}}(x_{2}),\dots,T^{I_{m}\rho}(x_{m}),\dots,T^{I_{n}}(x_{n}))
\eea
\bea
\delta T(C_{a}^{\mu\nu}(x_{1}), T^{I_{2}}(x_{2}),\dots,T^{I_{n}}(x_{n})) = 
\nonumber\\
\sum_{m=2}^{n} (- 1)^{s_{1}(m)}\partial_{\rho}^{m}T(C_{a}^{\mu\nu}(x_{1}),T^{I_{2}}(x_{2}),\dots,T^{I_{m}\rho}(x_{m}),\dots,T^{I_{n}}(x_{n}))
\eea
with
\be
s_{1}(m) \equiv \sum_{p=2}^{m} |I_{p}|
\ee
Then we generalize naturally the operator 
$
\delta^{\prime}
$
given by (\ref{dprime}) to chronological products; the formula is
\be
\delta^{\prime}T(A_{1}(x_{1}), \dots, A_{n}(x_{n})) =
\sum_{m=1}^{n}(- 1)^{s_{m}} T(A_{1}(x_{1}), \dots, \delta^{\prime}A_{m}(x_{m}),\dots,A_{n}(x_{n}))
\ee
Finally, we can define
\be
s^{\prime} = s - i \delta^{\prime} = d_{Q} - i(\delta + \delta^{\prime})
\ee
on any chronological product of the form
$
T(A_{1}(x_{1}), \dots, A_{n}(x_{n})) 
$
with
$
A_{1}, \dots, A_{n} 
$
of the type
$
T^{I}
$
or submonomials as
$
B_{a}^{\mu},
$
etc.

In the case 
$
k = 1
$
we have directly from (\ref{t1}) - (\ref{tmunu1}):
\bea
sT(T(x_{1})^{(1)}, T^{I_{2}}(x_{2}),\dots,T^{I_{n}}(x_{n})) = 
%\nonumber\\
: v_{a\mu}(x_{1})~s^{\prime}T(C_{a}^{\mu}(x_{1})^{(0)}, T^{I_{2}}(x_{2}),\dots,T^{I_{n}}(x_{n})):
\nonumber\\
+ {1\over 2}~: F_{a\nu\mu}(x_{1})~s^{\prime}T(E_{a}^{\mu\nu}(x_{1})^{(0)}, T^{I_{2}}(x_{2}),\dots,T^{I_{n}}(x_{n})):
\nonumber\\
- : u_{a}(x_{1})~s^{\prime}T(D_{a}(x_{1})^{(0)}, T^{I_{2}}(x_{2}),\dots,T^{I_{n}}(x_{n})) :
\nonumber\\
- : \partial_{\mu}\tilde{u}_{a}(x_{1})~s^{\prime}T(B_{a}^{\mu}(x_{1})^{(0)},T^{I_{2}}(x_{2}),\dots,T^{I_{n}}(x_{n})) :
\label{s1}
\eea
\bea
sT(T^{\mu}(x_{1})^{(1)}, T^{I_{2}}(x_{2}),\dots,T^{I_{n}}(x_{n}))  =
%\nonumber\\
- : u_{a}(x_{1})~s^{\prime}T(C_{a}^{\mu}(x_{1})^{(0)},T^{I_{2}}(x_{2}),\dots,T^{I_{n}}(x_{n})) :
\nonumber\\
+ : v_{a\mu}(x_{1})~s^{\prime}T(C_{a}^{\nu\mu}(x_{1})^{(0)}, T^{I_{2}}(x_{2}),\dots,T^{I_{n}}(x_{n})) :
\nonumber\\
+ : F_{a}^{\mu\nu}(x_{1})~s^{\prime}T(B_{a\nu}(x_{1})^{(0)}, T^{I_{2}}(x_{2}),\dots,T^{I_{n}}(x_{n})) :
\nonumber\\
+ : \partial^{\mu}\tilde{u}_{a}(x_{1})~s^{\prime}T(B_{a}(x_{1})^{(0)}, T^{I_{2}}(x_{2}),\dots,T^{I_{n}}(x_{n})) :
\label{smu1}
\eea
\bea
sT(T^{\mu\nu}(x_{1})^{(1)}, T^{I_{2}}(x_{2}),\dots,T^{I_{n}}(x_{n})) =
%\nonumber\\
: u_{a}(x_{1})~s^{\prime}T(C_{a}^{\mu\nu}(x_{1})^{(0)}, T^{I_{2}}(x_{2}),\dots,T^{I_{n}}(x_{n})):
\nonumber\\
+ : F_{a}^{\mu\nu}(x_{1})~T(B_{a}(x_{1})^{(0)}, T^{I_{2}}(x_{2}),\dots,T^{I_{n}}(x_{n})):
\label{smunu1}
\eea

Taking into account that we can replace $s$ by 
$
s^{\prime}
$
in the left hand sides of the preceding three equations, it results that we have a graded ``commutativity" between
$
s^{\prime}
$
and the projection
$
T^{I}(x_{1}) \rightarrow T^{I}(x_{1}) ^{(1)}.
$
This ``commutativity" is lost if we go to the case 
$
k = 2.
$
From (\ref{t2}) - (\ref{tmunu2}) we have:
\bea
s^{\prime}T(T(x_{1})^{(2)}, T^{I_{2}}(x_{2}),\dots,T^{I_{n}}(x_{n})) =
%\nonumber\\
: C_{a}^{\mu}(x_{1})~s^{\prime}T(v_{a\mu}(x_{1})^{(0)}, T^{I_{2}}(x_{2}),\dots,T^{I_{n}}(x_{n})):
\nonumber\\
+ {1\over 2}~: E_{a}^{\nu\mu}(x_{1})~ s^{\prime}T(F_{a\nu\mu}(x_{1})^{(0)}, T^{I_{2}}(x_{2}),\dots,T^{I_{n}}(x_{n})):
\nonumber\\
+ : D_{a}(x_{1})~ s^{\prime}T(u_{a}(x_{1})^{(0)}, T^{I_{2}}(x_{2}),\dots,T^{I_{n}}(x_{n})):
\nonumber\\
+ : B_{a}^{\mu}(x_{1})~\partial_{\mu}^{1} s^{\prime}T(\tilde{u}_{a}(x_{1})^{(0)}, T^{I_{2}}(x_{2}),\dots,T^{I_{n}}(x_{n})):
\nonumber\\
- i~: B_{a\mu}(x_{1}) \Box_{1} T(v_{a}^{\mu}(x_{1})^{(0)}, T^{I_{2}}(x_{2}),\dots,T^{I_{n}}(x_{n})):
\nonumber\\
+ i~: B_{a}(x_{1}) \Box_{1} T(\tilde{u}_{a}(x_{1})^{(0)}, T^{I_{2}}(x_{2}),\dots,T^{I_{n}}(x_{n})):
\label{s2}
\eea
\bea
s^{\prime}T(T^{\mu}(x_{1})^{(2)}, T^{I_{2}}(x_{2}),\dots,T^{I_{n}}(x_{n})) =
%\nonumber\\
: C_{a}^{\mu}(x_{1})~ s^{\prime}T(u_{a}(x_{1})^{(0)}, T^{I_{2}}(x_{2}),\dots,T^{I_{n}}(x_{n})):
\nonumber\\
- : C_{a}^{\nu\mu}(x_{1})~s^{\prime}T(v_{a\nu}(x_{1})^{(0)}, T^{I_{2}}(x_{2}),\dots,T^{I_{n}}(x_{n})):
\nonumber\\
+ : B_{a\nu}(x_{1})~s^{\prime}T(F_{a}^{\nu\mu}(x_{1})^{(0)}, T^{I_{2}}(x_{2}),\dots,T^{I_{n}}(x_{n})):
\nonumber\\
- : B_{a}(x_{1})~\partial^{\mu}_{1} s^{\prime}T(\tilde{u}_{a}(x_{1})^{(0)}, T^{I_{2}}(x_{2}),\dots,T^{I_{n}}(x_{n})):
\nonumber\\
+ i~: B_{a}(x_{1}) \Box_{1} T(v_{a}^{\mu}(x_{1})^{(0)}, T^{I_{2}}(x_{2}),\dots,T^{I_{n}}(x_{n})):
\label{smu2}
\eea
and
\bea
s^{\prime}T(T^{\mu\nu}(x_{1})^{(1)}, T^{I_{2}}(x_{2}),\dots,T^{I_{n}}(x_{n})) =
%\nonumber\\
: B_{a}(x_{1})~s^{\prime}T(F_{a}^{\mu\nu}(x_{1})^{(0)}, T^{I_{2}}(x_{2}),\dots,T^{I_{n}}(x_{n})):
\nonumber\\
- : C_{a}^{\mu\nu}(x_{1})~ s^{\prime}T(u(x_{1})^{(0)}, T^{I_{2}}(x_{2}),\dots,T^{I_{n}}(x_{n})):
\label{smunu2}
\eea
so in the first two relations there are supplementary $\delta$ terms because if we apply (\ref{linear}) to the extra-terms, the 
d'Alembert operator 
$
\Box_{1}
$
acts on the Feynman distributions 
$
D^{F}_{0}(x_{1} - x_{m})
$
and produces 
$
\delta(x_{1} - x_{m}).
$

We also have trivially
\be
sT(T^{I_{1}}(x_{1})^{(3)}, T^{I_{2}}(x_{2}),\dots,T^{I_{n}}(x_{n})) =
T^{I_{1}}(x_{1})~ sT(T^{I_{2}}(x_{2}),\dots,T^{I_{n}}(x_{n})).
\label{s3}
\ee

The final step is to express all relations (\ref{s1}) - (\ref{s3}) in a compact way, using Hopf algebra notations. First
we extend the operations
$
d_{Q}
$
and
$
\partial_{\mu}
$
to product of algebras through:
\be
d_{Q}(a \otimes b) \equiv d_{Q}a \otimes b + ( - 1)^{|a|} a \otimes d_{Q}b
\label{dQ2}
\ee
(for $a$ of fixed grading number) and
\be
\partial_{\mu} (a \otimes b) \equiv \partial_{\mu}a \otimes b + a \otimes \partial_{\mu}b
\label{part2}
\ee
respectively. Next we define 
\be
s \Delta T^{I} \equiv d_{Q} \Delta T^{I} - i~\partial_{\mu}\Delta T^{I\mu} 
\ee
and we can obtain by elementary computations:
\bea
s \Delta T = - i~(\Box v_{a}^{\mu} \otimes B_{a\mu} + B_{a\mu} \otimes \Box v_{a}^{\mu})
+ i~(\Box \tilde{u}_{a} \otimes B_{a} + B_{a} \otimes \Box \tilde{u}_{a})
\nonumber\\
s \Delta T^{\mu} = i~(\Box v_{a}^{\mu} \otimes B_{a} + B_{a} \otimes \Box v_{a}^{\mu})
\nonumber \\
s \Delta T^{\mu\nu} = 0
\eea
so we have on-shell
\be
s \Delta T^{I} \approxeq 0.
\ee

Suppose now that
$
A_{1},\cdots, A_{n} = T, T^{\mu}, T^{\mu\nu}
$
and, according to the previous relations, we have:
\be
s \Delta A_{l} = i~(B_{l}^{(1)} \otimes \Box B_{l}^{(2)} + \Box B_{l}^{(2)} \otimes B_{l}^{(1)})
\ee

With these notations we can write the relations (\ref{s1}) - (\ref{s3}) in a compact way as follows:
\bea
s^{\prime}T(A_{1}(x_{1}), A_{2}(x_{2}),\dots,A_{n}(x_{n})) = 
\sum \tau_{1} :A_{1}^{(1)}(x_{1})~s^{\prime}T_{0}(A_{1}^{(2)}(x_{1}), A_{2}(x_{2}),\dots,A_{n}(x_{n})): 
\nonumber\\
+ i\sum ~:B_{1}^{(1)}(x_{1})~\Box_{1}T_{0}(B_{1}^{(2)}(x_{1}), A_{2}(x_{2}),\dots,A_{n}(x_{n}))
\label{s-hopf-1}
\eea
where the expressions
$
T_{0}(A_{1}^{(1)}(x_{1}), A_{2}(x_{2}),\dots,A_{n}(x_{n})) 
$
are of Wick type only in
$
A_{2},\dots,A_{n}
$
and we have defined the sign
\be
\tau_{1} = ( - 1)^{|A_{1}^{(1)}|}.
\ee

Now we can use induction and iterate the procedure in the entries 
$
A_{2},\dots,A_{n}.
$
The final result is 
\bea
s^{\prime}T(A_{1}(x_{1}), A_{2}(x_{2}),\dots,A_{n}(x_{n}))
\nonumber\\
= \sum \tau_{n} \epsilon_{n} s^{\prime}T_{0}(A_{1}^{(2)}(x_{1}),\dots,A_{n}^{(2)}(x_{n})) :A_{1}^{(1)}(x_{1}) \dots A_{n}^{(1)}(x_{n}): 
\nonumber\\
+ i~\sum_{l=1}^{n} \tau^{\prime}_{l} \epsilon^{\prime}_{l,n}
\Box_{l}T_{0}(A_{1}^{(2)}(x_{1}), \dots, B_{l}^{(2)}(x_{l}),\dots,A_{n}(x_{n}))
:A_{1}^{(1)}(x_{1}) \dots B_{l}^{(1)}(x_{l}) \dots A_{n}^{(1)}(x_{n}):
\label{s-hopf-n}
\eea
where the sign
$
\epsilon_{n}
$
was defined before by (\ref{epsilon})
\be
\tau_{n} = ( - 1)^{\sum_{l=1}^{n} |A_{l}^{(1)}|}
\ee
\be
\tau^{\prime}_{l} = ( - 1)^{\sum_{p=1}^{l-1} |A_{l}|}
\ee
and 
$
\epsilon^{\prime}_{l,n}
$
is obtained from
$
\epsilon_{n}
$
making
$
A_{l}^{(1)} \rightarrow B_{l}^{(1)}, \quad A_{l}^{(2)} \rightarrow B_{l}^{(2)}.
$

We notice that in the right hand sides of the formula (\ref{s-hopf-n}) the operator
$
s^{\prime}
$
acts on numerical distributions so we have
$
s^{\prime}T_{0} = - i~(\delta + \delta^{\prime}).
$

It follows that we can reduce the gauge invariance condition (\ref{brst-n}) to numerical relations of the type
\be
(\delta + \delta^{\prime})T_{0}(A_{1}(x_{1}),\dots,A_{n}(x_{n})) = {\rm delta~terms}.
\ee
However, it is not trivial to determine the explicit form of the right hand side.
\newpage
\section{Second Order Gauge Invariance. Loop Contributions\label{loop}}
To illustrate the advantage of our approach we study in detail the second order case of the pure Yang-Mills model. We can iterate
the formulas (\ref{t2}) - (\ref{tmunu2}) or use directly (\ref{w-hopf-1}) to obtain:
\bea
T(T(x_{1})^{(1)}, T(x_{2})^{(1)}) =
%\nonumber\\
: v_{a\mu}(x_{1})~v_{b\nu}(x_{2}):~T(C_{a}^{\mu}(x_{1})^{(0)}, C_{b}^{\nu}(x_{2})^{(0)})
\nonumber\\
+ {1\over 4}~: F_{a\nu\mu}(x_{1})~F_{b\sigma\rho}(x_{2}):~T(E_{a}^{\mu\nu}(x_{1})^{(0)}, E_{b}^{\rho\sigma}(x_{2})^{(0)})
\nonumber\\
+ {1\over 2}~: v_{a\mu}(x_{1})~F_{b\sigma\rho}(x_{2}):~T(C_{a}^{\mu}(x_{1})^{(0)},E_{b}^{\rho\sigma}(x_{2})^{(0)})
+ ( 1 \longleftrightarrow 2)
\nonumber\\
- [ : u_{a}(x_{1})~\partial_{\mu}\tilde{u}_{b}(x_{2}):~T(D_{a}^{\mu}(x_{1})^{(0)}, B_{b}^{\mu}(x_{2})^{(0)}) + ( 1 \longleftrightarrow 2) ]
\label{t1-2}
\eea
\bea
T(T^{\mu}(x_{1})^{(1)}, T(x_{2})^{(1)}) =
%\nonumber\\
: u_{a}(x_{1})~v_{b\nu}(x_{2}):~T(C_{a}^{\mu}(x_{1})^{(0)}, C_{b}^{\nu}(x_{2})^{(0)})
\nonumber\\
+ {1\over 2}~: u_{a\mu}(x_{1})~F_{b\sigma\rho}(x_{2}):~T(C_{a}^{\nu\mu}(x_{1})^{(0)},E_{b}^{\rho\sigma}(x_{2})^{(0)})
\nonumber\\
- : v_{a\nu}(x_{1})~u_{b}(x_{2}):~T(C_{a}^{\nu\mu}(x_{1})^{(0)}, D_{b}^{\mu}(x_{2})^{(0)})
\nonumber\\
- : F_{a\nu\mu}(x_{1})~u_{b}(x_{2}):~T(B_{a\nu}(x_{1})^{(0)}, D_{b}^{\mu}(x_{2})^{(0)})
\label{tmu1-2}
\eea
\be
T(T^{\mu}(x_{1})^{(1)}, T^{\nu}(x_{2})^{(1)}) =
%\nonumber\\
: u_{a}(x_{1})~u_{b}(x_{2}):~T(C_{a}^{\mu\nu}(x_{1})^{(0)}, C_{b}^{\nu}(x_{2})^{(0)})
\label{tmu-nu-2}
\ee
and
\be
T(T^{\mu\nu}(x_{1})^{(1)}, T(x_{2})^{(1)}) =
%\nonumber\\
 : u_{a}(x_{1})~u_{b}(x_{2}):~T(C_{a}^{\mu\nu}(x_{1})^{(0)}, D_{b}(x_{2})^{(0)})
\label{tmunu1-2}
\ee
\be
T(T^{\mu\nu}(x_{1})^{(1)}, T^{\rho}(x_{2})^{(1)}) = 0.
\label{tmunurho-2}
\ee
The expressions of the type
$
T(C_{a}^{\mu}(x_{1})^{(0)}, C_{b}^{\nu}(x_{2})^{(0)}),
$
etc., are associated with $2$-loop contributions in the second order of the perturbation theory. We have now the following result:
\begin{thm}
The following relations are true:
\bea
sT(T(x_{1})^{(1)}, T(x_{2})^{(1)}) =
\nonumber\\
- [: u_{a}(x_{1})~v_{b\nu}(x_{2}):~s^{\prime}T(D_{a}(x_{1})^{(0)}, C_{b}^{\nu}(x_{2})^{(0)}) + ( 1 \longleftrightarrow 2)]
\nonumber\\
- {1\over 2}~: [ u_{a}(x_{1})~F_{b\sigma\rho}(x_{2}):~s^{\prime}T(D_{a}(x_{1})^{(0)},E_{b}^{\rho\sigma}(x_{2})^{(0)})
+ ( 1 \longleftrightarrow 2) ]
\label{st1-2}
\eea
\be
sT(T^{\mu}(x_{1})^{(1)}, T(x_{2})^{(1)}) =
\nonumber\\
: u_{a}(x_{1})~u_{b}(x_{2}):~s^{\prime}T(C_{a}^{\mu}(x_{1})^{(0)}, D_{b}(x_{2})^{(0)})
\label{stmu1-2}
\ee
\be
sT(T^{\mu}(x_{1})^{(1)}, T^{\nu}(x_{2})^{(1)}) = 0
\label{stmu-nu-2}
\ee
\be
sT(T^{\mu\nu}(x_{1})^{(1)}, T(x_{2})^{(1)}) = 0
\label{stmunu1-2}
\ee
\be
sT(T^{\mu\nu}(x_{1})^{(1)}, T^{\rho}(x_{2})^{(1)}) = 0.
\label{stmunurho-2}
\ee
\label{stt-loop}
\end{thm}
{\bf Proof:} We sketch to proof of the first relation. We have to compute the following expressions:
$
A = d_{Q}T(T(x_{1})^{(1)}, T(x_{2})^{(1)}), B = - i~\partial_{\mu}^{1}T(T^{\mu}(x_{1})^{(1)}, T(x_{2})^{(1)}),
C = - i~\partial_{\mu}^{2}T(T(x_{1})^{(1)}, T^{\mu}(x_{2})^{(1)})
$
and we use for this the relations (\ref{t1-2}) and (\ref{tmu1-2}). In the sum 
$
A + B + C
$
(which is the left hand side of (\ref{st1-2})), there are a lot of cancelations; we are left with
\bea
a = - i :u_{a}(x_{1})~v_{b\nu}(x_{2}): [  \partial_{\mu}^{1}T(C_{a}^{\mu}(x_{1})^{(0)}, C_{b}^{\nu}(x_{2})^{(0)})
+ \partial_{\mu}^{2}T(D_{a}(x_{1})^{(0)}, C_{b}^{\nu\mu}(x_{2})^{(0)})]
\nonumber\\
= i :u_{a}(x_{1})~v_{b\nu}(x_{2}):\delta T(D_{a}(x_{1})^{(0)}, C_{b}^{\nu}(x_{2})^{(0)})
\nonumber\\
= - :u_{a}(x_{1})~v_{b\nu}(x_{2}):~s^{\prime}T(D_{a}(x_{1})^{(0)}, C_{b}^{\nu}(x_{2})^{(0)})
\eea
(which is the first term from the right hand side of (\ref{st1-2})) and two more similar terms.
$\qed$

All the expressions in the right hand sides from the preceding theorem are in fact zero:
\begin{thm}
The following relations are true:
\bea
s^{\prime}T(D_{a}(x_{1})^{(0)}, C_{b}^{\nu}(x_{2})^{(0)}) = 0
\nonumber\\
s^{\prime}T(D_{a}(x_{1})^{(0)},E_{b}^{\rho\sigma}(x_{2})^{(0)}) = 0
\eea
\label{loop-DCE}
\end{thm}
{\bf Proof:}
The basic distributions used for the computations are:
\be
d_{0,0} \equiv  {1\over 2}~[(D_{0}^{(+)})^{2} - (D_{0}^{(-)})^{2} ]
\label{d00}
\ee
and associated ones:
\bea
d_{\mu\nu} \equiv D^{(+)}_{0} \partial_{\mu}\partial_{\nu}D^{(+)}_{0}
- D^{(-)}_{0} \partial_{\mu}\partial_{\nu}D^{(-)}_{0}, \qquad
%\nonumber\\
f_{\mu\nu} \equiv \partial_{\mu}D^{(+)}_{0} \partial_{\nu}D^{(+)}_{0}
- \partial_{\mu}D^{(-)}_{0} \partial_{\nu}D^{(-)}_{0}
\label{d+f}
\eea
All these distributions have causal support; in fact one can derive that
\bea
d_{\mu\nu}  = {2\over 3}~\left( \partial_{\mu}\partial_{\nu} - {1\over 4} \eta_{\mu\nu}~\square\right)d_{0,0}, \qquad
%\nonumber\\
f_{\mu\nu} = {1\over 3}~\left( \partial_{\mu}\partial_{\nu} + {1\over 2} \eta_{\mu\nu}~\square\right)d_{0,0}.
\label{d+f-d00}
\eea
The causal splitting of the distribution
$
d_{0,0}
$
gives a Feynman propagator
$
d^{F}_{0,0}
$
which is not unique because the degree of singularity of 
$
d_{0,0}
$
in $0$; any choice will be good. More important, we can obtain the corresponding Feynman propagators by
\bea
d^{F}_{\mu\nu}  = {2\over 3}~\left( \partial_{\mu}\partial_{\nu} - {1\over 4} \eta_{\mu\nu}~\square\right)d^{F}_{0,0}, \qquad
%\nonumber\\
f^{F}_{\mu\nu} = {1\over 3}~\left( \partial_{\mu}\partial_{\nu} + {1\over 2} \eta_{\mu\nu}~\square\right)d^{F}_{0,0}.
\label{d+f-F}
\eea

With these conventions, we can obtain
\bea
T(C_{a}^{\mu}(x_{1})^{(0)}, C_{b}^{\nu}(x_{2})^{(0)}) = {2 \over 3}~g_{ab}~(\partial^{\mu}\partial^{\nu} - \eta^{\mu\nu} \Box)d^{F}_{0,0}
\nonumber\\
T(C_{a}^{\mu}(x_{1})^{(0)},E_{b}^{\rho\sigma}(x_{2})^{(0)}) = {1 \over 2}~g_{ab}~
(\eta^{\mu\sigma} \partial^{\rho} - \eta^{\mu\rho} \partial^{\sigma})d^{F}_{0,0}
\nonumber\\
T(D_{a}(x_{1})^{(0)}, B_{b}^{\mu}(x_{2})^{(0)}) = - {1 \over 2}~g_{ab}~\partial^{\mu}d^{F}_{0,0}
\nonumber\\
T(D_{a}(x_{1})^{(0)}, C_{b}^{\mu\nu}(x_{2})^{(0)}) = 0 
\label{BB-loop}
\eea
where
\be
g_{ab} \equiv f_{acd}~f_{bcd}
\ee
and from here
\bea
\delta T(D_{a}(x_{1})^{(0)}, C_{b}^{\nu}(x_{2})^{(0)}) 
\nonumber\\
\equiv - [\partial_{\mu}^{1}T(C_{a}^{\mu}(x_{1})^{(0)}, C_{b}^{\nu}(x_{2})^{(0)})
- \partial_{\mu}^{2}T(D_{a}(x_{1})^{(0)}, C_{b}^{\nu\mu}(x_{2})^{(0)}] = 0
\nonumber\\
(\delta + \delta^{\prime}) T(D_{a}(x_{1})^{(0)}, E_{b}^{\rho\sigma}(x_{2})^{(0)}) =
\nonumber\\
\equiv - \partial_{\mu}^{1}T(C_{a}^{\mu}(x_{1})^{(0)}, E_{b}^{\rho\sigma}(x_{2})^{(0)}
- [ \partial^{\sigma}_{2}T(D_{a}(x_{1})^{(0)}, B_{b}^{\rho}(x_{2})^{(0)})  - (\rho \leftrightarrow \sigma) ]
\nonumber\\
- T(D_{a}(x_{1})^{(0)}, C_{b}^{\rho\sigma}(x_{2})^{(0)}) = 0
\eea
and this proves that there are no anomalies from the loop contributions in the second order of the perturbation theory.
$\qed$

We can extend the previous arguments to loop contributions associated to chronological products of Wick submonomials. 
The following formulas are true:
\bea
T(B_{a}^{\mu}(x_{1})^{(0)}, T(x_{2})) = - u_{b}(x_{2})~T(B_{a}^{\mu}(x_{1})^{(0)}, D_{b}^{(0)}(x_{2}))
\nonumber\\
\nonumber\\
T(B_{a}^{\mu}(x_{1})^{(0)}, T^{I}(x_{2})) = 0, \quad {\rm for} \quad |I| = 1,2
\nonumber\\
\nonumber\\
T(C_{a}^{\mu}(x_{1})^{(0)}, T(x_{2})) = v_{b\nu}(x_{2})~T(C_{a}^{\mu}(x_{1})^{(0)}, C_{b}^{\nu}(x_{1})(x_{2})^{(0)})
\nonumber\\
+ {1 \over 2}~ F_{b\sigma\rho}(x_{2})~T(C_{a}^{\mu}(x_{1})^{(0)}, E_{b}^{\rho\sigma}(x_{2})^{(0)})
\nonumber\\
\nonumber\\
T(C_{a}^{\mu}(x_{1})^{(0)}, T^{\nu}(x_{2})) = u_{b}(x_{2})~T(C_{a}^{\mu}(x_{1})^{(0)}, C_{b}^{\nu}(x_{1})(x_{2})^{(0)})
\nonumber\\
\nonumber\\
T(C_{a}^{\mu}(x_{1})^{(0)}, T^{\rho\sigma}(x_{2})) = 0
\nonumber\\
\nonumber\\
T(D_{a}(x_{1})^{(0)}, T(x_{2})) = - \partial_{\mu}\tilde{u}_{b}^{\mu}(x_{2})~T(D_{a}(x_{1})^{(0)}, B_{b}^{\mu}(x_{2})^{(0)})
\nonumber\\
\nonumber\\
T(D_{a}(x_{1})^{(0)}, T^{\mu}(x_{2})) = v_{b\nu}(x_{2})~T(D_{a}(x_{1})^{(0)}, C_{b}^{\nu\mu}(x_{1})(x_{2})^{(0)})
\nonumber\\
+ :F_{b}^{\mu\nu}(x_{2})~T(D_{a}(x_{1})^{(0)}, B_{b\nu}(x_{2})^{(0)}):
\nonumber\\
\nonumber\\
T(D_{a}(x_{1})^{(0)}, T^{\mu\nu}(x_{2})) = u_{b}(x_{2})~T(D_{a}(x_{1})^{(0)}, C_{b}^{\mu\nu}(x_{1})(x_{2})^{(0)})
\nonumber\\
\nonumber\\
T(E_{a}^{\mu\nu}(x_{1})^{(0)}, T(x_{2})) = v_{b\rho}(x_{2})~T(E_{a}^{\mu\nu}(x_{1})^{(0)}, C_{b}^{\rho}(x_{2})^{(0)}) 
\nonumber\\
+ {1 \over 2}~ :F_{b\sigma\rho}(x_{2})~T(E_{a}^{\mu\nu}(x_{1})^{(0)}, E_{b}^{\rho\sigma}(x_{2})^{(0)})
\nonumber\\
\nonumber\\
T(E_{a}^{\mu\nu}(x_{1})^{(0)}, T^{\rho}(x_{2})) = u_{b}(x_{2})~T(E_{a}^{\mu\nu}(x_{1})^{(0)}, C_{b}^{\rho}(x_{2})^{(0)}) 
\nonumber\\
\nonumber\\
T(E_{a}^{\mu\nu}(x_{1})^{(0)}, T^{\rho\sigma}(x_{2})) = 0
\nonumber\\
\nonumber\\
T(B_{a}(x_{1})^{(0)}, T^{I}(x_{2})) = 0,\quad \forall I
\nonumber\\
\nonumber\\
T(C_{a}^{\mu\nu}(x_{1})^{(0)}, T(x_{2})) = - u_{b}(x_{2})~T(C_{a}^{\mu\nu}(x_{1})^{(0)}, D_{b}(x_{2})^{(0)})
\nonumber\\
\nonumber\\
T(C_{a}^{\mu\nu}(x_{1})^{(0)}, T^{I}(x_{2})) = 0, \quad {\rm for} \quad |I| = 1,2.
\label{BCDE-T0}
\eea
Then we similar to theorem \ref{stt-loop}:
\begin{thm}
\bea
s^{\prime}T(B_{a}^{\mu}(x_{1})^{(0)}, T^{I}(x_{2})) = 0, \quad \forall I
\nonumber\\
\nonumber\\
s^{\prime}T(C_{a}^{\mu}(x_{1})^{(0)}, T(x_{2})) = i u_{b}(x_{2})~\delta T(C_{a}^{\mu}(x_{1})^{(0)}, D_{b}(x_{1})(x_{2})^{(0)}) 
\nonumber\\
\nonumber\\
s^{\prime}T(C_{a}^{\mu}(x_{1})^{(0)}, T^{I}(x_{2})) = 0 \quad {\rm for} \quad |I| = 1,2 
\nonumber\\
\nonumber\\
s^{\prime}T(D_{a}(x_{1})^{(0)}, T(x_{2})) = - i v_{b\nu}(x_{2})~\delta T(D_{a}(x_{1})^{(0)}, C_{b}^{\nu}(x_{2})^{(0)})
\nonumber\\
- {i \over 2}~ F_{b\sigma\rho}(x_{2})~(\delta + \delta^{\prime})T(D_{a}(x_{1})^{(0)}, E_{b}^{\rho\sigma}(x_{2})^{(0)})
\nonumber\\
\nonumber\\
s^{\prime}T(D_{a}(x_{1})^{(0)}, T^{\mu}(x_{2})) = - i u_{b}(x_{2})~\delta T(D_{a}(x_{1})^{(0)}, C_{b}^{\mu}(x_{1})(x_{2})^{(0)})
\nonumber\\
\nonumber\\
s^{\prime}T(D_{a}(x_{1})^{(0)}, T^{\mu\nu}(x_{2})) = 0
\nonumber\\
\nonumber\\
s^{\prime}T(E_{a}^{\mu\nu}(x_{1})^{(1)}, T(x_{2})) = 
i u_{b}(x_{2})~(\delta + \delta^{\prime})T(E_{a}^{\mu\nu}(x_{1})^{(0)}, D_{b}(x_{2})^{(0)}) 
\nonumber\\
\nonumber\\
s^{\prime}T(E_{a}^{\mu\nu}(x_{1})^{(1)}, T^{I}(x_{2})) = 0 \quad {\rm for} \quad |I| = 1,2 
\nonumber\\
\nonumber\\
s^{\prime}T(B_{a}(x_{1})^{(0)}, T^{I}(x_{2})) = 0, \quad \forall I
\nonumber\\
\nonumber\\
T(C_{a}^{\mu\nu}(x_{1})^{(1)}, T^{I}(x_{2})) = 0, \quad \forall I.
\label{sBCDE-T0}
\eea
\end{thm}
\newpage
Now, using theorem \ref{loop-DCE}, we have
\begin{thm}
\bea
s^{\prime}T(C_{a}^{\mu}(x_{1})^{(0)}, T^{I}(x_{2})) = 0 \quad \forall I 
\nonumber\\
\nonumber\\
s^{\prime}T(D_{a}(x_{1})^{(0)}, T^{I}(x_{2})) = 0 \quad \forall  I 
\nonumber\\
\nonumber\\
s^{\prime}T(E_{a}^{\mu\nu}(x_{1})^{(0)}, T^{I}(x_{2})) = 0 \quad \forall I.
\label{sBCDE-loop}
\eea
\end{thm}

Finally we have:
\begin{thm}
The following relation is true:
\be
s^{\prime}T(\xi_{a}\cdot T^{I}(x_{1})^{(0)}, \xi_{b}\cdot T^{J}(x_{2})^{(0)}) = 
- i~(\delta + \delta^{\prime})T(\xi_{a}\cdot T^{I}(x_{1})^{(0)}, \xi_{b}\cdot T^{J}(x_{2})^{(0)}) = 0.
\ee
\end{thm}
{\bf Proof} We use the relations (\ref{BB-loop}).
$\qed$
\newpage
\section{Second Order Gauge Invariance. Tree Contributions\label{tree}}

Now we study tree contributions. We can iterate (\ref{t1}) - (\ref{tmunu1}) or use directly (\ref{w-hopf-1}) to obtain:
\bea
T(T(x_{1})^{(2)}, T(x_{2})^{(2)}) =
%\nonumber\\
: C_{a}^{\mu}(x_{1})~C_{b}^{\nu}(x_{2}):~T(v_{a\mu}(x_{1})^{(0)}, v_{b\nu}(x_{2})^{(0)})
\nonumber\\
+ {1\over 4}~: E_{a}^{\nu\mu}(x_{1})~E_{b}^{\sigma\rho}(x_{2}):~T(F_{a\mu\nu}(x_{1})^{(0)}, F_{b\rho\sigma}(x_{2})^{(0)})
\nonumber\\
+ {1\over 2}~[ : E_{a}^{\mu\nu}(x_{1})~C_{b}^{\rho}(x_{2}):~T(F_{a\nu\mu}(x_{1})^{(0)},v_{b\rho}(x_{2})^{(0)})
+ ( 1 \longleftrightarrow 2) ]
\nonumber\\
- [ : D_{a}(x_{1})~B_{b}^{\mu}(x_{2}):~T(u_{a}(x_{1})^{(0)}, \tilde{u}_{b,\mu}(x_{2})^{(0)}) + ( 1 \longleftrightarrow 2)]
\label{t2-2}
\eea
\bea
T(T^{\mu}(x_{1})^{(2)}, T(x_{2})^{(2)}) =
%\nonumber\\
 : C_{a}^{\mu}(x_{1})~B_{b}^{\nu}(x_{2}):~T(u_{a,\nu}(x_{1})^{(0)}, \tilde{u}_{b}(x_{2})^{(0)})
\nonumber\\
+ : C_{a}^{\nu\mu}(x_{1})~C_{b}^{\rho}(x_{2}):~T(v_{a\nu}(x_{1})^{(0)},v_{b\rho}(x_{2})^{(0)})
\nonumber\\
+ {1 \over 2}~: C_{a}^{\nu\mu}(x_{1})~E_{b}^{\rho\sigma}(x_{2}):~T(v_{a\nu}(x_{1})^{(0)}, F_{b\sigma\rho}(x_{2})^{(0)})
\nonumber\\
- : B_{a\nu}(x_{1})~C_{b}^{\rho}(x_{2}):~T(F_{a}^{\nu\mu}(x_{1})^{(0)}, v_{b\rho}(x_{2})^{(0)})
\nonumber\\
- {1 \over 2}~: B_{a\nu}(x_{1})~E_{b}^{\rho\sigma}(x_{2}):~T(F_{a}^{\nu\mu}(x_{1})^{(0)}, F_{b\sigma\rho}(x_{2})^{(0)})
\nonumber\\
- : B_{a}(x_{1})~D_{b}(x_{2}):~T({\tilde{u}_{a,}~}^{\mu}(x_{1})^{(0)}, u_{b}(x_{2})^{(0)})
\label{tmu2-2}
\eea
\bea
T(T^{\mu}(x_{1})^{(2)}, T^{\nu}(x_{2})^{(2)}) =
%\nonumber\\
- : C_{a}^{\mu}(x_{1})~B_{b}(x_{2}):~T(u_{a}(x_{1})^{(0)}, \tilde{u}_{b,}^{\nu}(x_{2})^{(0)})
\nonumber\\
+ : C_{a}^{\rho\mu}(x_{1})~C_{b}^{\sigma\nu}(x_{2}):~T(v_{a\rho}(x_{1})^{(0)},v_{b\sigma}(x_{2})^{(0)})
\nonumber\\
- : C_{a}^{\rho\mu}(x_{1})~B_{b\sigma}(x_{2}):~T(v_{a\rho}(x_{1})^{(0)}, F_{b}^{\sigma\nu}(x_{2})^{(0)})
\nonumber\\
- : B_{a\rho}(x_{1})~C_{b}^{\sigma\nu}(x_{2}):~T(F_{a}^{\rho\mu}(x_{1})^{(0)}, v_{b\sigma}(x_{2})^{(0)})
\nonumber\\
+ : B_{a\rho}(x_{1})~B_{b\sigma}(x_{2}):~T(F_{a}^{\rho\mu}(x_{1})^{(0)}, F_{b}^{\sigma\nu}(x_{2})^{(0)})
\nonumber\\
- : B_{a}(x_{1})~C_{b}^{\nu}(x_{2}):~T({\tilde{u}_{a,}~}^{\mu}(x_{1})^{(0)}, u_{b}(x_{2})^{(0)})
\label{tmu-nu2-2}
\eea
\bea
T(T^{\mu\nu}(x_{1})^{(2)}, T(x_{2})^{(2)}) =
%\nonumber\\
 : B_{a}(x_{1})~C_{b}^{\rho}(x_{2}):~T(F^{\mu\nu}_{a}(x_{1})^{(0)}, v_{b\rho}(x_{2})^{(0)})
\nonumber\\
+ {1 \over 2}~: B_{a}(x_{1})~E_{b}^{\rho\sigma}(x_{2}):~T(F_{a}^{\mu\nu}(x_{1})^{(0)}, F_{b\sigma\rho}(x_{2})^{(0)})
\nonumber\\
+ : C_{a}^{\mu\nu}(x_{1})~B_{b}^{\rho}(x_{2}):~T(u_{a}(x_{1})^{(0)}, \tilde{u}_{b,\rho}(x_{2})^{(0)})
\label{tmunu2-2}
\eea
\bea
T(T^{\mu\nu}(x_{1})^{(2)}, T^{\rho}(x_{2})^{(2)}) =
%\nonumber\\
 : B_{a}(x_{1})~C_{b}^{\sigma\rho}(x_{2}):~T(F^{\mu\nu}_{a}(x_{1})^{(0)}, v_{b\sigma}(x_{2})^{(0)})
\nonumber\\
- : B_{a}(x_{1})~B_{b\sigma}(x_{2}):~T(F_{a}^{\mu\nu}(x_{1})^{(0)}, F_{b}^{\sigma\rho}(x_{2})^{(0)})
\nonumber\\
- : C_{a}^{\mu\nu}(x_{1})~B_{b}(x_{2}):~T(u_{a}(x_{1})^{(0)}, {\tilde{u}_{b,}~}^{\rho}(x_{2})^{(0)})
\label{tmunu-rho2-2}
\eea
\be
T(T^{\mu\nu}(x_{1})^{(2)}, T^{\rho\sigma}(x_{2})^{(2)}) =
%\nonumber\\
 : B_{a}(x_{1})~B_{b}(x_{2}):~T(F^{\mu\nu}_{a}(x_{1})^{(0)}, F_{b}^{\sigma\rho}(x_{2})^{(0)})
\label{tmunu-rhosigma2-2}
\ee
We must give the values of the chronological products
$
T(\xi^{(0)}_{p}(x_{1}), \xi^{(0)}_{q}(x_{2}))
$
which are not unique. For the pure Yang-Mills model the causal commutators of the basic fields are
\bea
D(v^{\mu}_{a}(x_{1}), v^{\nu}_{b}(x_{2})) \equiv [ v^{\mu}_{a}(x_{1}), v^{\nu}_{b}(x_{2}) ] 
= i~\eta^{\mu\nu}~\delta_{ab}~D_{0}(x_{1} - x_{2}),
\nonumber \\
D(u_{a}(x_{1}), \tilde{u}_{b}(x_{2})) \equiv [ u_{a}(x_{1}), \tilde{u}_{b}(x_{2}) ] = - i~\delta_{ab}~D_{0}(x_{1} - x_{2}),
\nonumber\\
D(\tilde{u}_{a}(x_{1}), u_{b}(x_{2})) \equiv [ \tilde{u}_{a}(x_{1}), u_{b}(x_{2}) ] = i~\delta_{ab}~D_{0}(x_{1} - x_{2}).
\label{comm-2-massless-vector}
\eea
where in the left hand side we have the graded commutator. The causal splitting 
$
D = D^{\rm adv} - D^{\rm ret}
$
is unique because the degree of singularity of
$
D_{0}
$
is 
$
\omega = - 2
$ 
and we obtain 
\bea
T(v^{\mu}_{a}(x_{1})^{(0)}, v^{\nu}_{b}(x_{2})^{(0)}) = i~\eta^{\mu\nu}~\delta_{ab}~D^{F}_{0}(x_{1} - x_{2}),
\nonumber \\
T( u_{a}(x_{1})^{(0)}, \tilde{u}_{b}(x_{2})^{(0)}) = - i~\delta_{ab}~D^{F}_{0}(x_{1} - x_{2}),
\nonumber\\
T( \tilde{u}_{a}(x_{1})^{(0)}, u_{b}(x_{2})^{(0)}) = i~\delta_{ab}~D^{F}_{0}(x_{1} - x_{2}).
\label{chr-2-massless-vector}
\eea
From the previous relations we also have uniquely:
\bea
T(\xi^{(0)}_{a,\mu}(x_{1}), \xi^{(0)}_{b}(x_{2})) = \partial_{\mu}^{1}T(\xi_{a}(x_{1}), \xi_{b}(x_{2}))
\nonumber\\
T(\xi^{(0)}_{a}(x_{1}), \xi^{(0)}_{b,\nu}(x_{2})) =  \partial_{\nu}^{2}D^{F}(\xi_{a}(x_{1}), \xi_{b}(x_{2})).
\eea
However the causal splitting of 
$
T(\xi_{a,\mu}(x_{1})^{(0)}, \xi_{b,\nu}(x_{2})^{(0)})
$
is not unique because the distribution has the degree of singularity 
$
\omega = 0.
$
This was noticed for the first time in \cite{ASD} and \cite{DKS}. A possible choice is the {\it canonical} splitting, 
following from (\ref{2-point-der}):
\be
T(\xi^{(0)}_{a,\mu}(x_{1}), \xi^{(0)}_{b,\nu}(x_{2})) = 
i \partial_{\mu}^{1}\partial_{\nu}^{2}T(\xi_{a}(x_{1})^{(0)}, \xi_{b}(x_{2})^{(0)})
\label{chr-3-massless-vector}
\ee
and this gives from (\ref{t2-2}) - (\ref{tmunu-rhosigma2-2})
\bea
T(T(x_{1})^{(2)}, T(x_{2})^{(2)}) =
%\nonumber\\
i~D^{F}_{0}( x_{1} - x_{2})~ : C_{a}^{\mu}(x_{1})~C_{a\mu}(x_{2}):
\nonumber\\
+ i~\partial_{\mu}D^{F}_{0}( x_{1} - x_{2})~[ : C_{a\nu}(x_{1})~E_{a}^{\mu\nu}(x_{2}): - : D_{a}(x_{1})~B_{a}^{\mu}(x_{2}):]
+ ( 1 \longleftrightarrow 2)
\nonumber\\
- i~\partial_{\mu}\partial_{\nu}D^{F}_{0}( x_{1} - x_{2})~: E_{a}^{\mu\rho}(x_{1})~{E_{a}^{\nu}}_{\rho}(x_{2}):
\label{t2-2a}
\eea
\bea
T(T^{\mu}(x_{1})^{(2)}, T(x_{2})^{(2)}) =
%\nonumber\\
- i~D^{F}_{0}( x_{1} - x_{2})~: C_{a}^{\mu\nu}(x_{1})~C_{b\nu}(x_{2}):
\nonumber\\
+ i~\partial_{\nu}D^{F}_{0}( x_{1} - x_{2})~
[ : C_{a}^{\mu}(x_{1})~B_{a}^{\nu}(x_{2}): - : C_{a}^{\mu\rho}(x_{1})~{E_{a}^{\nu}}_{\rho}(x_{2}): 
- : B_{a}^{\nu}(x_{1})~C_{a}^{\mu}(x_{2}):]
\nonumber\\
+ i~\partial^{\mu}D^{F}_{0}( x_{1} - x_{2})~ [ : B_{a\nu}(x_{1})~C_{a}^{\nu}(x_{2}): -  :B_{a}(x_{1})~D_{a}(x_{2}):]
\nonumber\\
+ i~\partial_{\nu}\partial_{\rho}D^{F}_{0}( x_{1} - x_{2})~: B_{a}^{\nu}(x_{1})~E_{a}^{\mu\rho}(x_{2}):
%\nonumber\\
+ i~\partial^{\mu}\partial_{\nu}D^{F}_{0}( x_{1} - x_{2})~: B_{a\rho}(x_{1})~E_{a}^{\nu\rho}(x_{2}):
\label{tmu2-2a}
\eea
\bea
T(T^{\mu}(x_{1})^{(2)}, T^{\nu}(x_{2})^{(2)}) =
%\nonumber\\
i~D^{F}_{0}( x_{1} - x_{2})~: C_{a}^{\mu\rho}(x_{1})~{C_{a}^{\nu}}_{\rho}(x_{2}):
\nonumber\\
+ i~\partial^{\mu}D^{F}_{0}( x_{1} - x_{2})~
[ : B_{a\rho}(x_{1})~C_{a}^{\rho\nu}(x_{2}): - : B_{a}(x_{1})~C_{a}^{\nu}(x_{2}): ]
\nonumber\\
- i~\partial^{\nu}D^{F}_{0}( x_{1} - x_{2})~ [ : C_{a}^{\mu}(x_{1})~B_{a}(x_{2}): +  :C_{a}^{\rho\mu}(x_{1})~B_{a\rho}(x_{2}):]
\nonumber\\
- i~\partial_{\rho}D^{F}_{0}( x_{1} - x_{2})~ 
[ : C_{a}^{\mu\nu}(x_{1})~B_{a}^{\rho}(x_{2}): + : B_{a}^{\rho}(x_{1})~C_{a}^{\mu\nu}(x_{2}): ]
\nonumber\\
- i [ \eta^{\mu\nu}~\partial_{\rho}\partial_{\sigma}D^{F}_{0}( x_{1} - x_{2})~: B_{a}^{\rho}(x_{1})~B_{a}^{\sigma}(x_{2}):
+  \partial^{\mu}\partial^{\nu}D^{F}_{0}( x_{1} - x_{2})~: B_{a\rho}(x_{1})~B_{a}^{\rho}(x_{2}): ]
\nonumber\\
+ i [ \partial^{\mu}\partial_{\rho}D^{F}_{0}( x_{1} - x_{2})~: B_{a}^{\nu}(x_{1})~B_{a}^{\rho}(x_{2}):
+  \partial^{\nu}\partial_{\rho}D^{F}_{0}( x_{1} - x_{2})~: B_{a}^{\rho}(x_{1})~B_{a}^{\mu}(x_{2}): ]
\label{tmu-nu2-2a}
\eea
\bea
T(T^{\mu\nu}(x_{1})^{(2)}, T(x_{2})^{(2)}) =
%\nonumber\\
 i~[ \partial^{\mu}D^{F}_{0}( x_{1} - x_{2})~: B_{a}(x_{1})~C_{a}^{\nu}(x_{2}): - (\mu \leftrightarrow \nu) ]
\nonumber\\
+ i~\partial_{\rho}D^{F}_{0}( x_{1} - x_{2})~ : C_{a}^{\mu\nu}(x_{1})~B_{a}^{\rho}(x_{2}):
\nonumber\\
- i~[ \partial^{\mu}\partial_{\rho}D^{F}_{0}( x_{1} - x_{2})~: B_{a}(x_{1})~E_{a}^{\nu\rho}(x_{2}): - (\mu \leftrightarrow \nu) ]
\label{tmunu2-2a}
\eea
\bea
T(T^{\mu\nu}(x_{1})^{(2)}, T^{\rho}(x_{2})^{(2)}) =
%\nonumber\\
 i~[ \partial^{\mu}D^{F}_{0}( x_{1} - x_{2})~: B_{a}(x_{1})~C_{a}^{\nu\rho}(x_{2}): - (\mu \leftrightarrow \nu) ]
\nonumber\\
- i~\partial^{\rho}D^{F}_{0}( x_{1} - x_{2})~ : C_{a}^{\mu\nu}(x_{1})~B_{a}(x_{2}):
\nonumber\\
- i [ \partial_{\mu}\partial_{\rho}D^{F}_{0}( x_{1} - x_{2})~: B_{a}(x_{1})~B_{a}^{\nu}(x_{2}): - (\mu \leftrightarrow \nu )]
\nonumber\\
- i [ \eta^{\mu\rho}~ \partial^{\nu}\partial_{\sigma}D^{F}_{0}( x_{1} - x_{2})~: B_{a}(x_{1})~B_{a}^{\sigma}(x_{2}): 
- (\mu \leftrightarrow \nu) ]
\label{tmunu-rho2-2a}
\eea
\bea
T(T^{\mu\nu}(x_{1})^{(2)}, T^{\rho\sigma}(x_{2})^{(2)}) =
%\nonumber\\
- i~( \eta^{\nu\sigma} \partial^{\mu}\partial^{\rho} - \eta^{\nu\rho} \partial^{\mu}\partial^{\sigma} 
%\nonumber\\
+ \eta^{\mu\rho} \partial^{\nu}\partial^{\sigma} - \eta^{\mu\sigma} \partial^{\nu}\partial^{\rho})D^{F}_{0}( x_{1} - x_{2})
\nonumber\\
: B_{a}(x_{1})~B_{a}(x_{2}): 
\label{tmunu-rhosigma2-2a}
\eea

From these formulas we can determine now if gauge invariance is true; in fact, we have anomalies, as it is well known:
\begin{thm}
The following formulas are true
\bea
s^{\prime}T(T(x_{1})^{(2)}, T(x_{2})^{(2)}) =
%\nonumber\\
2~\delta( x_{1} - x_{2})~ (: B_{a\mu}~C_{a}^{\mu}: - : B_{a}~D_{a}:)(x_{2})
\nonumber\\
+ \partial_{\mu}\delta( x_{1} - x_{2})~[ : B_{a\nu}(x_{1})~E_{a}^{\mu\nu}(x_{2}): - : E_{a}^{\mu\nu}(x_{1})~B_{a}^{\mu}(x_{2}):]
\label{st2-2a}
\eea
\bea
s^{\prime}T(T^{\mu}(x_{1})^{(2)}, T(x_{2})^{(2)}) =
%\nonumber\\
\delta( x_{1} - x_{2})~ (- 2: B_{a}~C_{a}^{\mu}: + : C^{\mu\nu}_{a}~B_{a\nu}:)(x_{2})
\nonumber\\
+ \partial_{\nu}\delta( x_{1} - x_{2})~[ : B_{a}(x_{1})~E_{a}^{\mu\nu}(x_{2}): + : B_{a}^{\nu}(x_{1})~B_{a}^{\mu}(x_{2}):]
\nonumber\\
- \partial^{\mu}\delta( x_{1} - x_{2})~ : B_{a\nu}(x_{1})~B_{a}^{\nu}(x_{2}):
\label{stmu2-2a}
\eea
\bea
s^{\prime}T(T^{\mu}(x_{1})^{(2)}, T^{\nu}(x_{2})^{(2)}) =
%\nonumber\\
- 2 \delta( x_{1} - x_{2})~: B_{a}~C_{a}^{\mu\nu}:(x_{2})
\nonumber\\
+ \partial^{\mu}\delta( x_{1} - x_{2})~: B_{a}^{\nu}(x_{1})~B_{a}(x_{2}): 
+ \partial^{\nu}\delta( x_{1} - x_{2})~: B_{a}(x_{1})~B^{\mu}_{a}(x_{2}):
\nonumber\\
- \eta^{\mu\nu}~\partial_{\rho}\delta( x_{1} - x_{2})~ [ : B_{a}(x_{1})~B_{a}^{\rho}(x_{2}): + : B_{a}^{\rho}(x_{1})~B_{a}(x_{2}):]
\label{stmu-nu2-2a}
\eea
\bea
s^{\prime}T(T^{\mu\nu}(x_{1})^{(2)}, T(x_{2})^{(2)}) =
%\nonumber\\
\delta( x_{1} - x_{2})~: C^{\mu\nu}_{a}~B_{a}:(x_{2})
\nonumber\\
+ \partial^{\mu}\delta( x_{1} - x_{2})~: B_{a}(x_{1})~B^{\nu}_{a}(x_{2}): - (\mu \leftrightarrow \nu)
\label{stmunu2-2a}
\eea
\bea
s^{\prime}T(T^{\mu\nu}(x_{1})^{(2)}, T^{\rho}(x_{2})^{(2)}) =
\eta^{\mu\rho}~\partial^{\nu}\delta( x_{1} - x_{2})~: B_{a}(x_{1})~B_{a}(x_{2}): - (\mu \leftrightarrow \nu)
\label{stmunu-rho2-2a}
\eea
\bea
s^{\prime}T(T^{\mu\nu}(x_{1})^{(2)}, T^{\rho\sigma}(x_{2})^{(2)}) = 0
\label{stmunu-rhosigma2-2a}
\eea
\label{sTT}
\end{thm}
{\bf Proof:} We consider for illustration only the first relation. We have to compute the expressions
$
A = d_{Q}T(T(x_{1})^{(2)}, T(x_{2})^{(2)}), B = - i\partial_{\mu}^{1}T(T^{\mu}(x_{1})^{(2)}, T(x_{2})^{(2)}),
C = - i\partial_{\mu}^{2}T(T(x_{1})^{(2)}, T^{\mu}(x_{2})^{(2)})
$
using the relations (\ref{t2-2a}) and  (\ref{tmu2-2a}). The expression in the left hand side of (\ref{st2-2a}) is
$
A + B + C.
$
There are a lot of cancelations and only the terms with 
$
\Box
$
acting on 
$
D^{F}_{0}( x_{1} - x_{2}), \partial_{\rho}D^{F}_{0}( x_{1} - x_{2})
$
survive. They give the right hand side of (\ref{st2-2a}).
$\qed$

The final result is:
\begin{thm}
We have
\be
sT(T^{I_{1}}(x_{1}), T^{I_{2}}(x_{2})) = s^{\prime}T(T^{I_{1}}(x_{1})^{(2)}, T^{I_{2}}(x_{2})^{(2)})
\ee
and the anomalies given in the previous theorem can be eliminated if and only if the constants
$
f_{abc}
$
verify the Jacobi identity
\be
f_{eab}~f_{ecd} + f_{ebc}~f_{ead} + f_{eca}~f_{ebd}  = 0
\label{Jacobi}
\ee
using the finite renormalizations:
\be
T(A_{1}(x_{1}), A_{2}(x_{2})) \rightarrow T^{\rm ren}(A_{1}(x_{1}), A_{2}(x_{2})) 
= T(A_{1}(x_{1}), A_{2}(x_{2})) + \delta (x_{1} - x_{2})~N(A_{1},A_{2})(x_{2})
\label{finiteN}
\ee
where
\bea
N(T,T) \equiv {i \over 2}~E_{a}^{\mu\nu}~E_{a\mu\nu}
\nonumber\\
N(T^{\mu},T) = N(T, T^{\mu}) \equiv - i~B_{a\nu}~E_{a}^{\mu\nu}
\nonumber\\
N(T^{\mu},T^{\nu}) \equiv i~B_{a}^{\mu}~B_{a}^{\nu}
\nonumber\\
N(T^{\mu\nu},T) = N(T, T^{\mu\nu}) \equiv  - i~B_{a}~E_{a}^{\mu\nu}
\nonumber\\
N(T^{\mu\nu},T^{\rho}) = N(T^{\rho}, T^{\mu\nu}) = 0
\nonumber \\
N(T^{\mu\nu},T^{\rho\sigma}) = 0
\label{N-TT}
\eea
The previous finite renormalizations can be obtained from (\ref{t2-2}) - (\ref{tmunu-rhosigma2-2}) performing the finite renormalization
\be
N(v_{a\mu,\nu},v_{b\rho,\sigma}) = {i \over 2} \eta_{\mu\rho}~\eta_{\nu\sigma}~\delta_{ab}.
\label{vv}
\ee
\label{ren-TT}
\end{thm}
{\bf Proof:} From (\ref{vv}) we obtain
\be
N(F_{a}^{\mu\nu},F_{b}^{\rho\sigma}) = i~(\eta^{\mu\rho}~\eta^{\nu\sigma} - \eta^{\nu\rho}~\eta^{\mu\sigma})~\delta_{ab}.
\label{N-FF}
\ee
We substitute this in (\ref{t2-2}) - (\ref{tmunu-rhosigma2-2}) and obtain the new contributions (\ref{N-TT}). It is useful to prove:
\be
N(T^{\mu\nu},T) =  - N(T^{\mu},T^{\nu}).
\ee
Then we compute the supplementary terms in
$
sT(T^{I_{1}}(x_{1}), T^{I_{2}}(x_{2})).
$

For instance we have:
\bea
sT^{\rm ren}(T(x_{1}), T(x_{2})) = sT(T(x_{1}), T(x_{2})) + d_{Q}[\delta(x_{1} - x_{2}) N(T,T)(x_{2})] 
\nonumber\\
- i~\partial_{\mu}^{1}[ \delta(x_{1} - x_{2}) N(T^{\mu},T)(x_{2})]
- i~\partial_{\mu}^{2}[ \delta(x_{1} - x_{2}) N(T,T^{\mu})(x_{2})]
\eea
Using (\ref{N-TT}) we arrive at
\be
sT^{\rm ren}(T(x_{1}), T(x_{2})) = sT(T(x_{1}), T(x_{2})) + \delta(x_{1} - x_{2}) R(T,T)(x_{2})
\ee
where
\be
R(T,T) \equiv d_{Q}N(T,T) - i~\partial_{\mu}N(T^{\mu},T).
\ee
If we substitute (\ref{st2-2a}) we obtain after some simplifications:
\be
sT^{\rm ren}(T(x_{1}), T(x_{2})) = \delta(x_{1} - x_{2}) {\cal A}(T,T)(x_{2}) 
\ee
where:
\be
{\cal A}(T,T) \equiv 2 ( : B_{a\mu}~C_{a}^{\mu}: - : B_{a}~D_{a}:) - : E_{a\mu\nu}~C_{a}^{\mu\nu}:
\ee
If we use Jacobi identity we obtain 
$
{\cal A} = 0.
$
One can prove the converse of this statement; suppose that
$
\delta(x_{1} - x_{2}) {\cal A}(T,T)(x_{2})
$
is a coboundary i.e. 
\bea
\delta(x_{1} - x_{2}) {\cal A}(T,T)(x_{2}) =
d_{Q}[\delta(x_{1} - x_{2}) N(x_{2})] 
\nonumber\\
- i~\partial_{\mu}^{1}[ \delta(x_{1} - x_{2}) N^{\mu}(x_{1})]
- i~\partial_{\mu}^{2}[ \delta(x_{1} - x_{2}) N^{\mu}(x_{2})]
\eea
with $N$ and 
$
N^{\mu}
$
constrained only by Lorentz covariance and canonical dimension
$
\omega \leq 4.
$
We obtain from here
\be
{\cal A}(T,T) = d_{Q}N - i~\partial_{\mu}N^{\mu}.
\ee
If we write arbitrary expressions for $N$ and 
$
N^{\mu}
$
we can prove from here that 
$
{\cal A}(T,T) = 0
$
and this leads to the Jacobi identity.
$
\qed
$
%\newpage

We now consider chronological products of the type
$
T(\xi_{a}\cdot T^{I}, T^{J}).
$

The following formulas are true:
\bea
T(B_{a}^{\mu}(x_{1})^{(1)}, A(x_{2})) = - f_{abc} [ :u_{b}(x_{1})~T(v_{c}^{\mu}(x_{1})^{(0)}, A(x_{2})): 
\nonumber\\
- :v_{b}^{\mu}(x_{1})^{(0)}~T(u_{c}(x_{1})^{(0)},  A(x_{2})): ]
\nonumber\\
\nonumber\\
T(C_{a\mu}(x_{1})^{(1)}, A(x_{2})) = f_{abc} [ :v_{b}^{\nu}(x_{1})~T(F_{c\nu\mu}(x_{1})^{(0)}, A(x_{2})): 
\nonumber\\
- :F_{b\nu\mu}(x_{1})^{(0)}~T(v_{c}^{\nu}(x_{1})^{(0)},  A(x_{2})):
\nonumber\\
- :u_{b}(x_{1})~T(\tilde{u}_{c,\mu}(x_{1})^{(0)},  A(x_{2})): 
- :\partial_{\mu}\tilde{u}_{b}(x_{1})~T(u_{c}(x_{1})^{(0)}, A(x_{2})): ]
\nonumber\\
\nonumber\\
T(D_{a}(x_{1})^{(1)}, A(x_{2})) = f_{abc} [ :v_{b}^{\mu}(x_{1})~T(\tilde{u}_{c,\mu}(x_{1})^{(0)}, A(x_{2})): 
\nonumber\\
- :\partial_{\mu}\tilde{u}_{b}(x_{1})~T(v_{c}^{\mu}(x_{1})^{(0)}, A(x_{2})): ]
\nonumber\\
\nonumber\\
T(E_{a}^{\mu\nu}(x_{1})^{(1)}, A(x_{2})) = f_{abc} [ :v_{b}^{\mu}(x_{1})~T(v_{c}^{\nu}(x_{1})^{(0)}, A(x_{2})): - (\mu \leftrightarrow \nu) ]
\nonumber\\
\nonumber\\
T(B_{a}(x_{1})^{(1)}, A(x_{2})) = f_{abc} :u_{b}(x_{1})~T(u_{c}(x_{1})^{(0)}, A(x_{2})): 
\nonumber\\
\nonumber\\
T(C_{a}^{\mu\nu}(x_{1})^{(1)}, A(x_{2})) = - f_{abc} [ :u_{b}(x_{1})~T(F_{c}^{\mu\nu}(x_{1})^{(0)}, A(x_{2})): 
\nonumber\\
- :F_{b}^{\mu\nu}(x_{1})^{(0)}~T(u_{c}(x_{1})^{(0)}, A(x_{2})): ]
\label{BCDE-T}
\eea
where 
$
A = T, T^{\mu}, T^{\mu\nu}.
$

To compute the right hand sides above we need the expressions
$
T(\xi_{a}(x_{1})^{(0)},  A(x_{2})). 
$
From our more precise form of the Wick theorem again we derive:
\bea
T(v_{a}^{\mu}(x_{1})^{(0)}, T(x_{2})) = C_{b\nu}(x_{2})~T(v_{a}^{\mu}(x_{1})^{(0)}, v_{b}^{\nu}(x_{2})^{(0)})
\nonumber\\
+ {1\over 2}~ E_{b\rho\sigma}(x_{2})~T(v_{a}^{\mu}(x_{1})^{(0)}, F_{b}^{\sigma\rho}(x_{2})^{(0)})
\nonumber\\
\nonumber\\
T(v_{a}^{\mu}(x_{1})^{(0)}, T^{\nu}(x_{2})) = C_{b}^{\rho\nu}(x_{2})~T(v_{a}^{\mu}(x_{1})^{(0)}, v_{b\rho}(x_{2})^{(0)})
\nonumber\\
- B_{b\rho}(x_{2})~T(v_{a}^{\mu}(x_{1})^{(0)}, F_{b}^{\rho\nu}(x_{2})^{(0)})
\nonumber\\
\nonumber\\
T(v_{a}^{\mu}(x_{1})^{(0)}, T^{\rho\sigma}(x_{2})) = B_{b}(x_{2})~T(v_{a}^{\mu}(x_{1})^{(0)}, F_{b}^{\rho\sigma}(x_{2})^{(0)})
\nonumber\\
\nonumber\\
T(F_{a}^{\mu\nu}(x_{1})^{(0)}, T(x_{2})) = C_{b\rho}(x_{2})~T(F_{a}^{\mu\nu}(x_{1})^{(0)}, v_{b}^{\rho}(x_{2})^{(0)})
\nonumber\\
+ {1\over 2}~ E_{b\rho\sigma}(x_{2})~T(F_{a}^{\mu\nu}(x_{1})^{(0)}, F_{b}^{\sigma\rho}(x_{2})^{(0)})
\nonumber\\
\nonumber\\
T(F_{a}^{\mu\nu}(x_{1})^{(0)}, T^{\rho}(x_{2})) = C_{b}^{\sigma\rho}(x_{2})~T(F_{a}^{\mu\nu}(x_{1})^{(0)}, v_{b\sigma}(x_{2})^{(0)})
\nonumber\\
- B_{b\sigma}(x_{2})~T(F_{a}^{\mu\nu}(x_{1})^{(0)}, F_{b}^{\rho\sigma}(x_{2})^{(0)})
\nonumber\\
\nonumber\\
T(F_{a}^{\mu\nu}(x_{1})^{(0)}, T^{\rho\sigma}(x_{2})) = B_{b}(x_{2})~T(F_{a}^{\mu\nu}(x_{1})^{(0)}, F_{b}^{\rho\sigma}(x_{2})^{(0)})
\nonumber\\
\nonumber\\
T(u_{a}(x_{1})^{(0)}, T(x_{2})) = B_{b}^{\mu}(x_{2})~T(u_{a}(x_{1})^{(0)}, \tilde{u}_{b,\mu}(x_{2})^{(0)})
\nonumber\\
\nonumber\\
T(u_{a}(x_{1})^{(0)}, T^{\mu}(x_{2})) = - B_{b}(x_{2})~T(u_{a}(x_{1})^{(0)}, {\tilde{u}_{b,}~}^{\mu}(x_{2})^{(0)})
\nonumber\\
\nonumber\\
T(u_{a}(x_{1})^{(0)}, T^{\mu\nu}(x_{2})) = 0
\nonumber\\
\nonumber\\
T(\tilde{u}_{a}(x_{1})^{(0)}, T(x_{2})) = D_{b}(x_{2})~T(\tilde{u}_{a}(x_{1})^{(0)}, u_{b}(x_{2})^{(0)})
\nonumber\\
\nonumber\\
T(\tilde{u}_{a}(x_{1})^{(0)}, T^{\mu}(x_{2})) = C^{\mu}_{b}(x_{2})~T(\tilde{u}_{a}(x_{1})^{(0)}, u_{b}(x_{2})^{(0)})
\nonumber\\
\nonumber\\
T(\tilde{u}_{a}(x_{1})^{(0)}, T^{\mu\nu}(x_{2})) = - C^{\mu\nu}_{b}(x_{2})~T(\tilde{u}_{a}(x_{1})^{(0)}, u_{b}(x_{2})^{(0)})
\label{xi-T}
\eea
%\newpage
The canonical splitting -view (\ref{chr-2-massless-vector}) - (\ref{chr-3-massless-vector}) -  leads to more precise forms:
\bea
T(v_{a}^{\mu}(x_{1})^{(0)}, T(x_{2})) = i~D_{0}^{F}(x_{1} - x_{2})~C_{a}^{\mu}(x_{2}) 
- i~\partial_{\nu}D_{0}^{F}(x_{1} - x_{2})~ E_{a}^{\mu\nu}(x_{2})
\nonumber\\
\nonumber\\
T(v_{a}^{\mu}(x_{1})^{(0)}, T^{\nu}(x_{2})) = i~D_{0}^{F}(x_{1} - x_{2})~C_{a}^{\mu\nu}(x_{2}) 
\nonumber\\
- i~\partial^{\nu}D_{0}^{F}(x_{1} - x_{2})~ B_{a}^{\mu}(x_{2})
+ i~\eta^{\mu\nu}~\partial_{\rho}D_{0}^{F}(x_{1} - x_{2})~ B_{a}^{\rho}(x_{2})
\nonumber\\
\nonumber\\
T(v_{a}^{\mu}(x_{1})^{(0)}, T^{\rho\sigma}(x_{2})) = 
- i~(\eta^{\mu\sigma}~\partial^{\rho} - \eta^{\mu\rho}~\partial^{\sigma})D_{0}^{F}(x_{1} - x_{2})~ B_{a}(x_{2})
\nonumber\\
\nonumber\\
T(F_{a}^{\mu\nu}(x_{1})^{(0)}, T(x_{2})) = - i~[ \partial^{\nu}D_{0}^{F}(x_{1} - x_{2})~ C_{a}^{\mu}(x_{2}) - (\mu \leftrightarrow \nu)]
\nonumber\\
+ i~[ \partial^{\mu}\partial_{\rho}D_{0}^{F}(x_{1} - x_{2})~ E_{a}^{\nu\rho}(x_{2})- (\mu \leftrightarrow \nu)]
\nonumber\\
\nonumber\\
T(F_{a}^{\mu\nu}(x_{1})^{(0)}, T^{\rho}(x_{2})) = 
- i~[ \partial^{\nu}D_{0}^{F}(x_{1} - x_{2})~ C_{a}^{\mu\rho}(x_{2}) - (\mu \leftrightarrow \nu)]
\nonumber\\
- i~[\eta^{\mu\rho}~\partial^{\nu}\partial_{\sigma}D_{0}^{F}(x_{1} - x_{2})~ B_{a}^{\sigma}(x_{2})- (\mu \leftrightarrow \nu)]
- i~[\partial^{\mu}\partial^{\rho}D_{0}^{F}(x_{1} - x_{2})~ B_{a}^{\nu}(x_{2})- (\mu \leftrightarrow \nu)]
\nonumber\\
\nonumber\\
T(F_{a}^{\mu\nu}(x_{1})^{(0)}, T^{\rho\sigma}(x_{2})) = 
\nonumber\\
i~(\eta^{\nu\sigma}~\partial^{\mu}\partial^{\rho} - \eta^{\mu\sigma}~\partial^{\nu}\partial^{\rho}
+ \eta^{\mu\rho}~\partial^{\nu}\partial^{\sigma} - \eta^{\nu\rho}~\partial^{\mu}\partial^{\sigma})D_{0}^{F}(x_{1} - x_{2})B_{a}(x_{2})
\nonumber\\
\nonumber\\
T(u_{a}(x_{1})^{(0)}, T(x_{2})) = i~\partial_{\mu}D_{0}^{F}(x_{1} - x_{2})~B_{a}^{\mu}(x_{2})
\nonumber\\
\nonumber\\
T(u_{a}(x_{1})^{(0)}, T^{\mu}(x_{2})) = -i \partial_{\mu}D_{0}^{F}(x_{1} - x_{2})~B_{a}(x_{2})
\nonumber\\
\nonumber\\
T(u_{a}(x_{1})^{(0)}, T^{\mu\nu}(x_{2})) = 0
\nonumber\\
\nonumber\\
T(\tilde{u}_{a}(x_{1})^{(0)}, T(x_{2})) = i~D_{0}^{F}(x_{1} - x_{2})~D_{a}(x_{2})
\nonumber\\
\nonumber\\
T(\tilde{u}_{a}(x_{1})^{(0)}, T^{\mu}(x_{2})) = i~D_{0}^{F}(x_{1} - x_{2})~C^{\mu}_{a}(x_{2})
\nonumber\\
\nonumber\\
T(\tilde{u}_{a}(x_{1})^{(0)}, T^{\mu\nu}(x_{2})) = - i~D_{0}^{F}(x_{1} - x_{2})~C^{\mu\nu}_{a}(x_{2})
\label{xi-T-canonical}
\eea
\newpage

If we substitute the previous relations in (\ref{BCDE-T}) we can obtain the canonical expressions.
\bea
T(B_{a}^{\mu}(x_{1})^{(1)}, T(x_{2})) = 
- i~f_{abc} \{ D_{0}^{F}(x_{1} - x_{2})~:u_{b}(x_{1})~C_{c}^{\mu}(x_{2}) :
\nonumber\\
- \partial_{\nu}D_{0}^{F}(x_{1} - x_{2})~ [ :u_{b}(x_{1})~E_{c}^{\mu\nu}(x_{2}): + :v^{\mu}_{b}(x_{1})~B_{c}^{\nu}(x_{2}): ] \}
\nonumber\\
\nonumber\\
T(B_{a}^{\mu}(x_{1})^{(1)}, T^{\nu}(x_{2})) = - i~f_{abc} \{ D_{0}^{F}(x_{1} - x_{2})~:u_{b}(x_{1})~C_{c}^{\mu\nu}(x_{2}): 
\nonumber\\
+ \partial_{\rho}D_{0}^{F}(x_{1} - x_{2})~ [ \eta^{\mu\nu}~:u_{b}(x_{1})~B_{c}^{\rho}(x_{2}): 
- \eta^{\nu\rho}~:u_{b}(x_{1})~B_{c}^{\mu}(x_{2}): + \eta^{\nu\rho}~:v^{\mu}_{b}(x_{1})~B_{c}(x_{2}): ] \}
\nonumber\\
\nonumber\\
T(B_{a}^{\mu}(x_{1})^{(1)}, T^{\rho\sigma}(x_{2})) = - i~f_{abc} 
(\eta^{\mu\rho}\partial^{\sigma} - \eta^{\mu\sigma}\partial^{\rho})D_{0}^{F}(x_{1} - x_{2})~:u_{b}(x_{1})~B_{c}(x_{2}): 
\label{BT}
\eea
\bea
T(C_{a}^{\mu}(x_{1})^{(1)}, T(x_{2})) = - i~f_{abc} \{ D_{0}^{F}(x_{1} - x_{2})~:F^{\nu\mu}_{b}(x_{1})~C_{c\nu}(x_{2}): 
\nonumber\\
+ \partial^{\mu}D_{0}^{F}(x_{1} - x_{2})~ [ :v_{b\nu}(x_{1})~C_{c}^{\nu}(x_{2}): + :u_{b}(x_{1})~D_{c}(x_{2}): ] 
\nonumber\\
+ \partial_{\nu}D_{0}^{F}(x_{1} - x_{2})~ [ - :v_{b}^{\nu}(x_{1})~C_{c}^{\mu}(x_{2}): + :{F^{\mu}_{b}}_{\rho}(x_{1})~C_{c}^{\rho\nu}(x_{2}):
+ :\partial^{\mu}\tilde{u}_{b}(x_{1})~B_{c}^{\nu}(x_{2}):] 
\nonumber\\
+ \partial_{\nu}\partial_{\rho}D_{0}^{F}(x_{1} - x_{2})~ :v_{b}^{\nu}(x_{1})~E_{c}^{\mu\rho}(x_{2}):
- \partial^{\mu}\partial_{\rho}D_{0}^{F}(x_{1} - x_{2})~ :v_{b\nu}(x_{1})~E_{c}^{\nu\rho}(x_{2}): \}
\nonumber\\
\nonumber\\
T(C_{a}^{\mu}(x_{1})^{(1)}, T^{\nu}(x_{2})) =  - i~f_{abc} \{ D_{0}^{F}(x_{1} - x_{2})~:F^{\mu\rho}_{b}(x_{1})~{C_{c}^{\nu}}_{\rho}(x_{2}):
\nonumber\\
+ \partial^{\mu}D_{0}^{F}(x_{1} - x_{2})~ [ :v_{b\rho}(x_{1})~C_{c}^{\rho\nu}(x_{2}): + :u_{b}(x_{1})~C_{c}^{\nu}(x_{2}): ] 
\nonumber\\
- \partial^{\nu}D_{0}^{F}(x_{1} - x_{2})~ [ :F_{b}^{\rho\mu}(x_{1})~B_{c\rho}(x_{2}): + :\partial^{\mu}\tilde{u}_{b}(x_{1})~B_{c}(x_{2}):] 
\nonumber\\
+ \partial_{\rho}D_{0}^{F}(x_{1} - x_{2})~ [ - :v_{b}^{\rho}(x_{1})~C_{c}^{\mu\nu}(x_{2}): + :F^{\nu\mu}_{b}(x_{1})~B_{c}^{\rho}(x_{2}): ]
\nonumber\\
+ \partial^{\mu}\partial_{\rho}D_{0}^{F}(x_{1} - x_{2})~ :v_{b}^{\nu}(x_{1})~B_{c}^{\rho}(x_{2}):
+ \partial^{\nu}\partial_{\rho}D_{0}^{F}(x_{1} - x_{2})~ :v_{b}^{\rho}(x_{1})~B_{c}^{\mu}(x_{2}):
\nonumber\\
- \eta^{\mu\nu}~\partial_{\rho}\partial_{\sigma}D_{0}^{F}(x_{1} - x_{2})~ :v_{b}^{\rho}(x_{1})~B_{c}^{\sigma}(x_{2}):
- \partial^{\mu}\partial^{\nu}D_{0}^{F}(x_{1} - x_{2})~ :v_{b}^{\rho}(x_{1})~B_{c\rho}(x_{2}): \}
\nonumber\\
\nonumber\\
T(C_{a}^{\mu}(x_{1})^{(1)}, T^{\rho\sigma}(x_{2})) = - i~f_{abc} 
%\nonumber\\
~[ \partial^{\sigma}D_{0}^{F}(x_{1} - x_{2})~ :F_{b}^{\rho\mu}(x_{1})~B_{c}(x_{2}): 
\nonumber\\
- \partial^{\rho}D_{0}^{F}(x_{1} - x_{2})~ :F_{b}^{\sigma\mu}(x_{1})~B_{c}(x_{2}):  
\nonumber\\
- \partial^{\mu}D_{0}^{F}(x_{1} - x_{2})~ :u_{b}(x_{1})~C_{c}^{\rho\sigma}(x_{2}):
\nonumber\\
+ (\eta^{\mu\sigma}~\partial^{\rho}\partial_{\nu} - \eta^{\mu\rho}~\partial^{\sigma}\partial_{\nu})D_{0}^{F}(x_{1} - x_{2})~
:v_{b}^{\nu}(x_{1})~B_{c}(x_{2}):
\nonumber\\
+ \partial^{\mu}\partial^{\sigma}D_{0}^{F}(x_{1} - x_{2})~ :v_{b}^{\rho}(x_{1})~B_{c}(x_{2}):
- \partial^{\mu}\partial^{\rho}D_{0}^{F}(x_{1} - x_{2})~ :v_{b}^{\sigma}(x_{1})~B_{c}(x_{2}): ]
\label{CT}
\eea
\bea
T(D_{a}(x_{1})^{(1)}, T(x_{2})) = - i~f_{abc} \{ D_{0}^{F}(x_{1} - x_{2})~:\partial_{\mu}\tilde{u}_{b}(x_{1})~C_{c}^{\mu}(x_{2}) :
\nonumber\\
+ \partial_{\mu}D_{0}^{F}(x_{1} - x_{2})~ [ - :v^{\mu}_{b}(x_{1})~D_{c}(x_{2}): 
+ :\partial_{\nu}\tilde{u}^{\mu}_{b}(x_{1})~E_{c}^{\mu\nu}(x_{2}): ] \}
\nonumber\\
\nonumber\\
T(D_{a}(x_{1})^{(1)}, T^{\mu}(x_{2})) = - i~f_{abc} \{ - D_{0}^{F}(x_{1} - x_{2})~:\partial_{\nu}\tilde{u}_{b}(x_{1})~C_{c}^{\mu\nu}(x_{2}) :
\nonumber\\
+ \partial_{\nu}D_{0}^{F}(x_{1} - x_{2})~ [ - :v^{\nu}_{b}(x_{1})~C^{\mu}_{c}(x_{2}): 
+ :\partial^{\mu}\tilde{u}^{\mu}_{b}(x_{1})~B_{c}^{\nu}(x_{2}): 
- \eta^{\mu\nu}~ :\partial_{\rho}\tilde{u}^{\mu}_{b}(x_{1})~B_{c}^{\rho}(x_{2}):] \}
\nonumber\\
\nonumber\\
T(D_{a}(x_{1})^{(1)}, T^{\rho\sigma}(x_{2})) = 
- i~f_{abc} \{ - \partial_{\mu}D_{0}^{F}(x_{1} - x_{2})~:v^{\mu}_{b}(x_{1})~C_{c}^{\rho\sigma}(x_{2}) :
\nonumber\\
+ \partial^{\sigma}D_{0}^{F}(x_{1} - x_{2})~ :\partial^{\rho}\tilde{u}_{b}(x_{1})~B^{\mu}_{c}(x_{2}): 
- \partial^{\rho}D_{0}^{F}(x_{1} - x_{2})~ :\partial^{\sigma}\tilde{u}_{b}(x_{1})~B^{\mu}_{c}(x_{2}): \}
\label{DT}
\eea
\bea
T(E^{\mu\nu}_{a}(x_{1})^{(1)}, T(x_{2})) = 
\nonumber\\
- i~f_{abc} \{ D_{0}^{F}(x_{1} - x_{2})~[ :v^{\nu}_{b}(x_{1})~C_{c}^{\mu}(x_{2}): - :v^{\mu}_{b}(x_{1})~C_{c}^{\nu}(x_{2}):]
\nonumber\\
+ \partial_{\rho}D_{0}^{F}(x_{1} - x_{2})~ [ :v^{\mu}_{b}(x_{1})~E^{\nu\rho}_{c}(x_{2}): 
- v^{\nu}_{b}(x_{1})~E_{c}^{\mu\rho}(x_{2}): ] \}
\nonumber\\
\nonumber\\
T(E^{\mu\nu}_{a}(x_{1})^{(1)}, T^{\rho}(x_{2})) = 
- i~f_{abc} \{ D_{0}^{F}(x_{1} - x_{2})~:v^{\nu}_{b}(x_{1})~C_{c}^{\mu\rho}(x_{2}): 
\nonumber\\
+ \partial_{\sigma}D_{0}^{F}(x_{1} - x_{2})~ [ - \eta^{\nu\rho} :v^{\mu}_{b}(x_{1})~B^{\sigma}_{c}(x_{2}): 
+ \eta^{\rho\sigma}  v^{\mu}_{b}(x_{1})~B_{c}^{\nu}(x_{2}): ] \} - ( \mu \leftrightarrow \nu)
\nonumber\\
\nonumber\\
T(E^{\mu\nu}_{a}(x_{1})^{(1)}, T^{\rho\sigma}(x_{2})) = 
\nonumber\\
- i~f_{abc} [ ( \eta^{\nu\sigma}~\partial^{\rho} - \eta^{\nu\rho}~\partial^{\sigma})D_{0}^{F}(x_{1} - x_{2})~
:v^{\mu}_{b}(x_{1})~B_{c}(x_{2}):  - ( \mu \leftrightarrow \nu) ]
\label{ET}
\eea
\bea
T(B_{a}(x_{1})^{(1)}, T(x_{2})) = i~f_{abc} \partial_{\mu}D_{0}^{F}(x_{1} - x_{2})~ :u_{b}(x_{1})~B^{\mu}_{c}(x_{2}): 
\nonumber\\
\nonumber\\
T(B_{a}(x_{1})^{(1)}, T^{\mu}(x_{2})) = - i~f_{abc} \partial^{\mu}D_{0}^{F}(x_{1} - x_{2})~ :u_{b}(x_{1})~B_{c}(x_{2}): 
\nonumber\\
\nonumber\\
T(B_{a}(x_{1})^{(1)}, T^{\rho\sigma}(x_{2})) = 0
\label{B2T}
\eea
\bea
T(C_{a}^{\mu\nu}(x_{1})^{(1)}, T(x_{2})) = 
- i~f_{abc} [ \partial^{\mu}D_{0}^{F}(x_{1} - x_{2})~ :u_{b}(x_{1})~C_{c}^{\nu}(x_{2}): 
\nonumber\\
- \partial^{\nu}D_{0}^{F}(x_{1} - x_{2})~ :u_{b}(x_{1})~C_{c}^{\mu}(x_{2}):
- \partial_{\rho}D_{0}^{F}(x_{1} - x_{2})~ :F^{\mu\nu}_{b}(x_{1})~B_{c}^{\rho}(x_{2}): 
\nonumber\\
- \partial^{\mu}\partial_{\rho}D_{0}^{F}(x_{1} - x_{2})~ :u_{b}(x_{1})~E_{c}^{\nu\rho}(x_{2}): 
+ \partial^{\nu}\partial_{\rho}D_{0}^{F}(x_{1} - x_{2})~ :u_{b}(x_{1})~E_{c}^{\mu\rho}(x_{2}): ]
\nonumber\\
\nonumber\\
T(C_{a}^{\mu\nu}(x_{1})^{(1)}, T^{\rho}(x_{2})) = 
- i~f_{abc} [ \partial^{\mu}D_{0}^{F}(x_{1} - x_{2})~ :u_{b}(x_{1})~C_{c}^{\nu\rho}(x_{2}): 
\nonumber\\
- \partial_{\nu}D_{0}^{F}(x_{1} - x_{2})~ :u_{b}~C_{c}^{\mu\rho}(x_{2}):
\nonumber\\
+ \partial^{\rho}D_{0}^{F}(x_{1} - x_{2})~ :F_{b}^{\mu\nu}(x_{1})~B_{c}(x_{2}): 
\nonumber\\
+ (\eta^{\nu\rho}~\partial^{\mu}\partial_{\sigma} - \eta^{\mu\rho}~\partial^{\nu}\partial_{\sigma})D_{0}^{F}(x_{1} - x_{2})
~ :u_{b}(x_{1})~B_{c}^{\sigma}(x_{2}):
\nonumber\\
-\partial^{\mu}\partial^{\rho}D_{0}^{F}(x_{1} - x_{2})~ :u_{b}(x_{1})~B_{c}^{\nu}(x_{2}):
+ \partial^{\nu}\partial^{\rho}D_{0}^{F}(x_{1} - x_{2})~ :u_{b}(x_{1})~B_{c}^{\mu}(x_{2}): ]
\nonumber\\
\nonumber\\
T(C_{a}^{\mu\nu}(x_{1})^{(1)}, T^{\rho\sigma}(x_{2})) = 
- i~f_{abc} 
(\eta^{\mu\sigma}~\partial^{\nu}\partial^{\rho} - \eta^{\nu\sigma}~\partial^{\mu}\partial^{\rho}
\nonumber\\
+ \eta^{\nu\rho}~\partial^{\mu}\partial^{\sigma} - \eta^{\mu\rho}~\partial^{\nu}\partial^{\sigma})D_{0}^{F}(x_{1} - x_{2})~
:u_{b}(x_{1})~B_{c}(x_{2}):
\label{C2T}
\eea
\newpage

We can obtain in a similar way to theorem \ref{sTT} the following result.
\begin{thm}
The following formulas are true
\bea
s^{\prime}T(B_{a}^{\mu}(x_{1})^{(1)}, T^{I}(x_{2})) = 0
\label{sb1t}
\eea
\bea
s^{\prime}T(C_{a}^{\mu}(x_{1})^{(1)}, T(x_{2})) =
%\nonumber\\
\delta( x_{1} - x_{2})~ f_{abc}~(:u_{b}~C_{c}^{\mu}: + : F_{b}^{\mu\nu}~B_{c\nu}: - :\partial^{\mu}\tilde{u}_{b} B_{c}:)(x_{2})
\nonumber\\
- \partial_{\nu}\delta( x_{1} - x_{2})~f_{abc}~[ :u_{b}(x_{1})~E_{c}^{\mu\nu}(x_{2}): - : v_{b}^{\nu}(x_{1})~B_{c}^{\mu}(x_{2}):]
\nonumber\\
- \partial^{\mu}\delta( x_{1} - x_{2})~f_{abc}~ : v_{b}^{\nu}(x_{1})~B_{c\nu}(x_{2}):
\label{sc1t}
\eea
\bea
s^{\prime}T(C^{\mu}_{a}(x_{1})^{(1)}, T^{\nu}(x_{2})) =
%\nonumber\\
\delta( x_{1} - x_{2})~f_{abc}~(:u_{a}~C_{c}^{\mu\nu}: - : F_{b}^{\mu\nu}~B_{c}:)(x_{2})
\nonumber\\
+ \partial_{\rho}\delta( x_{1} - x_{2})~\eta^{\mu\nu}~f_{abc}~
[ : u_{b}(x_{1})~B_{c}^{\rho}(x_{2}): -  : v_{b}^{\rho}(x_{1})~B_{c}(x_{2}):]
\nonumber\\
- \partial^{\nu}\delta( x_{1} - x_{2})~f_{abc}~: u_{b}(x_{1})~B^{\mu}_{c}(x_{2}):
+ \partial^{\mu}\delta( x_{1} - x_{2})~ f_{abc}~ : v_{b}^{\nu}(x_{1})~B_{c}(x_{2}):
\label{sc2t}
\eea
\bea
s^{\prime}T(C^{\mu}(x_{1})^{(1)}, T^{\rho\sigma}(x_{2})) =
(\eta^{\mu\rho} \partial^{\sigma} - \eta^{\mu\sigma} \partial^{\rho})\delta( x_{1} - x_{2})~f_{abc}~ : u_{b}(x_{1})~B_{c}(x_{2}):
\label{sc3t}
\eea
\bea
s^{\prime}T(D(x_{1})^{(1)}, T(x_{2})) =
\delta( x_{1} - x_{2})~ f_{abc}~(:v_{b\mu}~C_{c}^{\mu}: + : u_{b}~D_{c}: + :\partial_{\mu}\tilde{u}_{b} B^{\mu}_{c}:)(x_{2})
\nonumber\\
+ \partial_{\mu}\delta( x_{1} - x_{2})~f_{abc}~: v_{b\nu}(x_{1})~E^{\mu\nu}_{c}(x_{2}):
\label{sd1}
\eea
\bea
s^{\prime}T(D_{a}(x_{1})^{(1)}, T^{\mu}(x_{2})) =
%\nonumber\\
\delta( x_{1} - x_{2})~ f_{abc}~(:v_{b\nu}~C_{c}^{\nu\mu}: + : u_{b}~C_{c}^{\mu}: - :\partial^{\mu}\tilde{u}_{b} B_{c}:)(x_{2})
\nonumber\\
+ \partial_{\nu}\delta( x_{1} - x_{2})~f_{abc}~ :v^{\mu}_{b}(x_{1})~B_{c}^{\nu}(x_{2}):
\nonumber\\
- \partial^{\mu}\delta( x_{1} - x_{2})~f_{abc}~ : v_{b}^{\nu}(x_{1})~B_{c\nu}(x_{2}):
\label{sd2t}
\eea
\bea
s^{\prime}T(D(x_{1})^{(1)}, T^{\rho\sigma}(x_{2})) =
- \delta( x_{1} - x_{2})~ f_{abc}~:u_{b}~C_{c}^{\rho\sigma}:(x_{2})
\nonumber\\
+ \partial^{\sigma}\delta( x_{1} - x_{2})~f_{abc}~ : v^{\rho}_{b}(x_{1})~B_{c}(x_{2}): - (\rho \leftrightarrow \sigma)
\label{sd3t}
\eea
\bea
s^{\prime}T(E_{a}^{\mu\nu}(x_{1})^{(1)}, T(x_{2})) =
\delta( x_{1} - x_{2})~ f_{abc}~:v^{\mu}_{b}~B^{\nu}_{c}:(x_{2}) - (\mu \leftrightarrow \nu)
\label{se1t}
\eea
\bea
s^{\prime}T(E_{a}^{\mu\nu}(x_{1})^{(1)}, T^{\rho}(x_{2})) =
\delta( x_{1} - x_{2})~ f_{abc}~\eta^{\mu\rho} :v^{\nu}_{b}~B_{c}:(x_{2}) - (\mu \leftrightarrow \nu)
\label{se2t}
\eea
\bea
s^{\prime}T(E_{a}^{\mu\nu}(x_{1})^{(1)}, T^{\rho\sigma}(x_{2})) = 0
\label{se3t}
\eea
\bea
s^{\prime}T(B_{a}(x_{1})^{(1)}, T^{I}(x_{2})) = 0
\label{sb2t}
\eea
\bea
s^{\prime}T(C_{a}^{\mu\nu}(x_{1})^{(1)}, T(x_{2})) =
%\nonumber\\
- \delta( x_{1} - x_{2})~ f_{abc}~ : F_{b}^{\mu\nu}~B_{c}:(x_{2})
\nonumber\\
+ \partial^{\mu}\delta( x_{1} - x_{2})~f_{abc}~ : u_{b}(x_{1})~B_{c}^{\nu}(x_{2}): - (\mu \leftrightarrow \nu)
\label{sc4t}
\eea
\bea
s^{\prime}T(C_{a}^{\mu\nu}(x_{1})^{(1)}, T^{\rho}(x_{2})) =
(\eta^{\mu\rho}\partial^{\nu} - \eta^{\nu\rho}\partial^{\mu})\delta( x_{1} - x_{2})~f_{abc}~ : u_{b}(x_{1})~B_{c}(x_{2}):
\label{sc5t}
\eea
\bea
s^{\prime}T(C_{a}^{\mu\nu}(x_{1})^{(1)}, T^{\rho\sigma}(x_{2})) = 0
\label{sc6t}
\eea
\end{thm}
\newpage
Next we investigate if we can remove the anomalies from the previous theorem can be removed with finite renormalizations. As theorem
\ref{ren-TT}
\begin{thm}
We have
\be
sT(\xi_{a}\cdot T^{I_{1}}(x_{1}), T^{I_{2}}(x_{2})) = s^{\prime}T(\xi_{a}\cdot T^{I_{1}}(x_{1})^{(1)}, T^{I_{2}}(x_{2})^{(2)})
\ee
and the anomalies can be removed using finite renormalizations of the type (\ref{finiteN}):
\bea
T^{\rm ren}(A_{1}(x_{1}), A_{2}(x_{2})) 
= T(A_{1}(x_{1}), A_{2}(x_{2})) + \delta (x_{1} - x_{2})~N(A_{1},A_{2})(x_{2})
\nonumber
\eea
where the non-trivial expressions $N$ are:
\bea
N(C_{a}^{\mu},T) \equiv i f_{abc} ~:v_{b\nu}~E_{c}^{\mu\nu}:
\nonumber\\
N(C_{a}^{\mu\nu},T) \equiv i f_{abc}~:u_{b}~E_{c}^{\mu\nu}:
\nonumber\\
N(C_{a}^{\mu},T^{\nu}) \equiv i~f_{abc}~( :v_{b}^{\nu}~B_{c}^{\mu}: - \eta^{\mu\nu}~ :v_{b\rho}~B_{c}^{\rho}:)
\nonumber\\
N(C_{a}^{\mu\nu},T^{\rho}) \equiv - i~f_{abc}~(\eta^{\mu\rho}~ :u_{b}~B_{c}^{\nu}: - \eta^{\nu\rho}~ :u_{b}~B_{c}^{\mu}:)
\nonumber\\
N(C_{a}^{\mu},T^{\rho\sigma}) \equiv  - i~f_{abc}~(\eta^{\mu\rho} :v^{\sigma}_{b}~B_{c}: - \eta^{\nu\rho}~ :v^{\rho}_{b}~B_{c}:)
\label{N-BCDE-T}
\eea
These finite renormalizations can be obtained using the finite renormalization (\ref{vv}) in the expressions (\ref{xi-T}) and are unique.
\label{ren-BCDE-T}
\end{thm}
{\bf Proof:} Is quite similar to the proof theorem \ref{ren-TT}. First we obtain (\ref{N-BCDE-T}) by substituting 
(\ref{vv}) in (\ref{BCDE-T}) + (\ref{xi-T}). Next, we compute the supplementary terms  coming from the finite renormalizations. For 
instance:
\bea
s^{\prime}T^{\rm ren}(C_{a}^{\mu}(x_{1}), T(x_{2})) = s^{\prime}T(C_{a}^{\mu}(x_{1}), T(x_{2}))
\nonumber\\
+ \delta(x_{1} - x_{2})~R(C_{a}^{\mu},T)(x_{2}) + i \partial_{\nu}\delta(x_{1} - x_{2})~R^{\nu}(C_{a}^{\mu},T)(x_{2})
\eea
where
\bea
R(C_{a}^{\mu},T) \equiv d_{Q}N(C_{a}^{\mu},T) - i \partial_{\nu}N(C_{a}^{\mu},T^{\nu})
\nonumber\\
R^{\nu}(C_{a}^{\mu},T) \equiv i~[ - N(C_{a}^{\mu\nu},T) + N(C_{a}^{\mu},T^{\nu}) ]
\eea
These expresions can be computed using the formulas from the statement. If we substitute in (\ref{sc1t}) one gets after some 
computations:
\be
s^{\prime}T^{\rm ren}(C_{a}^{\mu}(x_{1}), T(x_{2})) = \delta(x_{1} - x_{2})~{\cal A}(C_{a}^{\mu},T)(x_{2})
\ee
where
\be
{\cal A}(C_{a}^{\mu},T) \equiv f_{abc}~(- :v_{b\nu} C_{c}^{\mu\nu}: +  :u_{b} C_{c}^{\mu}: + :F_{b}^{\mu\nu} B_{c\nu}: 
- :\partial^{\mu}\tilde{u}_{b} B_{c}:).
\ee
But using Jacobi identity, the equality
$
{\cal A}(C_{a}^{\mu},T) = 0
$
follows easily. The same line of argument must be used for the other cases. By some computations one can prove the uniqueness
of the finite renormalizations (\ref{N-BCDE-T}).
$\qed$
\newpage
Now we can address the question of Wick property. We have the finite renormalizations from theorem \ref{ren-TT} and the 
preceding theorem. If the Wick property is preserved, the finite renormalizations should verify identities of the type
(\ref{consistency}). This true according to
\begin{thm}
The finite renormalizations (\ref{N-TT}) and (\ref{N-BCDE-T}) preserve Wick expansion property. Explicitly we have:
\bea
v_{a}^{\mu}\cdot N(T,T) = 2 N(C_{a}^{\mu},T)
\nonumber\\
u_{a}\cdot N(T^{\mu},T) = - N(C_{a}^{\mu},T) - N(T^{\mu},D_{a})
\nonumber\\
v^{\nu}_{a}\cdot N(T^{\mu},T) = N(C_{a}^{\nu\mu},T) + N(T^{\mu},C^{\nu}_{a})
\nonumber\\
u_{a}\cdot N(T^{\mu},T^{\nu}) = N(C_{a}^{\mu},T^{\nu}) - N(T^{\mu},C^{\nu}_{a})
\nonumber\\
v^{\rho}_{a}\cdot N(T^{\mu},T^{\nu}) = N(C_{a}^{\rho\mu},T^{\nu}) + N(T^{\mu},C^{\rho\nu}_{a})
\nonumber\\
u_{a}\cdot N(T^{\mu\nu},T) = - N(C_{a}^{\mu\nu},T) + N(T^{\mu\nu},D_{a})
\nonumber\\
v^{\rho}_{a}\cdot N(T^{\mu\nu},T) = N(T^{\mu\nu},C^{\rho}_{a})
\label{consistency1}
\eea
and
\be
N(B_{a}^{\mu},T^{I}) = N(D_{a},T^{I}) = N(E_{a}^{\mu\nu},T^{I}) = 0, \quad \forall I.
\label{consistency2}
\ee
\end{thm}
{\bf Proof:} 
We start with the relation
\bea
~[v_{a}^{\mu}(y), T(T(x_{1}),T(x_{2}))] = 
\nonumber\\
i~[ D_{0}(y - x_{1})~T(C_{a}^{\mu}(x_{1}),T(x_{2})) 
- \partial_{\nu}D_{0}(y - x_{1})~T(E_{a}^{\mu\nu}(x_{1}),T(x_{2}))] + ( 1 \leftrightarrow 2)
\eea
following from Wick expansion property (\ref{comm-wick}). If we consider finite renormalizations of the type 
(\ref{N-TT}) and (\ref{N-BCDE-T}) we will get new terms in the left and right hand of the preceding identity. The identity is 
preserved {\it iff} we have the first relation from (\ref{consistency1}) and one of the relation (\ref{consistency2}). 
In the same way we obtain
\bea
~[\tilde{u}_{a}(y), T(T^{\mu}(x_{1}),T(x_{2}))] = 
\nonumber\\
- i~D_{0}(y - x_{1})~T(C_{a}^{\mu}(x_{1}),T(x_{2})) - i~D_{0}(y - x_{2})~T(T^{\mu}(x_{1}),D_{a}(x_{2}))
\eea
and the preservation of this relation leads to the second relation from (\ref{consistency1}). From
\bea
~[v_{a}^{\nu}(y), T(T^{\mu}(x_{1}),T(x_{2}))] = 
i~D_{0}(y - x_{1})~T(C_{a}^{\nu\mu}(x_{1}),T(x_{2})) 
\nonumber\\
- i\partial^{\mu}D_{0}(y - x_{1})~T(B_{a}^{\nu}(x_{1}),T(x_{2}))
+ i~\eta^{\mu\nu}~\partial^{\rho}D_{0}(y - x_{1})~T(B_{a}^{\rho}(x_{1}),T(x_{2}))
\nonumber\\
+ i~D_{0}(y - x_{2})~T(T^{\mu}(x_{1}),C_{a}^{\nu}(x_{2})) 
- i~\partial_{\rho}D_{0}(y - x_{2})~T(T^{\mu}(x_{1}),E_{a}^{\nu\rho}(x_{2}))
\eea
and we obtain the third relation from (\ref{consistency1}). From 
\bea
~[\tilde{u}_{a}(y), T(T^{\mu}(x_{1}),T^{\nu}(x_{2}))] = 
\nonumber\\
i~D_{0}(y - x_{1})~T(C_{a}^{\mu}(x_{1}),T^{\nu}(x_{2})) - i~D_{0}(y - x_{2})~T(T^{\mu}(x_{1}),C^{\nu}_{a}(x_{2}))
\eea
we obtain the fourth relation from (\ref{consistency1}). From
\bea
~[v_{a}^{\rho}(y), T(T^{\mu}(x_{1}),T^{\nu}(x_{2}))] = 
i~D_{0}(y - x_{1})~T(C_{a}^{\rho\mu}(x_{1}),T^{\nu}(x_{2})) 
\nonumber\\
- i\partial^{\mu}D_{0}(y - x_{1})~T(B_{a}^{\rho}(x_{1}),T^{\nu}(x_{2}))
+ i~\eta^{\mu\rho}~\partial_{\sigma}D_{0}(y - x_{1})~T(B_{a}^{\sigma}(x_{1}),T(x_{2}))
\nonumber\\
+ i~D_{0}(y - x_{2})~T(T^{\mu}(x_{1}),C_{a}^{\rho\nu}(x_{2})) 
\nonumber\\
- i~\partial^{\nu}D_{0}(y - x_{2})~T(T^{\mu}(x_{1}),B_{a}^{\rho}(x_{2}))
+ i~\eta^{\nu\rho}~\partial_{\sigma}D_{0}(y - x_{2})~T(T^{\mu}(x_{1}),B_{a}^{\sigma}(x_{2}))
\eea
we get the fifth relation from (\ref{consistency1}). From
\bea
~[\tilde{u}_{a}(y), T(T^{\mu\nu}(x_{1}),T(x_{2}))] = 
\nonumber\\
- i~D_{0}(y - x_{1})~T(C_{a}^{\mu\nu}(x_{1}),T(x_{2})) + i~D_{0}(y - x_{2})~T(T^{\mu\nu}(x_{1}),D_{a}(x_{2}))
\eea
we obtain the sixth relation from (\ref{consistency1}). Finally, from
\bea
~[v_{a}^{\rho}(y), T(T^{\mu\nu}(x_{1}),T(x_{2}))] = 
i~(\eta^{\mu\rho} \partial^{\nu} - \eta^{\nu\rho} \partial^{\mu})D_{0}(y - x_{1})~T(B_{a}(x_{1}),T(x_{2})) 
\nonumber\\
+ i~D_{0}(y - x_{2})~T(T^{\mu\nu}(x_{1}),C_{a}^{\rho}(x_{2})) 
\nonumber\\
- i~\partial_{\sigma}D_{0}(y - x_{2})~T(T^{\mu\nu}(x_{1}),E_{a}^{\rho\sigma}(x_{2}))
\eea
we get the last relation from the (\ref{consistency1}). 

It remains to consider the explicit form of the finite renormalizations (\ref{N-TT}) and (\ref{N-BCDE-T}) and prove that the 
identities (\ref{consistency1}) are true.
$\qed$

The anomalies cannot be always removed. We provide an example. 
\begin{thm}
The following formulas are true
\bea
sT(v_{a}^{\mu}(x_{1}),T(x_{2})) = \delta(x_{1} - x_{2})~B_{a}^{\mu}(x_{2})
\nonumber\\
sT(v_{a}^{\mu}(x_{1}),T^{\nu}(x_{2})) = - \delta(x_{1} - x_{2})~\eta^{\mu\nu}~B_{a}(x_{2})
\nonumber\\
sT(v_{a}^{\mu}(x_{1}),T^{\rho\sigma}(x_{2})) = 0. 
\eea
\bea
sT(F_{a}^{\mu\nu}(x_{1}),T(x_{2})) = \partial^{\mu}\delta(x_{1} - x_{2})~B_{a}^{\nu}(x_{2}) - (\mu \leftrightarrow \nu)
\nonumber\\
sT(F_{a}^{\mu\nu}(x_{1}),T^{\rho}(x_{2})) = (\eta^{\mu\rho}\partial^{\nu} -  \eta^{\nu\rho}\partial^{\mu})\delta(x_{1} - x_{2})~B_{a}(x_{2})
\nonumber\\
sT(F_{a}^{\mu\nu}(x_{1}),T^{\rho\sigma}(x_{2})) = 0. 
\eea
\bea
sT(u_{a}(x_{1}),T(x_{2})) = - \delta(x_{1} - x_{2})~B_{a}(x_{2})
\nonumber\\
sT(u_{a}(x_{1}),T^{\nu}(x_{2})) = 0
\nonumber\\
sT(u_{a}(x_{1}),T^{\rho\sigma}(x_{2})) = 0. 
\eea
\be
sT(\tilde{u}_{a}(x_{1}),T^{I}(x_{2})) = 0, \quad \forall I.
\ee
\end{thm}
{\bf Proof:} We illustrate the first identity. We have by definition
\bea
sT(v_{a}^{\mu}(x_{1}),T(x_{2})) = d_{Q}T(v_{a}^{\mu}(x_{1}),T(x_{2})) - i~\partial^{\mu}_{1}T(u_{a}(x_{1}),T(x_{2}))
- i~\partial_{\nu}^{2}T(v_{a}^{\mu}(x_{1}),T^{\nu}(x_{2}))
\eea
so, using the formulas from (\ref{xi-T-canonical}) we obtain the first formula from the statement.
$\qed$

\begin{thm}
From (\ref{vv}) we obtain also the finite renormalizations
\bea
N(F_{a}^{\mu\nu},T) = - i~E_{a}^{\mu\nu}
\nonumber\\
N(F_{a}^{\mu\nu},T^{\rho}) = i~(\eta^{\mu\rho} B_{a}^{\nu} - \eta^{\nu\rho} B_{a}^{\mu})
\nonumber\\
N(F_{a}^{\mu\nu},T^{\rho\sigma}) = i~(\eta^{\mu\rho} \eta^{\nu\sigma} -  \eta^{\nu\rho} \eta^{\mu\sigma})B_{a}
\label{N-xiT}
\eea
Using these finite renormalizations we obtain 
\be
sT^{\rm ren}T(A_{1}(x_{1}), A_{2}(x_{2})) = \delta(x_{1} - x_{2})~{\cal A}(A_{1}, A_{2})(x_{2})
\ee
for 
$
A_{1} = \xi_{a} = v_{a}^{\mu}, u_{a}, \tilde{u}_{a}
$
etc., and
$
A_{2} = T^{I}
$
with the non-zero anomalies:
\bea
{\cal A}(v_{a}^{\mu}, T) = B_{a}^{\mu}
\nonumber\\
{\cal A}(v_{a}^{\mu}, T^{\nu}) = - \eta^{\mu\nu}~B_{a}
\nonumber\\
{\cal A}(F_{a}^{\mu\nu}, T) = C_{a}^{\mu\nu}
\nonumber\\
{\cal A}(u_{a}, T) = - B_{a}.
\eea
These anomalies cannot be removed by other finite renormalizations.
\end{thm}
{\bf Proof:} We consider: 
\bea
sT^{\rm ren}(F_{a}^{\mu\nu}(x_{1}),T(x_{2})) = sT(F_{a}^{\mu\nu}(x_{1}),T(x_{2})) 
\nonumber\\
+ d_{Q}[\delta(x_{1} - x_{2}) N(F_{a}^{\mu\nu},T)(x_{2})] - i~\partial_{\rho}^{2}[\delta(x_{1} - x_{2}) N(F_{a}^{\mu\nu},T^{\rho})(x_{2})]
\eea
and use the finite renormalizations (\ref{N-xiT}). We obtain the third relation from the statement.

It is rather simple to prove that the anomalies from the statement cannot be removed by other finite renormalizations.
$\qed$

\newpage
We continue the analysis of the tree contributions of chronological products involving Wick submonomials. We first have:
\bea
T(v_{a}^{\mu}(x_{1})^{(0)}, B_{b}^{\nu}(x_{2})) = - f_{bcd}~:u_{c}(x_{2})~T(v_{a}^{\mu}(x_{1})^{(0)}, v^{\nu}_{d}(x_{2})^{(0)}): 
\nonumber\\
\nonumber\\
T(v_{a}^{\mu}(x_{1})^{(0)}, C_{b}^{\nu}(x_{2})) = f_{bcd}~[ :F_{d}^{\rho\nu}(x_{2})~T(v_{a}^{\mu}(x_{1})^{(0)}, v_{c\rho}(x_{2})^{(0)}): 
\nonumber\\
+ :v_{c\rho}(x_{2})~T(v_{a}^{\mu}(x_{1})^{(0)}, F_{d}^{\rho\nu}(x_{2})^{(0)}): 
\nonumber\\
\nonumber\\
T(v_{a}^{\mu}(x_{1})^{(0)}, D_{b}(x_{2})) = f_{bcd}~:\partial_{\nu}\tilde{u}_{c}(x_{2})~T(v_{a}^{\mu}(x_{1})^{(0)}, v^{\nu}_{c}(x_{2})^{(0)}): 
\nonumber\\
\nonumber\\
T(v_{a}^{\mu}(x_{1})^{(0)}, E_{b}^{\rho\sigma}(x_{2})) = 
f_{bcd}~[ :v_{d}^{\sigma}(x_{2})~T(v_{a}^{\mu}(x_{1})^{(0)}, v_{c}^{\rho}(x_{2})^{(0)}): 
\nonumber\\
+ :v_{c}^{\rho}(x_{2})~T(v_{a}^{\mu}(x_{1})^{(0)}, v_{d}^{\sigma}(x_{2})^{(0)}):]
\nonumber\\
\nonumber\\
T(v_{a}^{\mu}(x_{1})^{(0)}, C_{b}^{\rho\sigma}(x_{2})) = - f_{bcd}~:u_{c}(x_{2})~T(v_{a}^{\mu}(x_{1})^{(0)}, F_{d}^{\rho\sigma}(x_{2})^{(0)}): 
\nonumber\\
\nonumber\\
T(F_{a}^{\mu\nu}(x_{1})^{(0)}, B_{b}^{\rho}(x_{2})) = - f_{bcd}~:u_{c}(x_{2})~T(F_{a}^{\mu\nu}(x_{1})^{(0)}, v^{\rho}_{d}(x_{2})^{(0)}): 
\nonumber\\
\nonumber\\
T(F_{a}^{\mu\nu}(x_{1})^{(0)}, C_{b}^{\rho}(x_{2})) = 
f_{bcd}~[ :F_{d}^{\sigma\rho}(x_{2})~T(F_{a}^{\mu\nu}(x_{1})^{(0)}, v_{c\sigma}(x_{2})^{(0)}): 
\nonumber\\
+ :v_{c\sigma}(x_{2})~T(F_{a}^{\mu\nu}(x_{1})^{(0)}, F_{d}^{\sigma\rho}(x_{2})^{(0)}): 
\nonumber\\
\nonumber\\
T(F_{a}^{\mu\nu}(x_{1})^{(0)}, D_{b}(x_{2})) = 
f_{bcd}~:\partial_{\rho}\tilde{u}_{c}(x_{2})~T(F_{a}^{\mu\nu}(x_{1})^{(0)}, v^{\rho}_{c}(x_{2})^{(0)}): 
\nonumber\\
\nonumber\\
T(F_{a}^{\mu\nu}(x_{1})^{(0)}, E_{b}^{\rho\sigma}(x_{2})) = 
f_{bcd}~[ :v_{d}^{\sigma}(x_{2})~T(F_{a}^{\mu\nu}(x_{1})^{(0)}, v_{c}^{\rho}(x_{2})^{(0)}): 
\nonumber\\
+ :v_{c}^{\rho}(x_{2})~T(F_{a}^{\mu\nu}(x_{1})^{(0)}, v_{d}^{\sigma}(x_{2})^{(0)}):]
\nonumber\\
\nonumber\\
T(F_{a}^{\mu\nu}(x_{1})^{(0)}, C_{b}^{\rho\sigma}(x_{2})) = 
- f_{bcd}~:u_{c}(x_{2})~T(F_{a}^{\mu\nu}(x_{1})^{(0)}, F_{d}^{\rho\sigma}(x_{2})^{(0)}): 
\nonumber\\
\nonumber\\
T(u_{a}(x_{1})^{(0)}, C_{b\mu}(x_{2})) = f_{bcd}~:u_{c}(x_{2})~T(u_{a}(x_{1})^{(0)}, \tilde{u}_{d,\mu}(x_{2})^{(0)}): 
\nonumber\\
\nonumber\\
T(u_{a}(x_{1})^{(0)}, D_{b}(x_{2})) = f_{bcd}~:v_{c}^{\mu}(x_{2})~T(u_{a}(x_{1})^{(0)}, \tilde{u}_{d,\mu}(x_{2})^{(0)}): 
\nonumber\\
\nonumber\\
T(\tilde{u}_{a}(x_{1})^{(0)}, B_{b}^{\mu}(x_{2})) = - f_{bcd}~:v_{d}^{\mu}(x_{2})~T(\tilde{u}_{a}(x_{1})^{(0)}, u_{c}(x_{2})^{(0)}): 
\nonumber\\
\nonumber\\
T(\tilde{u}_{a}(x_{1})^{(0)}, C_{b}^{\mu}(x_{2})) = - 
f_{bcd}~:\partial^{\mu}\tilde{u}_{d}(x_{2})~T(\tilde{u}_{a}(x_{1})^{(0)}, u_{c}(x_{2})^{(0)}): 
\nonumber\\
\nonumber\\
T(\tilde{u}_{a}(x_{1})^{(0)}, B_{b}(x_{2})) = f_{bcd}~:u_{d}(x_{2})~T(\tilde{u}_{a}(x_{1})^{(0)}, u_{c}(x_{2})^{(0)}): 
\nonumber\\
\nonumber\\
T(\tilde{u}_{a}(x_{1})^{(0)}, C_{b}^{\mu\nu}(x_{2})) = - f_{bcd}~:F_{d}^{\mu\nu}(x_{2})~T(\tilde{u}_{a}(x_{1})^{(0)}, u_{c}(x_{2})^{(0)}): 
\label{xi-BCDE}
\eea

If we use the canonical splitting we arrive at the more precise forms:
\bea
T(v_{a}^{\mu}(x_{1})^{(0)}, B_{b}^{\nu}(x_{2})) = - i~D_{0}^{F}(x_{1} -x_{2})~\eta^{\mu\nu}~f_{abc}~u_{c}(x_{2}) 
\nonumber\\
\nonumber\\
T(v_{a}^{\mu}(x_{1})^{(0)}, C_{b}^{\nu}(x_{2})) = - i~f_{abc}~[ D_{0}^{F}(x_{1} - x_{2})~F_{c}^{\mu\nu}(x_{2}) 
\nonumber\\
+ \eta^{\mu\nu}~\partial_{\rho}D_{0}^{F}(x_{1} - x_{2})~v_{c}^{\rho}(x_{2}) - \partial^{\nu}D_{0}^{F}(x_{1} - x_{2})~v_{c}^{\mu}(x_{2}) ] 
\nonumber\\
\nonumber\\
T(v_{a}^{\mu}(x_{1})^{(0)}, D_{b}(x_{2})) = - i~D_{0}^{F}(x_{1} - x_{2})~f_{abc}~\partial^{\mu}\tilde{u}_{c}(x_{2})
\nonumber\\
\nonumber\\
T(v_{a}^{\mu}(x_{1})^{(0)}, E_{b}^{\rho\sigma}(x_{2})) 
= - i~f_{abc}~D_{0}^{F}(x_{1} - x_{2})~(\eta^{\mu\rho}~v_{c}^{\sigma} - \eta^{\mu\sigma}~v_{c}^{\rho})(x_{2}) 
\nonumber\\
\nonumber\\
T(v_{a}^{\mu}(x_{1})^{(0)}, C_{b}^{\rho\sigma}(x_{2})) =
i~f_{abc}~(\eta^{\mu\sigma}~\partial^{\rho} - \eta^{\mu\rho}~\partial^{\sigma})D_{0}^{F}(x_{1} - x_{2})~u_{c}(x_{2}) 
\nonumber\\
\nonumber\\
T(F_{a}^{\mu\nu}(x_{1})^{(0)}, B_{b}^{\rho}(x_{2})) = 
i~f_{abc}~(\eta^{\mu\rho}~\partial^{\nu} - \eta^{\nu\rho}~\partial^{\mu})D_{0}^{F}(x_{1} - x_{2})~u_{c}(x_{2}) 
\nonumber\\
\nonumber\\
T(F_{a}^{\mu\nu}(x_{1})^{(0)}, C_{b}^{\rho}(x_{2})) =  i~f_{abc}~[ \partial^{\nu}D_{0}^{F}(x_{1} - x_{2})~F_{c}^{\mu\rho}(x_{2}) 
\nonumber\\
+ \partial^{\mu}\partial^{\rho}D_{0}^{F}(x_{1} - x_{2})~v_{c}^{\nu}(x_{2})
+ \eta^{\mu\rho}~\partial^{\nu}\partial_{\sigma}D_{0}^{F}(x_{1} - x_{2})~v_{c}^{\sigma}(x_{2}) ] - (\mu \leftrightarrow \nu)
\nonumber\\
\nonumber\\
T(F_{a}^{\mu\nu}(x_{1})^{(0)}, D_{b}(x_{2})) = i~f_{abc}~\partial^{\nu}D_{0}^{F}(x_{1} - x_{2})~\partial^{\mu}\tilde{u}_{c}(x_{2}) 
- (\mu \leftrightarrow \nu)
\nonumber\\
\nonumber\\
T(F_{a}^{\mu\nu}(x_{1})^{(0)}, E_{b}^{\rho\sigma}(x_{2})) = 
i~f_{abc}~[ \eta^{\mu\rho}~\partial^{\nu}D_{0}^{F}(x_{1} - x_{2})~v_{c}^{\sigma}(x_{2}) -  (\mu \leftrightarrow \nu)]
- (\rho \leftrightarrow \sigma)
\nonumber\\
\nonumber\\
T(F_{a}^{\mu\nu}(x_{1})^{(0)}, C_{b}^{\rho\sigma}(x_{2})) = 
\nonumber\\
i~f_{abc}~(\eta^{\nu\sigma}~\partial^{\mu}\partial_{\rho} - \eta^{\mu\sigma}~\partial^{\nu}\partial_{\rho}
+ \eta^{\mu\rho}~\partial^{\nu}\partial_{\sigma} - \eta^{\nu\rho}~\partial^{\mu}\partial_{\sigma})D_{0}^{F}(x_{1} - x_{2})~u_{c}(x_{2}) 
\nonumber\\
\nonumber\\
T(u_{a}(x_{1})^{(0)}, C_{b}^{\mu}(x_{2})) = i~f_{abc}~\partial^{\mu}D_{0}^{F}(x_{1} - x_{2})~u_{c}(x_{2}) 
\nonumber\\
\nonumber\\
T(u_{a}^{\mu}(x_{1})^{(0)}, D_{b}(x_{2})) = i~f_{abc}~\partial_{\mu}D_{0}^{F}(x_{1} - x_{2})~v_{c}^{\mu}(x_{2}) 
\nonumber\\
\nonumber\\
T(\tilde{u}_{a}(x_{1})^{(0)}, B_{b}^{\mu}(x_{2})) = i~f_{abc}~D_{0}^{F}(x_{1} - x_{2})~v_{c}^{\mu}(x_{2}) 
\nonumber\\
\nonumber\\
T(\tilde{u}_{a}(x_{1})^{(0)}, C_{b}^{\mu}(x_{2})) = i~f_{abc}~D_{0}^{F}(x_{1} - x_{2})~\partial^{\mu}\tilde{u}_{c}(x_{2}) 
\nonumber\\
\nonumber\\
T(\tilde{u}_{a}(x_{1})^{(0)}, B_{b}(x_{2})) = - i~f_{abc}~D_{0}^{F}(x_{1} - x_{2})~u_{c}(x_{2}) 
\nonumber\\
\nonumber\\
T(\tilde{u}_{a}(x_{1})^{(0)}, C_{b}^{\mu\nu}(x_{2})) =   i~f_{abc}~D_{0}^{F}(x_{1} - x_{2})~F_{c}^{\mu\nu}(x_{2}) 
\label{xi-BCDE-canonical}
\eea

Now we consider expresions of the type
$
T(\xi_{a}\cdot T^{I},\xi_{b}\cdot T^{J})
$
and we have with Wick theorem:
\bea
T(B_{a}^{\mu}(x_{1})^{(1)}, B_{b}^{\nu}(x_{2})) = - f_{adc}~:u_{d}(x_{1})~T(v_{c}^{\mu}(x_{1})^{(0)}, B_{b}^{\nu}(x_{2})):
\nonumber\\
\nonumber\\
T(B_{a}^{\mu}(x_{1})^{(1)}, C_{b}^{\nu}(x_{2})) = 
- f_{adc}~[ :u_{d}(x_{1})~T(v_{c}^{\mu}(x_{1})^{(0)}, C_{b}^{\nu}(x_{2})):
\nonumber\\
- :v_{d}^{\mu}(x_{1})~T(u_{c}(x_{1})^{(0)}, C_{b}^{\nu}(x_{2})):]
\nonumber\\
\nonumber\\
T(B_{a}^{\mu}(x_{1})^{(1)}, D_{b}(x_{2})) = 
- f_{adc}~[ :u_{d}(x_{1})~T(v_{c}^{\mu}(x_{1})^{(0)}, D_{b}(x_{2})):
\nonumber\\
- :v_{d}^{\mu}(x_{1})~T(u_{c}(x_{1})^{(0)}, D_{b}(x_{2})):]
\nonumber\\
\nonumber\\
T(B_{a}^{\mu}(x_{1})^{(1)}, E_{b}^{\rho\sigma}(x_{2})) = - f_{adc}~:u_{d}(x_{1})~T(v_{c}^{\mu}(x_{1})^{(0)}, E_{b}^{\rho\sigma}(x_{2})):
\nonumber\\
\nonumber\\
T(B_{a}^{\mu}(x_{1})^{(1)}, B_{b}(x_{2})) = 0 
\nonumber\\
\nonumber\\
T(B_{a}^{\mu}(x_{1})^{(1)}, C_{b}^{\rho\sigma}(x_{2})) = - f_{adc}~:u_{d}(x_{1})~T(v_{c}^{\mu}(x_{1})^{(0)}, C_{b}^{\rho\sigma}(x_{2})):
\nonumber\\
\nonumber\\
T(C_{a}^{\mu}(x_{1})^{(1)}, C_{b}^{\nu}(x_{2})) = 
f_{adc}~[ :v_{d\rho}(x_{1})~T(F_{c}^{\rho\mu}(x_{1})^{(0)}, C_{b}^{\nu}(x_{2})):
\nonumber\\
- :F_{d}^{\rho\mu}(x_{1})~T(v_{c\rho}(x_{1})^{(0)}, C_{b}^{\nu}(x_{2})):
- :u_{d}(x_{1})~T({\tilde{u}_{c,}~}^{\mu}(x_{1})^{(0)}, C_{b}^{\nu}(x_{2})):
\nonumber\\
- :\partial^{\mu}\tilde{u}_{c}(x_{1}) T(u_{c}(x_{1})^{(0)}, C_{b}^{\nu}(x_{2})):]
\nonumber\\
\nonumber\\
T(C_{a}^{\mu}(x_{1})^{(1)}, D_{b}(x_{2})) = 
f_{adc}~[ :v_{d\nu}(x_{1})~T(F_{c}^{\nu\mu}(x_{1})^{(0)}, D_{b}(x_{2})):
\nonumber\\
- :F_{d}^{\nu\mu}(x_{1})~T(v_{c\nu}(x_{1})^{(0)}, D_{b}(x_{2})):
- :\partial^{\mu}\tilde{u}_{d}(x_{1})~T(u_{c}(x_{1})^{(0)}, D_{b}(x_{2})):]
\nonumber\\
\nonumber\\
T(C_{a}^{\mu}(x_{1})^{(1)}, E_{b}^{\rho\sigma}(x_{2})) = 
f_{adc}~[ :v_{d\nu}(x_{1})~T(F_{c}^{\nu\mu}(x_{1})^{(0)}, E_{b}^{\rho\sigma}(x_{2})):
\nonumber\\
- :F_{d}^{\nu\mu}(x_{1})~T(v_{c\nu}(x_{1})^{(0)}, E_{b}^{\rho\sigma}(x_{2})):]
\nonumber\\
\nonumber\\
T(C_{a}^{\mu}(x_{1})^{(1)}, B_{b}(x_{2})) = - f_{adc}~:u_{d}(x_{1})~T({\tilde{u}_{c,}~}^{\mu}(x_{1})^{(0)}, B_{b}(x_{2})):
\nonumber\\
\nonumber\\
T(C_{a}^{\mu}(x_{1})^{(1)}, C_{b}^{\rho\sigma}(x_{2})) = 
f_{adc}~[ :v_{d\nu}(x_{1})~T(F_{c}^{\nu\mu}(x_{1})^{(0)}, C_{b}^{\rho\sigma}(x_{2})):
\nonumber\\
- :F_{d}^{\nu\mu}(x_{1})~T(v_{c\nu}(x_{1})^{(0)}, C_{b}^{\rho\sigma}(x_{2})):
- :u_{d}(x_{1})~T({\tilde{u}_{c,}~}^{\mu}(x_{1})^{(0)}, C_{b}^{\rho\sigma}(x_{2})):]
\nonumber\\
\nonumber\\
T(D_{a}(x_{1})^{(1)}, D_{b}(x_{2})) = - f_{adc}~:\partial_{\mu}\tilde{u}_{d}(x_{1})~T(v_{c}^{\mu}(x_{1})^{(0)}, D_{b}(x_{2})):
\nonumber\\
\nonumber\\
T(D_{a}(x_{1})^{(1)}, E_{b}^{\rho\sigma}(x_{2})) = - f_{adc}~:\partial_{\mu}\tilde{u}_{d}(x_{1})~
T(v_{c}^{\mu}(x_{1})^{(0)}, E_{b}^{\rho\sigma}(x_{2})):
\nonumber\\
\nonumber\\
T(D_{a}(x_{1})^{(1)}, B_{b}(x_{2})) = f_{adc}~:v_{d}^{\mu}(x_{1})~T({\tilde{u}_{c,}~}^{\mu}(x_{1})^{(0)}, B_{b}(x_{2})):
\nonumber\\
\nonumber\\
T(D_{a}(x_{1})^{(1)}, C_{b}^{\rho\sigma}(x_{2})) = 
f_{adc}~[ :v_{d}^{\mu}(x_{1})~T(\tilde{u}_{c,\mu}(x_{1})^{(0)}, C_{b}^{\rho\sigma}(x_{2})):
\nonumber\\
- :\partial_{\mu}\tilde{u}_{d}(x_{1})~T(v_{c}^{\mu}(x_{1})^{(0)}, C_{b}^{\rho\sigma}(x_{2})): ]
\nonumber\\
\nonumber\\
T(E_{a}^{\mu\nu}(x_{1})^{(1)}, E_{b}^{\rho\sigma}(x_{2})) = 
f_{adc}~:v_{d}^{\mu}(x_{1})~T(v_{c}^{\nu}(x_{1})^{(0)}, E_{b}^{\rho\sigma}(x_{2})): - (\mu \leftrightarrow \nu)
\nonumber\\
\nonumber\\
T(E_{a}^{\mu\nu}(x_{1})^{(1)}, B_{b}(x_{2})) = 0
\nonumber\\
\nonumber\\
T(E_{a}^{\mu\nu}(x_{1})^{(1)}, C_{b}^{\rho\sigma}(x_{2})) = 
f_{adc}~:v_{d}^{\mu}(x_{1})~T(v_{c}^{\nu}(x_{1})^{(0)}, C_{b}^{\rho\sigma}(x_{2})): - (\mu \leftrightarrow \nu)
\nonumber\\
\nonumber\\
\nonumber\\
T(B_{a}(x_{1})^{(1)}, B_{b}(x_{2})) = 0
\nonumber\\
\nonumber\\
T(B_{a}(x_{1})^{(1)}, C_{b}^{\mu\nu}(x_{2})) = 0
\nonumber\\
\nonumber\\
T(C_{a}^{\mu\nu}(x_{1})^{(1)}, C_{b}^{\rho\sigma}(x_{2})) = 
- f_{adc}~:u_{d}(x_{1})~T(F_{c}^{\mu\nu}(x_{1})^{(0)}, E_{b}^{\rho\sigma}(x_{2})): 
\eea

Using the canonical splitting we have:
\bea
T(B_{a}^{\mu}(x_{1})^{(1)}, B_{b}^{\nu}(x_{2})^{(1)}) = i~\eta^{\mu\nu}~f_{ace}~f_{bde}~D_{0}^{F}(x_{1} - x_{2})~:u_{c}(x_{1}) u_{d}(x_{2}):
\nonumber\\
\nonumber\\
T(B_{a}^{\mu}(x_{1})^{(1)}, C_{b}^{\nu}(x_{2})^{(1)}) = i~f_{ace}~f_{bde}~
\{ D_{0}^{F}(x_{1} - x_{2})~:u_{c}(x_{1}) F_{d}^{\mu\nu}(x_{2})):
\nonumber\\
+ \eta^{\mu\nu}~\partial_{\rho}D_{0}^{F}(x_{1} - x_{2})~:u_{c}(x_{1}) v_{d}^{\rho}(x_{2}):
+ \partial^{\nu}D_{0}^{F}(x_{1} - x_{2})~[ :v_{c}^{\mu}(x_{1}) u_{d}(x_{2}): - :u_{c}(x_{1}) v_{d}^{\mu}(x_{2}): ]\}
\nonumber\\
\nonumber\\
T(B_{a}^{\mu}(x_{1})^{(1)}, D_{b}(x_{2})^{(1)}) = 
i~f_{ace}~f_{bde}~[ D_{0}^{F}(x_{1} - x_{2})~:u_{c}(x_{1}) \partial^{\mu}\tilde{u}_{d}(x_{2})):
\nonumber\\
+ \partial_{\nu}D_{0}^{F}(x_{1} - x_{2})~:v_{c}^{\mu}(x_{1}) v_{d}^{\nu}(x_{2}):
\nonumber\\
\nonumber\\
T(B_{a}^{\mu}(x_{1})^{(1)}, E_{b}^{\rho\sigma}(x_{2})^{(1)}) =
i~f_{ace}~f_{bde}~D_{0}^{F}(x_{1} - x_{2})~[ \eta^{\mu\rho}~:u_{c}(x_{1}) v_{d}^{\sigma}(x_{2})): - (\rho \leftrightarrow \sigma) ]
\nonumber\\
\nonumber\\
T(B_{a}^{\mu}(x_{1})^{(1)}, B_{b}(x_{2})^{(1)}) = 0 
\nonumber\\
\nonumber\\
T(B_{a}^{\mu}(x_{1})^{(1)}, C_{b}^{\rho\sigma}(x_{2})^{(1)}) = 
i~f_{ace}~f_{bde}~(\eta^{\mu\rho}~\partial^{\sigma} - \eta^{\mu\sigma}~\partial^{\rho})D_{0}^{F}(x_{1} - x_{2})~:u_{c}(x_{1}) u_{d}(x_{2})):
\nonumber\\
\nonumber\\
T(C_{a}^{\mu}(x_{1})^{(1)}, C_{b}^{\nu}(x_{2})^{(1)}) = i~f_{ace}~f_{bde}~
\{ D_{0}^{F}(x_{1} - x_{2})~:F_{c}^{\mu\rho}(x_{1}) {F_{d}^{\nu}}_{\rho}(x_{2})):
\nonumber\\
+ \partial^{\mu}D_{0}^{F}(x_{1} - x_{2})~[ :v_{c\rho}(x_{1}) F_{d}^{\rho\nu}(x_{2}): - :u_{c}(x_{1}) \partial^{\nu}\tilde{u}_{d}(x_{2}):]
\nonumber\\
- \partial^{\nu}D_{0}^{F}(x_{1} - x_{2})~[ :F_{c}^{\rho\nu}(x_{1}) v_{d\rho}(x_{2}): + :\partial^{\nu}\tilde{u}_{c}(x_{1}) u_{d}(x_{2}):]
\nonumber\\
- \partial_{\rho}D_{0}^{F}(x_{1} - x_{2})~[ :F_{c}^{\mu\nu}(x_{1}) v_{d}^{\rho}(x_{2}): + :v_{c}^{\rho}(x_{1}) F_{d}^{\mu\nu}(x_{2}):]
\nonumber\\
- \partial^{\mu}\partial^{\nu}D_{0}^{F}(x_{1} - x_{2})~:v_{c\rho}(x_{1}) v_{d}^{\rho}(x_{2}): 
- \eta^{\mu\nu}~\partial_{\rho}\partial_{\sigma}D_{0}^{F}(x_{1} - x_{2})~:v_{c}^{\rho}(x_{1}) v_{d}^{\sigma}(x_{2}):
\nonumber\\
+ \partial^{\nu}\partial_{\rho}D_{0}^{F}(x_{1} - x_{2})~:v_{c}^{\rho}(x_{1}) v_{d}^{\mu}(x_{2}):
+ \partial^{\mu}\partial_{\rho}D_{0}^{F}(x_{1} - x_{2})~:v_{c}^{\nu}(x_{1}) v_{d}^{\rho}(x_{2}): \}
\nonumber\\
\nonumber\\
T(C_{a}^{\mu}(x_{1})^{(1)}, D_{b}(x_{2})^{(1)}) = i~f_{ace}~f_{bde}~
\{ - D_{0}^{F}(x_{1} - x_{2})~:F_{c}^{\mu\nu}(x_{1}) \partial_{\nu}\tilde{u}_{d}(x_{2}):
\nonumber\\
+ \partial^{\mu}D_{0}^{F}(x_{1} - x_{2})~:v_{c}^{\nu}(x_{1}) \partial_{\nu}\tilde{u}_{d}(x_{2}): 
\nonumber\\
- \partial_{\nu}D_{0}^{F}(x_{1} - x_{2})~[ :v_{c}^{\nu}(x_{1}) \partial^{\mu}\tilde{u}_{d}(x_{2}): 
+ :\partial^{\mu}\tilde{u}_{c}(x_{1}) v_{d}^{\nu}(x_{2}):]
\nonumber\\
\nonumber\\
T(C_{a}^{\mu}(x_{1})^{(1)}, E_{b}^{\rho\sigma}(x_{2})^{(1)}) = i~f_{ace}~f_{bde}~
\{ D_{0}^{F}(x_{1} - x_{2})~[ :F_{c}^{\rho\mu}(x_{1}) v_{d}^{\sigma}(x_{2}): -  :F_{c}^{\sigma\mu}(x_{1}) v_{d}^{\rho}(x_{2}): ]
\nonumber\\
+ \partial^{\mu}D_{0}^{F}(x_{1} - x_{2})~[ :v_{c}^{\rho}(x_{1}) v_{d}^{\sigma}(x_{2}): - :v_{c}^{\sigma}(x_{1}) v_{d}^{\rho}(x_{2}): ] 
\nonumber\\
- \partial_{\nu}D_{0}^{F}(x_{1} - x_{2})~[ \eta^{\mu\sigma}~:v_{c}^{\nu}(x_{1}) v_{d}^{\rho}(x_{2}): 
- \eta^{\mu\rho}~:v_{c}^{\nu}(x_{1}) v_{d}^{\sigma}(x_{2}):]\}
\nonumber\\
\nonumber\\
T(C_{a}^{\mu}(x_{1})^{(1)}, B_{b}(x_{2})^{(1)}) = i~f_{ace}~f_{bde}~\partial^{\mu}D_{0}^{F}(x_{1} - x_{2})~:u_{c}(x_{1}) u_{d}(x_{2}): 
\nonumber\\
\nonumber\\
T(C_{a}^{\mu}(x_{1})^{(1)}, C_{b}^{\rho\sigma}(x_{2})^{(1)}) = i~f_{ace}~f_{bde}~
[ - \partial^{\mu}D_{0}^{F}(x_{1} - x_{2})~:u_{c}(x_{1}) F_{d}^{\rho\sigma}(x_{2}):
\nonumber\\
+ \partial^{\rho}D_{0}^{F}(x_{1} - x_{2})~:F_{c}^{\mu\sigma}(x_{1}) u_{d}(x_{2}): 
- \partial^{\sigma}D_{0}^{F}(x_{1} - x_{2})~:F_{c}^{\mu\rho}(x_{1}) u_{d}(x_{2}): 
\nonumber\\
+ (\eta^{\mu\sigma}~\partial^{\rho}\partial_{\nu} - \eta^{\mu\rho}~\partial^{\sigma}\partial_{\nu})D_{0}^{F}(x_{1} - x_{2})~
:v_{c}^{\nu}(x_{1}) u_{d}(x_{2}): 
\nonumber\\
+ \partial^{\mu}\partial^{\sigma}D_{0}^{F}(x_{1} - x_{2}) :v_{c}^{\rho}(x_{1}) u_{d}(x_{2}): 
- \partial^{\mu}\partial^{\rho}D_{0}^{F}(x_{1} - x_{2}) :v_{c}^{\sigma}(x_{1}) u_{d}(x_{2}):]
\nonumber\\
\nonumber\\
T(D_{a}(x_{1})^{(1)}, D_{b}(x_{2})^{(1)}) = i~f_{ace}~f_{bde}~D_{0}^{F}(x_{1} - x_{2})~
:\partial_{\mu}\tilde{u}_{c}(x_{1}) \partial^{\mu}\tilde{u}_{d}(x_{2}): 
\nonumber\\
\nonumber\\
T(D_{a}(x_{1})^{(1)}, E_{b}^{\rho\sigma}(x_{2})^{(1)}) = i~f_{ace}~f_{bde}~D_{0}^{F}(x_{1} - x_{2})~
[ :\partial^{\rho}\tilde{u}_{c}(x_{1}) v_{d}^{\sigma}(x_{2}): -  :\partial^{\sigma}\tilde{u}_{c}(x_{1}) v_{d}^{\rho}(x_{2}): ]
\nonumber\\
\nonumber\\
T(D_{a}(x_{1})^{(1)}, B_{b}(x_{2})^{(1)}) = - i~f_{ace}~f_{bde}~\partial_{\mu}D_{0}^{F}(x_{1} - x_{2})~:v_{c}^{\mu}(x_{1}) u_{d}(x_{2}): 
\nonumber\\
\nonumber\\
T(D_{a}(x_{1})^{(1)}, C_{b}^{\rho\sigma}(x_{2})^{(1)}) = i~f_{ace}~f_{bde}~
[ \partial^{\mu}D_{0}^{F}(x_{1} - x_{2})~:v_{c}^{\mu}(x_{1}) F_{d}^{\rho\sigma}(x_{2}):
\nonumber\\
+ \partial^{\sigma}D_{0}^{F}(x_{1} - x_{2})~:\partial^{\rho}\tilde{u}_{c}(x_{1}) u_{d}(x_{2}): 
- \partial^{\rho}D_{0}^{F}(x_{1} - x_{2})~:\partial^{\sigma}\tilde{u}_{c}(x_{1}) u_{d}(x_{2}): ]
\nonumber\\
\nonumber\\
T(E_{a}^{\mu\nu}(x_{1})^{(1)}, E_{b}^{\rho\sigma}(x_{2})^{(1)}) = - i~f_{ace}~f_{bde}~D_{0}^{F}(x_{1} - x_{2})~
\nonumber\\
~[ \eta^{\nu\rho}~:v_{c}^{\mu}(x_{1}) v_{d}^{\sigma}(x_{2}): -  \eta^{\nu\sigma}~:v_{c}^{\mu}(x_{1}) v_{d}^{\rho}(x_{2}):
%\nonumber\\
+ \eta^{\mu\sigma}~:v_{c}^{\nu}(x_{1}) v_{d}^{\rho}(x_{2}): - \eta^{\mu\rho}~:v_{c}^{\nu}(x_{1}) v_{d}^{\sigma}(x_{2}):]
\nonumber\\
\nonumber\\
T(E_{a}^{\mu\nu}(x_{1})^{(1)}, B_{b}(x_{2})^{(1)}) = 0
\nonumber\\
\nonumber\\
T(E_{a}^{\mu\nu}(x_{1})^{(1)}, C_{b}^{\rho\sigma}(x_{2})^{(1)}) = i~f_{ace}~f_{bde}~
\nonumber\\
~[ (\eta^{\nu\sigma}\partial^{\rho} - \eta^{\nu\rho}\partial^{\sigma})D_{0}^{F}(x_{1} - x_{2})~:v_{c}^{\mu}(x_{1}) u_{d}(x_{2}): 
\nonumber\\
- (\eta^{\mu\sigma}\partial^{\rho} - \eta^{\mu\rho}\partial^{\sigma})D_{0}^{F}(x_{1} - x_{2})~:v_{c}^{\nu}(x_{1}) u_{d}(x_{2}): ]
\nonumber\\
\nonumber\\
T(B_{a}(x_{1})^{(1)}, B_{b}(x_{2})^{(1)}) = 0
\nonumber\\
\nonumber\\
T(B_{a}(x_{1})^{(1)}, C_{b}^{\mu\nu}(x_{2})^{(1)}) = 0
\nonumber\\
\nonumber\\
T(C_{a}^{\mu\nu}(x_{1})^{(1)}, C_{b}^{\rho\sigma}(x_{2})^{(1)}) = - i~f_{ace}~f_{bde}~
\nonumber\\
~(\eta^{\nu\sigma}\partial^{\mu}\partial^{\rho} - \eta^{\mu\sigma}\partial^{\nu}\partial^{\rho}
+ \eta^{\mu\rho}\partial^{\nu}\partial^{\sigma} - \eta^{\nu\rho}\partial^{\mu}\partial^{\sigma}))D_{0}^{F}(x_{1} - x_{2})~
:u_{c}(x_{1}) u_{d}(x_{2}): 
\eea

We can use the previous expresions to compute the corresponding anomalies. We have:
\begin{thm}
The following relations are true:
\bea
s^{\prime}T(B_{a}^{\mu}(x_{1})^{(1)}, C_{b}^{\mu}(x_{2})^{(1)}) = \delta( x_{1} - x_{2})~\eta^{\mu\nu}~f_{ace}~f_{bde}~(:u_{c} u_{d}:)(x_{2})
\label{sbc}
\eea
\bea
s^{\prime}T(B_{a}^{\mu}(x_{1})^{(2)}, D_{b}(x_{2})^{(1)}) = \delta( x_{1} - x_{2})~f_{ace}~f_{bde}~
(:v_{c}^{\mu} u_{d}: - :u_{c} v_{d}^{\mu}:)(x_{2})
\label{sbd}
\eea
\bea
s^{\prime}T(C^{\mu}_{a}(x_{1})^{(2)}, C_{b}^{\nu}(x_{2})^{(1)}) = \delta( x_{1} - x_{2})~f_{ace}~f_{bde}~
(:F_{c}^{\mu\nu} u_{d}: - :u_{c} F_{d}^{\mu\nu}:)(x_{2})
\nonumber\\
+ \eta^{\mu\nu}~f_{ace}~f_{bde}~\partial_{\rho}\delta( x_{1} - x_{2})~
[:v_{c}^{\rho}(x_{1}) u_{d}(x_{2}): - :u_{c}(x_{1}) v_{d}^{\rho}(x_{2}): ]
\nonumber\\
- f_{ace}~f_{bde}~[ \partial^{\mu}\delta( x_{1} - x_{2})~:v_{c}^{\nu}(x_{1}) u_{d}(x_{2}): 
- \partial^{\nu}\delta( x_{1} - x_{2})~:u_{c}(x_{1}) v_{d}^{\mu}(x_{2}): 
\label{scc}
\eea
\bea
s^{\prime}T(C_{a}^{\mu}(x_{1})^{(2)}, D_{b}(x_{2})^{(1)}) = - \delta( x_{1} - x_{2})~f_{ace}~f_{bde}~
(:u_{c} \partial^{\mu}\tilde{u}_{d}: + :F_{c}^{\nu\mu}v_{d\nu}: +  :\partial^{\mu}\tilde{u}_{c} u_{d}:)(x_{2})
\nonumber\\
- f_{ace}~f_{bde}~\partial^{\mu}\delta( x_{1} - x_{2})~:v_{c\rho}(x_{1}) v_{d}^{\rho}(x_{2}): 
+ f_{ace}~f_{bde}~\partial^{\nu}\delta( x_{1} - x_{2})~:v_{c}^{\nu}(x_{1}) v_{d}^{\mu}(x_{2}): 
\label{scd}
\eea
\bea
s^{\prime}T(C^{\mu}(x_{1})^{(2)}, E_{b}^{\rho\sigma}(x_{2})^{(1)}) = 
\delta( x_{1} - x_{2})~f_{ace}~f_{bde}~
(\eta^{\mu\sigma}~:u_{c} v_{d}^{\rho}: - \eta^{\mu\rho}~:u_{c} v_{d}^{\sigma}:)(x_{2})
\label{sce}
\eea
\bea
s^{\prime}T(C_{a}^{\mu}(x_{1})^{(2)}, C_{b}^{\rho\sigma}(x_{2})^{(1)}) = f_{ace}~f_{bde}~
(\eta^{\mu\sigma}\partial^{\rho} - \eta^{\mu\rho}\partial^{\sigma})\delta( x_{1} - x_{2}) :u_{c}(x_{1}) u_{d}(x_{2}): 
\label{scc1}
\eea
\bea
s^{\prime}T(D(x_{1})^{(2)}, D_{b}(x_{2})^{(1)}) = \delta( x_{1} - x_{2})~f_{ace}~f_{bde}~
(:\partial^{\mu}\tilde{u}_{c} v_{d}^{\mu}: - :v_{c}^{\mu}\partial^{\mu}\tilde{u}_{d}:)(x_{2})
\label{sdd}
\eea
\bea
s^{\prime}T(D_{a}(x_{1})^{(2)}, E_{b}^{\rho\sigma}(x_{2})^{(1)}) = \delta( x_{1} - x_{2})~f_{ace}~f_{bde}~
(- :v_{c}^{\rho} v_{d}^{\sigma}: + :v_{c}^{\sigma}v_{d}^{\rho}:)(x_{2})
\label{sde}
\eea
\bea
s^{\prime}T(D_{a}(x_{1})^{(2)}, B_{b}(x_{2})^{(1)}) = - \delta( x_{1} - x_{2})~f_{ace}~f_{bde}~( :u_{c} u_{d}:)(x_{2})
\label{sdb}
\eea
\bea
s^{\prime}T(D_{a}(x_{1})^{(2)}, C_{b}^{\rho\sigma}(x_{2})^{(1)}) = \delta( x_{1} - x_{2})~f_{ace}~f_{bde}~
(:u_{c} F_{d}^{\rho\sigma}:)(x_{2})
\nonumber\\
- f_{ace}~f_{bde}~[ \partial^{\rho}\delta( x_{1} - x_{2})~:v_{c}^{\sigma}(x_{1}) u_{d}(x_{2}): - ( \rho \leftrightarrow \sigma) ]
\label{sdc}
\eea

We also have:
\bea
s^{\prime}T(v_{a}^{\mu}(x_{1}), C_{b}^{\nu}(x_{2})) = \delta( x_{1} - x_{2})~\eta^{\mu\nu}~f_{abc}~u_{c}(x_{2})
\nonumber\\
s^{\prime}T(v_{a}^{\mu}(x_{1}), D_{b}(x_{2})) = \delta( x_{1} - x_{2})~f_{abc}~v_{c}^{\mu}(x_{2})
\nonumber\\
s^{\prime}T(F_{a}^{\mu\nu}(x_{1}), C_{b}^{\rho}(x_{2})) = 
(\eta^{\nu\rho}\partial^{\mu} - \eta^{\mu\rho}\partial^{\nu})\delta( x_{1} - x_{2})~f_{abc}~u_{c}(x_{2})
\nonumber\\
s^{\prime}T(F_{a}^{\mu\nu}(x_{1}), D_{b}^{\rho}(x_{2})) = 
\partial^{\mu}\delta( x_{1} - x_{2})~f_{abc}~v_{c}^{\nu}(x_{2}) - (\mu \leftrightarrow \nu)
\nonumber\\
s^{\prime}T(u_{a}(x_{1}), D_{b}(x_{2})) = - \delta( x_{1} - x_{2})~\eta^{\mu\nu}~f_{abc}~u_{c}(x_{2})
\label{s-xi-BCDE}
\eea
\end{thm}
Now we investigate if we can impose gauge invariance for expresions of the type
$
T(\xi_{a}\cdot T^{I},\xi_{b}\cdot T^{J})
$
and 
$
T(\xi_{a},\xi_{b}\cdot T^{J})
$;
the answer is negative.
\begin{thm}
The finite renormalization (\ref{vv}) induces also:
\bea
N(F_{a}^{\mu\nu},C_{b}^{\rho}) = i~f_{abc}~(\eta^{\nu\rho} v_{c}^{\mu} - \eta^{\mu\rho} v_{c}^{\nu})
\nonumber\\
N(F_{a}^{\mu\nu},C_{b}^{\rho\sigma}) = - i~f_{abc}~(\eta^{\mu\rho} \eta^{\nu\sigma} - \eta^{\nu\rho} \eta^{\mu\sigma}) u_{c}
\label{N-xiC}
\eea
and
\bea
N(C_{a}^{\mu},C_{b}^{\nu}) = i~f_{ace} f_{bde}~(\eta^{\mu\nu} :v_{c\rho} v_{d}^{\rho}: - :v_{c}^{\nu} v_{d}^{\mu}:)
\nonumber\\
N(C_{a}^{\mu},C_{b}^{\rho\sigma}) = i~f_{ace} f_{bde}~(\eta^{\mu\rho} :v_{c}^{\sigma} u_{d}: - \eta^{\mu\sigma} :v_{c}^{\rho} u_{d}:)
\nonumber\\
N(C_{a}^{\mu\nu},C_{b}^{\rho\sigma}) = i~f_{ace} f_{bde}~(\eta^{\mu\rho} \eta^{\nu\sigma} -  \eta^{\nu\rho} \eta^{\mu\sigma}):u_{c} u_{d}:
\label{N-CC}
\eea

If we perform there finite renormalizations the expresions from the previous theorem go in to:
\be
s^{\prime}T^{\rm ren}T(A_{1}(x_{1}),A_{2}(x_{2}) = \delta( x_{1} - x_{2})~{\cal A}(A_{1},A_{2})(x_{2})
\ee
where 
$
A_{1},A_{2}
$
are Wick submonomials of the type
$
\xi_{a}\cdot T^{I}.
$
The non-trivial expresions
$
{\cal A}(A_{1},A_{2})
$
are:
\bea
{\cal A}(C_{a}^{\mu},C_{b}^{\nu}) = f_{abc}~C_{c}^{\mu\nu}
\nonumber\\
{\cal A}(C_{a}^{\mu},D_{b}) = - f_{abc}~C_{c}^{\mu}
\nonumber\\
{\cal A}(C_{a}^{\mu},E_{b}^{\rho\sigma}) = f_{abc}~( \eta^{\mu\rho} B_{c}^{\sigma} - \eta^{\mu\sigma} B_{c}^{\rho})
\nonumber\\
{\cal A}(D_{a},D_{b}) = - f_{abc}~D_{c}
\nonumber\\
{\cal A}(D_{a},E_{b}^{\rho\sigma}) = - f_{abc}~E_{c}^{\rho\sigma}
\nonumber\\
{\cal A}(D_{a},B_{b}) = - f_{abc}~B_{c}
\nonumber\\
{\cal A}(D_{a},C_{b}^{\rho\sigma}) = - f_{abc}~C_{c}^{\rho\sigma}
\nonumber\\
{\cal A}(B_{a}^{\mu},C_{b}^{\nu}) = \eta^{\mu\nu}~f_{abc}~B_{c}
\nonumber\\
{\cal A}(B_{a}^{\mu},D_{b}) = f_{abc}~B_{c}^{\mu}.
\label{ano-BB}
\eea
In the case when 
$
A_{1} = v_{a}^{\mu}, u_{a}, \tilde{u}_{a}
$
and
$
A_{2} = B_{b}^{\mu},C_{b}^{\mu}, D_{b}, E_{b}^{\rho\sigma}, B_{b}, C_{b}^{\mu\nu}
$
the non-trivial expresions
$
{\cal A}(A_{1},A_{2})
$
are:
\bea
{\cal A}(v_{a}^{\mu},C_{b}^{\nu}) = f_{abc}~\eta^{\mu\nu}~u_{c}
\nonumber\\
{\cal A}(v_{a}^{\mu},D_{b}) = f_{abc}~v_{c}^{\mu}
\nonumber\\
{\cal A}(F_{a}^{\mu\nu},D_{b}) = f_{abc}~F_{c}^{\mu\nu}
\nonumber\\
{\cal A}(u_{a},D_{b}) = - f_{abc}~u_{c}.
\eea
These anomalies cannot be removed by other finite renormalizations. 
\end{thm}
{\bf Proof:} 
We illustrate the proof considering the first case from the statement. We have
\bea
s^{\prime}T^{\rm ren}(C_{a}^{\mu}(x_{1}), C_{b}^{\nu}(x_{2})) = s^{\prime}T(C_{a}^{\mu}(x_{1}), T(x_{2}))
\nonumber\\
+ \delta(x_{1} - x_{2})~R(C_{a}^{\mu},C_{b}^{\nu})(x_{2}) + i \partial_{\nu}\delta(x_{1} - x_{2})~R^{\rho}(C_{a}^{\mu},C_{b}^{\nu})(x_{2})
\eea
where
\bea
R(C_{a}^{\mu},C_{b}^{\nu}) \equiv d_{Q}N(C_{a}^{\mu},C_{b}^{\nu}) - i \partial_{\rho}N(C_{a}^{\mu},C_{b}^{\nu\rho})
\nonumber\\
R^{\rho}(C_{a}^{\mu},C_{b}^{\nu}) \equiv i~[ - N(C_{a}^{\mu\rho},C_{b}^{\nu}) + N(C_{a}^{\mu},C_{b}^{\nu\rho}) ]
\eea
Using the expresion
$
s^{\prime}T(C_{a}^{\mu}(x_{1}), T(x_{2}))
$
from the previous theorem and the expresions $N$ from the statement we obtain after some computations the first relation from 
(\ref{ano-BB}). The rest of the relations can be obtained in the same way. 

Next we prove that the anomaly from the first relation from (\ref{ano-BB}) cannot be removed. The most general finite renormalizations
would be:
\bea
N(C_{a}^{\mu}(x_{1}),C_{b}^{\nu}(x_{2})) = i~\delta(x_{1} - x_{2})~R_{ab}^{\mu\nu}(x_{2})
\nonumber\\
N(C_{a}^{\mu\nu}(x_{1}),C_{b}^{\nu}(x_{2})) = i~\delta(x_{1} - x_{2})~R_{ab}^{\mu\nu,\rho}(x_{2})
\eea
where the generic forms are:
\bea
R_{ab}^{\mu\nu} = f_{abcd}^{(1)}~:v_{c}^{\mu} v_{d}^{\nu}: + f_{abcd}^{(2)}~\eta^{\mu\nu} :v_{c\rho} v_{d}^{\rho}:
+ f_{abcd}^{(3)}~\eta^{\mu\nu} :u_{c} \tilde{u}_{d}:
\nonumber\\
R_{ab}^{\mu\nu,\rho} = f_{abcd}^{(4)}~( \eta^{\mu\rho} :u_{c} v_{d}^{\nu}: -\eta^{\nu\rho} :u_{c} v_{d}^{\mu}:) 
\eea
with 
\be
f_{abcd}^{(1)} = f_{badc}^{(1)}, \quad f_{abcd}^{(2)} = f_{bacd}^{(2)} = f_{abdc}^{(2)}, \quad f_{abcd}^{(3)} = f_{bacd}^{(3)}.
\ee
We insert everything in the equation
\bea
\delta( x_{1} - x_{2})~{\cal A}(C_{a}^{\mu},C_{b}^{\nu})(x_{2}) = 
\nonumber\\
d_{Q}NC_{a}^{\mu}(x_{1}),C_{b}^{\nu}(x_{2}))
- i \partial_{\rho}^{1}N(C_{a}^{\mu\rho}(x_{1}),C_{b}^{\nu}(x_{2}))  - i \partial_{\rho}^{2}N(C_{a}^{\mu}(x_{1}),C_{b}^{\nu\rho}(x_{2}))
\eea
and after some computations we obtain 
$
f_{abc} = 0
$
which is not possible.
$\qed$
%\newpage

However, the Wick property is preserved.
\begin{thm}
The following relations are true:
The finite renormalizations (\ref{N-TT}) and (\ref{N-BCDE-T}) preserve Wick expansion property. Explicitly we have:
\bea
v_{b}^{\nu}\cdot N(C_{a}^{\mu},T) = f_{abc}~N(F_{c}^{\nu\mu},T) + N(C_{a}^{\mu},C_{b}^{\nu})
\nonumber\\
v_{b}^{\rho}\cdot N(C_{a}^{\mu},T^{\nu}) = f_{abc}~N(F_{c}^{\rho\mu},T^{\nu}) + N(C_{a}^{\mu},C_{b}^{\rho\nu})
\nonumber\\
u_{b}\cdot N(C_{a}^{\mu},T^{\nu}) = N(C_{a}^{\mu},C_{b}^{\nu})
\nonumber\\
v_{b}^{\rho}\cdot N(C_{a}^{\mu\nu},T) = N(C_{a}^{\nu\mu},C^{\rho}_{b})
\nonumber\\
u_{b}\cdot N(C_{a}^{\mu\nu},T) = f_{abc}~N(F_{c}^{\mu\nu},T) + N(C_{a}^{\mu\nu},D_{b})
\nonumber\\
v_{b}^{\sigma}\cdot N(C_{a}^{\mu\nu},T^{\rho}) = N(C_{a}^{\mu\nu},C_{b}^{\sigma\rho})
\nonumber\\
u_{b}\cdot N(C_{a}^{\mu\nu},T^{\rho}) = - f_{abc}~N(F_{c}^{\mu\nu},T^{\rho}) - N(C_{a}^{\mu\nu},C_{b}^{\rho})
\nonumber\\
v_{b}^{\rho}\cdot N(C_{a}^{\mu},T^{\rho\sigma}) = f_{abc}~N(F_{c}^{\nu\mu},T^{\rho\sigma})
\nonumber\\
u_{b}\cdot N(C_{a}^{\mu},T^{\rho\sigma}) = - N(C_{a}^{\mu},C_{b}^{\rho\sigma})
\label{consistency3}
\eea
and
\be
N(v_{a}^{\mu},T^{I}) = 0,\quad N(u_{a},T^{I}) = 0, \quad N(\tilde{u}_{a},T^{I}) = 0.
\label{consistency4}
\ee

We also have:
\bea
v_{c}^{\rho}\cdot N(C_{a}^{\mu},C_{b}^{\nu}) = f_{acd}~N(F_{d}^{\rho\mu},C_{b}^{\nu}) + f_{bcd}~N(C_{a}^{\mu},F_{d}^{\rho\nu})
\nonumber\\
v_{c}^{\lambda}\cdot N(C_{a}^{\mu\nu},C_{b}^{\rho}) = f_{bcd}~N(C_{a}^{\mu\nu},F_{d}^{\lambda\rho})
\nonumber\\
u_{c}\cdot N(C_{a}^{\mu\nu},C_{b\rho}) = f_{acd}~N(F_{d}^{\mu\nu},C_{b\rho}) - f_{bcd}~N(C_{a}^{\mu\nu},\tilde{u}_{d,\rho})
\nonumber\\
u_{c}\cdot N(C_{a}^{\mu\nu},C_{b}^{\rho\sigma}) = 
- f_{acd}~N(F_{d}^{\mu\nu},C_{b}^{\rho\sigma}) + f_{bcd}~N(C_{a}^{\mu\nu},F_{d}^{\rho\sigma})
\label{consistency5}
\eea

\bea
v_{b}^{\rho}\cdot N(F_{a}^{\mu\nu},T) = N(F_{a}^{\mu\nu},C_{b}^{\rho})
\nonumber\\
v_{b}^{\sigma}\cdot N(F_{a}^{\mu\nu},T^{\rho}) = - N(F_{a}^{\mu\nu},C_{b}^{\rho\sigma})
\nonumber\\
u_{b}\cdot N(F_{a}^{\mu\nu},T^{\rho}) = - N(F_{a}^{\mu\nu},C_{b}^{\rho})
\nonumber\\
u_{b}\cdot N(F_{a}^{\mu\nu},T^{\rho\sigma}) = 0.
\label{consistency6}
\eea
and
\bea
v_{c}^{\sigma}\cdot N(F_{a}^{\mu\nu},C_{b}^{\rho}) = -  f_{bcd}~N(F_{a}^{\mu\nu},F_{d}^{\rho\sigma})
\nonumber\\
u_{c}\cdot N(F_{a}^{\mu\nu},C_{b}^{\rho\sigma}) =  f_{bcd}~N(F_{a}^{\mu\nu},F_{b}^{\rho\sigma})
\label{consistency7}
\eea
\end{thm}
{\bf Proof:} 
We start with the relation
\bea
~[v_{b}^{\mu}(y), T(C_{a}^{\mu}(x_{1}),T(x_{2}))] = 
\nonumber\\
i~f_{abc}~D_{0}(y - x_{1})~T(F_{c}^{\nu\mu}(x_{1}),T(x_{2})) 
\nonumber\\
- i~f_{abc}~\partial_{\mu}D_{0}(y - x_{1})~[ \eta^{\mu\rho}~T(v_{c}^{\nu}(x_{1}),T(x_{2})) - \eta^{\mu\nu}~T(v_{c}^{\rho}(x_{1}),T(x_{2})) ]
\nonumber\\
+ i~D_{0}(y - x_{2})~T(C_{a}^{\mu}(x_{1}),C_{b}^{\nu}(x_{2}))
- i~\partial_{\rho}D_{0}(y - x_{1})~T(C_{a}^{\mu}(x_{1}),E_{b}^{\nu\rho}(x_{2}))
\eea
following from Wick expansion property (\ref{comm-wick}). If we consider finite renormalizations of the type 
(\ref{N-BCDE-T}), (\ref{N-xiT}) and  (\ref{N-CC}),we will get new terms in the left and right hand of the preceding identity. The identity is 
preserved {\it iff} we have the first relation from (\ref{consistency3}) and some relations from  
(\ref{consistency2}) and (\ref{consistency4}). 

The rest of the consistency relations are going in the same way.
$\qed$
\newpage
\section{Finite Renormalizations}

We have proved in Theorem \ref{ren-TT} that the anomalies can be eliminated by adding to the chronological products some 
quasi-local operators. However, the chronological products have still some arbitrariness: in principle, we can add other
quasi-local operators in such a way that gauge invariance in the second order is preserved. This arbitrariness is described by
\begin{thm}
The finite renormalizations are of the type
\be
R(T^{I}(x_{1}),T^{J}(x_{2})) = \delta(x_{1} - x_{2})~N(T^{I},T^{J})(x_{2}) 
\label{R}
\ee
where the polynomials
$
N(T^{I},T^{J})
$
verify the symmetry property:
\be
N(T^{I},T^{J}) = (-1)^{|I||J|}~N(T^{J},T^{I})
\ee
and
\be
gh(N(T^{I},T^{J})) = |I| + |J|, \qquad \omega(N(T^{I},T^{J})) \leq 4.
\ee

These finite renormalizations do not produce anomalies iff there exists expresions
$
N^{I}
$
such that
\be
N(T^{I},T^{J}) = N^{JI}
\label{NI}
\ee
and
\be
d_{Q}N^{I} = i~d_{\mu}N^{I\mu}.
\label{sN}
\ee
\label{R-finite}
\end{thm}
{\bf Proof:} The general form of the finite anomaly (\ref{R}) follows form Bogoliubov axioms (power counting and ghost number
asignement). We now want that this redefinition of the chronological products does not create new anomalies i.e. the expresion
\bea
sR(T^{I}(x_{1}),T^{J}(x_{2})) \equiv d_{Q}R(T^{I}(x_{1}),T^{J}(x_{2}))
\nonumber\\
- i~\partial_{\mu}^{1}R(T^{I\mu}(x_{1}),T^{J}(x_{2}))
- i~(-1)^{|I|}~\partial_{\mu}^{2}R(T^{I}(x_{1}),T^{J\mu}(x_{2}))
\eea
is null. We insert (\ref{R}) and by direct computation we have:
\bea
sR(T^{I}(x_{1}),T^{J}(x_{2})) = \delta(x_{1} - x_{2})~[ d_{Q}N(T^{I},T^{J}) - i~(-1)^{|I|}~\partial_{\mu}N(T^{I},T^{J\mu})](x_{2})
\nonumber\\
- i~\partial_{\mu}\delta(x_{1} - x_{2})~[ N(T^{I\mu},T^{J}) - (-1)^{|I|}~N(T^{I},T^{J\mu})](x_{2}) 
\label{sR}
\eea
and this expresion is null iff
\bea
d_{Q}N(T^{I},T^{J}) - i~(-1)^{|I|}~\partial_{\mu}N(T^{I},T^{J\mu}) = 0
\nonumber\\
N(T^{I\mu},T^{J}) - (-1)^{|I|}~N(T^{I},T^{J\mu})
\label{sRa}
\eea
We define
\be
N^{I} \equiv N(T^{\emptyset},T^{I})
\ee
and we can use the second relation (\ref{sRa}) to prove (\ref{NI}) by induction. The relation (\ref{sN}) follows immediatly 
from the first relation (\ref{sRa}).
$\qed$

It remains to investigate the solution of equation (\ref{sN}) in the pure Yang-Mills case, with the restrictions
\be
gh(N^{I}) = |I|, \qquad \omega(N^{I}) \leq 4.
\ee
In canonical dimension 
$
\omega = 4
$
we have the tri-linear solution (\ref{Tint}) and one can easily prove that there are no bilinear and quadri-linear solutions. In
canonical dimension
$
\omega = 3
$
there are no solution and, finally, in canonical dimension
$
\omega = 2
$
we find the solution
\bea
N = f_{ab}~\Bigl( {1\over 2}~v_{a}^{\mu}~v_{b\mu} + u_{a}~\tilde{u}_{b}\Bigl)
\nonumber\\
N^{\mu} = f_{ab}~u_{a}~v_{b}^{\mu}
\eea
with 
$
f_{ab}
$
symmetric in 
$
a \leftrightarrow b.
$
However we cannot find 
$
N^{\mu\nu}
$
such that
\be
d_{Q}N^{\mu} = i~d_{\nu}N^{\mu\nu}
\ee
so we remanin only with the tri-linear solution (\ref{Tint}). It follows that we have only one constant arbitrariness in the
second order of the perturbation theory. In fact, this result remains true in all orders of the perturbation theory.

%\newpage
\section{Conclusions}
In further papers we will extend the Hopf structure and the Wick expansion property to the case of the general Yang-Mills model 
(including massive vector Bosons and Dirac fields) and to gravity. An interesting point would be to see if our Hopf version of 
the Wick expansion property is connected with the Feynman graph version.


\newpage
\begin{thebibliography}{99}

\bibitem{ASD} A. Aste, G. Scharf, M. Duetsch, 
``{\it On gauge invariance and spontaneous symmetry breaking}", 
arXiv: 9705216, J. Phys. A: Math. Gen. {\bf 30} (1997) 5785 - 5792

\bibitem{BS}
N. N. Bogoliubov, D. Shirkov,
``{\it Introduction to the Theory of Quantized Fields}",
John Wiley and Sons, 1976 (3rd edition)

\bibitem{BLOT}
N. N. Bogolubov, A. A. Logunov, A.I. Oksak, I. Todorov, 
``{\it General Principles of Quantum Field Theory}", Kluwer 1989

\bibitem{BFFO}
C. Brouder, B. Fauser, A. Frabetti, R. Oeckl,
``{\it Quantum field theory and Hopf algebra cohomology}", 
arXiv:hep-th/0311253, J. Phys. {\bf A 37} (2004) 5895 - 5927

\bibitem{B}
C. Brouder, ``{\it Quantum field theory meets Hopf algebra}",
arXiv: 0611153v3, Math. Nachr. {\bf 282} (2009) 1664 - 1690

\bibitem{D}
M. D\"utsch, ``{\it From Classical Field Theory to Perturbative Quantum Field Theory}", 
Progress in Mathematical Physics {\bf 74}, Springer 2019

\bibitem{DF}
M. D\"utsch, K. Fredenhagen,
``{\it Algebraic Quantum Field Theory, Perturbation Theory,
and the Loop Expansion}",
arXiv: hep-th/0001129, Commun. Math. Phys. {\bf 219} (2001) 5 - 30

\bibitem{DKS} M. Duetsch, F. Krahe, G. Scharf, 
``{\it Scalar QED Revisited}",
Il Nuovo Cimento {\bf 106 A} (3) (1993) 277 - 307

\bibitem{EG}
H. Epstein, V. Glaser,
``{\it The R\^ole of Locality in Perturbation Theory}",
Ann. Inst. H. Poincar\'e {\bf 19 A} (1973) 211-295

\bibitem{Gl}
V. Glaser,
``{\it Electrodynamique Quantique}",
L'enseignement du 3e cycle de la physique en Suisse Romande (CICP), Semestre
d'hiver 1972/73

%\bibitem{cohomology}
%D. R. Grigore, 
%``{\it Cohomological Aspects of Gauge Invariance in the Causal Approach}", 
%arxiv: hep-th/0711.3986, Romanian Journ. Phys. {\bf 55} (2010) 386-438

\bibitem{ward}
D. R. Grigore, ``{\it Ward Identities and Renormalization of General Gauge Theories}",
Journ. Phys. {\bf A 37} (2004) 2803-2834

\bibitem{algebra}
D. R. Grigore,
``{\it A Generalization of Gauge Invariance}", 
arXiv: hep-th/1612.04998, Journal of Mathematical Physics {\bf 58} (2017) 082303

\bibitem{ano-free}
D. R. Grigore, ``{\it Anomaly-Free Gauge Models: A Causal Approach}", 
arXiv: hep-th/1804.08276, Romanian Journ. Phys. {\bf 64} (2019) 102

%\bibitem{third order} Dan Radu Grigore,
%``{\it Third Order Anomalies in the Causal Approach}",
%hep-th/1910.10640, Romanian Journ. Phys. {\bf 65} (2020) 114 

\bibitem{H}
K. Hepp, ``{\it Renormalization Theory}", in ``{\it Statistical Mechanics and Quantum Field Theory}" pp. 429 - 500,
(Les Houches 1970), C. DeWitt-Morette, Raymond Stora (eds.), Gordon and Breach 1971

\bibitem{K1} 
D. Kreimer, ``{\it On the Hopf Algebra Structure of Perturbative Quantum Field Theories}",
Adv. Theor. Math. Phys. {\bf 2} (1998) 303 - 334

\bibitem{K2}
D. Kreimer,
``{\it Locality, QED and Classical Electrodynamics}",
Ann. Phys. (Leipzig) {\bf 7} (1998) 687 - 694

\bibitem{K3}
D. Kreimer,
``{\it Anatomy of a Gauge Theory}",
Annals of Physics {\bf 321} (2006) 2757 - 2781

\bibitem{P}
J. Polchinski, ``{\it Renormalization and Effective Lagrangians}",
Nucl. Phys. {\bf B 231} (1984) 269 - 295

\bibitem{PS}
G. Popineau, R. Stora, 
``{\it A Pedagogical Remark on the Main Theorem of Perturbative Renormalization Theory}", 
Nuclear Physics {\bf B 912} (2016) 70 - 78

\bibitem{S}
M. Salmhofer, ``{\it Renormalization: An Introduction}", (Theoretical and Mathematical Physics) Springer 1999

\bibitem{Sc1}
G. Scharf,
``{\it Finite Quantum Electrodynamics: The Causal Approach}",
(second edition) Springer, 1995; (third edition) Dover, 2014

\bibitem{Sc2}
G. Scharf,
``{\it Quantum Gauge Theories. A True Ghost Story}",
John Wiley, 2001,
``{\it Quantum Gauge Theories - Spin One and Two}",
Google books, 2010
and
``{\it Gauge Field Theories: Spin One and Spin Two, 100 Years After General Relativity}", Dover 2016

\bibitem{Sto1}
R. Stora,
``{\it Lagrangian Field Theory}",
Les Houches lectures, Gordon and Breach, N.Y., 1971, 
C. De Witt, C. Itzykson eds.

\bibitem{St1}
O. Steinmann,
``{\it Perturbation Expansions in Axiomatic Field Theory}",
Lect. Notes in Phys. {\bf 11}, Springer, 1971

\bibitem{Su1}
W. D. van Suijlekom
``{\it The Hopf Algebra of Feynman Graphs in Quantum Electrodynamics}", 
Letters in Mathematical Physics {\bf 77} (2006) 265 - 281

\bibitem{Su2}
W. D. van Suijlekom,
``{\it Renormalization of Gauge Fields: A Hopf Algebra Approach}",
Commun. Math. Phys. {\bf 276} (2007) 773 - 798

%\bibitem{Z-J}
%J. Zinn-Justin, ``{\it Renormalization of Gauge Theories}", in
%``{\it Trends in Elementary Particle Theory}", (International Summer Institute on Theoretical Physics in Bonn 1974), pp. 2 - 39,
%H. Rollnik, K. Dietz (eds.), Springer 1975

\bibitem{WG}
A. S. Wightman, L. G\aa rding,
``{\it Fields as Operator-Valued Distributions in Relativistic Quantum Field
Theory}", Arkiv Fysik {\bf 28} (1965) 129-184

\end{thebibliography}
\end{document}